\documentclass[a4paper,fleqn,usenatbib]{mnras}

\usepackage[T1]{fontenc}
\usepackage{ae,aecompl}

\usepackage{graphicx}	% Including figure files
\usepackage{amsmath}	% Advanced maths commands
\usepackage{amssymb}	% Extra maths symbols

\usepackage{multirow}
\usepackage{longtable}

\newcommand\eg{{\it e.g.} }

\newcommand\ie{{\it i.e.} }

\newcommand{\firsttable}{% no arguments
\begin{table}
\begin{center}

\begin{tabular}{|ccccccc|}
\hline

$\beta$ & $a_{\rm pl}$  & $M_{\rm pl}$        &$I_i$          & Ejected & Accreted\\
        & (au)         & $M_\oplus$         &radians    & &&\\
\hline
0.010 &  1.0 &            1 &   0.0 &            0$\pm$           2 &           17$\pm$           4\\
0.010 &  1.0 &            9 &   0.0 &           21$\pm$           4 &           61$\pm$           7\\
0.010 &  1.0 &           30 &   0.0 &          207$\pm$          14 &           85$\pm$           9\\
0.010 &  1.0 &          100 &   0.0 &         1654$\pm$          40 &          119$\pm$          10\\
0.010 &  1.0 &          317 &   0.0 &         2514$\pm$          50 &           62$\pm$           7\\
0.010 &  1.0 &          634 &   0.0 &         2720$\pm$          52 &           97$\pm$           9\\
0.010 &  1.0 &          951 &   0.0 &         2628$\pm$          51 &           68$\pm$           8\\
0.010 &  1.0 &         1585 &   0.0 &         2760$\pm$          52 &            4$\pm$           2\\
0.010 &  1.0 &         3170 &   0.0 &         2955$\pm$          54 &           18$\pm$           4\\
0.044 &  1.0 &            1 &   0.0 &            0$\pm$           2 &           25$\pm$           5\\
0.044 &  1.0 &            3 &   0.0 &            0$\pm$           2 &           14$\pm$           3\\
0.044 &  1.0 &            9 &   0.0 &            5$\pm$           2 &           24$\pm$           4\\
0.044 &  1.0 &           30 &   0.0 &          223$\pm$          14 &          101$\pm$          10\\
0.044 &  1.0 &          100 &   0.0 &         1157$\pm$          34 &           78$\pm$           8\\
0.044 &  1.0 &          200 &   0.0 &         2212$\pm$          47 &           75$\pm$           8\\
0.044 &  1.0 &          317 &   0.0 &         2655$\pm$          51 &           67$\pm$           8\\
0.044 &  1.0 &          634 &   0.0 &         2754$\pm$          52 &          109$\pm$          10\\
0.044 &  1.0 &          951 &   0.0 &         2665$\pm$          51 &          154$\pm$          12\\
0.044 &  1.0 &         1585 &   0.0 &         2753$\pm$          52 &           52$\pm$           7\\
0.044 &  1.0 &         3170 &   0.0 &         2940$\pm$          54 &           20$\pm$           4\\
0.100 &  1.0 &            1 &   0.0 &            0$\pm$           2 &            8$\pm$           2\\
0.100 &  1.0 &            9 &   0.0 &            5$\pm$           2 &           13$\pm$           3\\
0.100 &  1.0 &           30 &   0.0 &          139$\pm$          11 &           49$\pm$           7\\
0.100 &  1.0 &          100 &   0.0 &         1406$\pm$          37 &           64$\pm$           8\\
0.100 &  1.0 &          317 &   0.0 &         2555$\pm$          50 &           47$\pm$           6\\
0.100 &  1.0 &          634 &   0.0 &         2797$\pm$          52 &           62$\pm$           7\\
0.100 &  1.0 &          951 &   0.0 &         2642$\pm$          51 &          269$\pm$          16\\
0.100 &  1.0 &          951 &   0.0 &         2683$\pm$          51 &          234$\pm$          15\\
0.100 &  1.0 &         1585 &   0.0 &         2737$\pm$          52 &          179$\pm$          13\\
0.100 &  1.0 &         3170 &   0.0 &         2929$\pm$          54 &           59$\pm$           7\\

\hline
\end{tabular}

\caption{The results of the N-body simulations (see \S \ref{sec:simulations}) for {\it low} initial inclinations ($I_i=0$rad) and $M_*=1M_\odot$. }
\label{tab:results_lowinc}

\end{center}
\end{table}}

\newcommand{\secondtable}{% no arguments
\onecolumn
\begin{longtable}{ccc ccc}%{| p{.20\textwidth} | p{.80\textwidth} |} 

\hline

$\beta$ & $a_{\rm pl}$  & $M_{\rm pl}$        &$I_i$          & Ejected & Accreted  \\
        & (au)         & $M_\oplus$         &radians    & &\\
\hline

0.010 &  1.0 &            1 &   0.0 &            0$\pm$           2 &           17$\pm$           4\\
0.010 &  1.0 &            9 &   0.0 &           21$\pm$           4 &           61$\pm$           7\\
0.010 &  1.0 &           30 &   0.0 &          207$\pm$          14 &           85$\pm$           9\\
0.010 &  1.0 &          100 &   0.0 &         1654$\pm$          40 &          119$\pm$          10\\
0.010 &  1.0 &          317 &   0.0 &         2514$\pm$          50 &           62$\pm$           7\\
0.010 &  1.0 &          634 &   0.0 &         2720$\pm$          52 &           97$\pm$           9\\
0.010 &  1.0 &          951 &   0.0 &         2628$\pm$          51 &           68$\pm$           8\\
0.010 &  1.0 &         1585 &   0.0 &         2760$\pm$          52 &            4$\pm$           2\\
0.010 &  1.0 &         3170 &   0.0 &         2955$\pm$          54 &           18$\pm$           4\\

\hline
0.044 &  1.0 &            1 &   0.0 &            0$\pm$           2 &           25$\pm$           5\\
0.044 &  1.0 &            3 &   0.0 &            0$\pm$           2 &           14$\pm$           3\\
0.044 &  1.0 &            9 &   0.0 &            5$\pm$           2 &           24$\pm$           4\\
0.044 &  1.0 &           30 &   0.0 &          223$\pm$          14 &          101$\pm$          10\\
0.044 &  1.0 &          100 &   0.0 &         1157$\pm$          34 &           78$\pm$           8\\
0.044 &  1.0 &          200 &   0.0 &         2212$\pm$          47 &           75$\pm$           8\\
0.044 &  1.0 &          317 &   0.0 &         2655$\pm$          51 &           67$\pm$           8\\
0.044 &  1.0 &          634 &   0.0 &         2754$\pm$          52 &          109$\pm$          10\\
0.044 &  1.0 &          951 &   0.0 &         2665$\pm$          51 &          154$\pm$          12\\
0.044 &  1.0 &         1585 &   0.0 &         2753$\pm$          52 &           52$\pm$           7\\
0.044 &  1.0 &         3170 &   0.0 &         2940$\pm$          54 &           20$\pm$           4\\
\hline

0.100 &  1.0 &            1 &   0.0 &            0$\pm$           2 &            8$\pm$           2\\
0.100 &  1.0 &            9 &   0.0 &            5$\pm$           2 &           13$\pm$           3\\
0.100 &  1.0 &           30 &   0.0 &          139$\pm$          11 &           49$\pm$           7\\
0.100 &  1.0 &          100 &   0.0 &         1406$\pm$          37 &           64$\pm$           8\\
0.100 &  1.0 &          317 &   0.0 &         2555$\pm$          50 &           47$\pm$           6\\
0.100 &  1.0 &          634 &   0.0 &         2797$\pm$          52 &           62$\pm$           7\\
0.100 &  1.0 &          951 &   0.0 &         2642$\pm$          51 &          269$\pm$          16\\
0.100 &  1.0 &          951 &   0.0 &         2683$\pm$          51 &          234$\pm$          15\\
0.100 &  1.0 &         1585 &   0.0 &         2737$\pm$          52 &          179$\pm$          13\\
0.100 &  1.0 &         3170 &   0.0 &         2929$\pm$          54 &           59$\pm$           7\\
\hline
0.100 &  0.1 &          317 &   0.3 &          860$\pm$          29 &          922$\pm$          30\\
0.100 &  0.1 &           30 &   0.3 &            2$\pm$           5 &           89$\pm$           9\\
0.100 &  0.1 &          100 &   0.3 &          126$\pm$          11 &          272$\pm$          16\\
0.100 &  0.1 &          317 &   0.3 &         1320$\pm$          36 &          547$\pm$          23\\
0.100 &  0.1 &          634 &   0.3 &         2011$\pm$          44 &          595$\pm$          24\\
\hline
0.100 &  0.5 &            1 &   0.3 &            0$\pm$           2 &            1$\pm$           3\\
0.100 &  0.5 &           30 &   0.3 &            1$\pm$           3 &           18$\pm$           4\\
0.100 &  0.5 &          100 &   0.3 &          229$\pm$          15 &           91$\pm$           9\\
0.100 &  0.5 &          317 &   0.3 &         1766$\pm$          42 &          100$\pm$          10\\
\hline

0.002 &  1.0 &           30 &   0.3 &          417$\pm$          20 &          243$\pm$          15\\
0.002 &  1.0 &          100 &   0.3 &         2163$\pm$          46 &          114$\pm$          10\\
0.002 &  1.0 &          317 &   0.3 &         2747$\pm$          52 &          112$\pm$          10\\
\hline

0.005 &  1.0 &           30 &   0.3 &          287$\pm$          16 &          177$\pm$          13\\
0.005 &  1.0 &          100 &   0.3 &         1940$\pm$          44 &          131$\pm$          11\\
0.005 &  1.0 &          317 &   0.3 &         2639$\pm$          51 &          132$\pm$          11\\
\hline
0.010 &  1.0 &            1 &   0.3 &            0$\pm$           2 &            0$\pm$           2\\
0.010 &  1.0 &            9 &   0.3 &            7$\pm$           2 &           21$\pm$           4\\
0.010 &  1.0 &           30 &   0.3 &          241$\pm$          15 &          122$\pm$          11\\
0.010 &  1.0 &          100 &   0.3 &         1601$\pm$          40 &          132$\pm$          11\\
0.010 &  1.0 &          200 &   0.3 &         2354$\pm$          48 &          114$\pm$          10\\
0.010 &  1.0 &          317 &   0.3 &         2582$\pm$          50 &          110$\pm$          10$^\dagger$\\
0.010 &  1.0 &          951 &   0.3 &         2764$\pm$          52 &           80$\pm$           8$^\dagger$\\
\hline

0.044 &  1.0 &            1 &   0.3 &            0$\pm$           2 &            0$\pm$           2\\
0.044 &  1.0 &            9 &   0.3 &            0$\pm$           2 &            8$\pm$           2\\
0.044 &  1.0 &           30 &   0.3 &          177$\pm$          13 &           41$\pm$           6\\
0.044 &  1.0 &          100 &   0.3 &         1213$\pm$          34 &           85$\pm$           9\\
0.044 &  1.0 &          200 &   0.3 &         2244$\pm$          47 &           80$\pm$           8\\
0.044 &  1.0 &          317 &   0.3 &         2585$\pm$          50 &           79$\pm$           8\\
0.044 &  1.0 &          951 &   0.3 &         2767$\pm$          52 &           93$\pm$           9\\
0.044 &  1.0 &         1585 &   0.3 &         2818$\pm$          53 &           49$\pm$           7\\
0.044 &  1.0 &         3170 &   0.3 &         3064$\pm$          55 &            3$\pm$           1\\
\hline

0.100 &  1.0 &            1 &   0.3 &            0$\pm$           2 &            0$\pm$           2\\
0.100 &  1.0 &            3 &   0.3 &            0$\pm$           2 &            0$\pm$           2\\
0.100 &  1.0 &            9 &   0.3 &            0$\pm$           2 &            2$\pm$           3\\
0.100 &  1.0 &           30 &   0.3 &          108$\pm$          10 &           11$\pm$           3\\
0.100 &  1.0 &          100 &   0.3 &         1144$\pm$          33 &           36$\pm$           6\\
0.100 &  1.0 &          200 &   0.3 &         2095$\pm$          45 &           49$\pm$           7\\
0.100 &  1.0 &          317 &   0.3 &         2523$\pm$          50 &           56$\pm$           7\\
0.100 &  1.0 &          951 &   0.3 &         2904$\pm$          53 &           53$\pm$           7\\
0.100 &  1.0 &         1585 &   0.3 &         2960$\pm$          54 &           26$\pm$           5\\
\hline
0.400 &  1.0 &            1 &   0.3 &            0$\pm$           2 &            0$\pm$           2\\
0.400 &  1.0 &          100 &   0.3 &           91$\pm$           9 &            9$\pm$           3\\
0.400 &  1.0 &          317 &   0.3 &         1438$\pm$          37 &           16$\pm$           4\\
\hline
0.100 & 10.0 &            1 &   0.3 &            0$\pm$           2 &            0$\pm$           2\\
0.100 & 10.0 &           30 &   0.3 &          634$\pm$          25 &            4$\pm$           2\\
0.100 & 10.0 &          100 &   0.3 &         2038$\pm$          45 &            9$\pm$           3\\
0.100 & 10.0 &          317 &   0.3 &         2916$\pm$          54 &           12$\pm$           3\\
\hline
0.100 & 50.0 &            1 &   0.3 &            1$\pm$           3 &            0$\pm$           2\\
0.100 & 50.0 &           30 &   0.3 &         1070$\pm$          32 &            0$\pm$           2\\
0.100 & 50.0 &          100 &   0.3 &         2516$\pm$          50 &            3$\pm$           1\\

\hline
\\
\\

\caption{The results of the N-body simulations (see \S \ref{sec:simulations}). All simulations were for $M_*=1M_\odot$ and $N=3,100$ particles and all those particles not ejected or accreted made it past the planet, except those marked by $^\dagger$ where a single long-lived particle trapped in the 2:1 resonance exterior to the planet after 500Myr. Errors are $\frac{\sqrt{N}}{N}$ unles $N<3$, in which case Poisson statistics from \citet{Gehrels1986} are used. }
\label{tab:results}

\end{longtable}

\twocolumn
}

\newcommand{\thirdtable}{% no arguments
\begin{table}
\begin{center}

\caption{The results of the N-body simulations (see \S \ref{sec:simulations}) for $a_{\rm pl}=1$au, $\beta=0.1$ and $I_i=0.3$, varying the stellar mass. }
\label{tab:results_mstar}

\begin{tabular}{|ccccccc|}
\hline

$M_*$       & $M_{\rm pl}$ & Ejected & Accreted & Total Particles \\
       $M_{\odot}$ & $M_\oplus$  & &&\\

\hline
0.010&  100&       3100$\pm$          55 &            0$\pm$           0&        3100\\
0.100&   100&      3100$\pm$          55 &            0$\pm$           0&        3100\\
2.000&    100&      121$\pm$          11 &           48$\pm$           6&        3100\\
5.000&     100&       1$\pm$           1 &           33$\pm$           5&        3100\\
\hline
0.010&    1 &     1981$\pm$          44 &            8$\pm$           2&        2310\\
0.100&     1 &     85$\pm$           9 &            0$\pm$           0&        3100\\
2.000&      1 &     0$\pm$           0 &            0$\pm$           0&        3100\\
5.000&        1 &   0$\pm$           0 &            0$\pm$           0&        3100\\

\hline
\end{tabular}

\end{center}
\end{table}}

\newcommand{\highecctable}{% no arguments
\begin{table}
\begin{tabular}{|ccc ccc|}
\hline

$\beta$ & $a_{\rm pl}$  & $M_{\rm pl}$        &$e_i$          & Ejected & Accreted  \\
        & (au)         & $M_\oplus$         &radians    & &\\
\hline

0.100 &            1 &          100 &   0.1 &         1118$\pm$          33 &   
        64$\pm$           8\\
0.100 &            1 &            10 &   0.1 &            0$\pm$           0 &   
         3$\pm$           1\\
0.100 &            1 &            1 &   0.1 &            0$\pm$           0 &   
         0$\pm$           0\\
0.100 &            1 &          317 &   0.1 &         2517$\pm$          50 &   
        80$\pm$           8\\

\hline
0.100 &            1 &          100 &   0.4 &         1074$\pm$          32 &   
        77$\pm$           8\\
0.100 &            1 &            10 &   0.4 &            2$\pm$           1 &   
         4$\pm$           2\\
0.100 &            1 &            1 &   0.4 &            0$\pm$           0 &   
         0$\pm$           0\\
0.100 &            1 &          317 &   0.4 &         2428$\pm$          49 &   
       109$\pm$          10\\

\hline
0.100 &            1 &          100 &   0.5 &         1231$\pm$          35 &   
        71$\pm$           8\\
0.100 &            1 &            10 &   0.5&            6$\pm$           2 &   
         1$\pm$           1\\
0.100 &            1 &            1 &   0.5 &            0$\pm$           0 &   
         0$\pm$           0\\
0.100 &            1 &          317 &   0.5 &         1358$\pm$          36 &   
        41$\pm$           6\\

\hline

\\
\\

\end{tabular}

\caption{The results of a sub-set of the N-body simulations (see \S
  \ref{sec:simulations}), where initial eccentricites of the particles
  are varied. }
\label{tab:highecc}

\end{table}
}

\title[ ]{Using warm dust to constrain unseen planets}

\author[A. Bonsor et al.]{\parbox{\textwidth}{Amy Bonsor$^{1}$\thanks{E-mail: abonsor@ast.cam.ac.uk}, Mark C. Wyatt$^{1}$, Quentin Kral$^{1}$, Grant Kennedy$^{2,3}$, Andrew Shannon$^{4,5}$ and Steve Ertel$^6$}\vspace{0.4cm}\\
\parbox{\textwidth}{
$^{1}$Institute of Astronomy, University of Cambridge, Madingley Road, Cambridge, CB3 0HA, UK\\
$^{2}$Department of Physics, University of Warwick, Gibbet Hill Road, Coventry, CV4 7AL, UK\\
$^{3}$Centre for Exoplanets and Habitability, University of Warwick, Gibbet Hill Road, Coventry, CV4 7AL, UK\\
$^{4}$Department of Astronomy $\&$~Astrophysics, The Pennsylvania State University, State College, PA, USA\\
$^{5}$Center for Exoplanets and Habitable Worlds, The Pennsylvania State University, State College, PA, USA\\
$^6$Steward Observatory, Department of Astronomy, University of Arizona, 993 N. Cherry Ave, Tucson, AZ 85721, USA
}}

\date{Accepted XXX. Received YYY; in original form ZZZ}

\pubyear{2015}

\begin{document}
\label{firstpage}
\pagerange{\pageref{firstpage}--\pageref{lastpage}}
\maketitle

\begin{abstract}

Cold outer debris belts orbit a significant fraction of stars, many of which are planet-hosts. Radiative forces from the star lead to dust particles leaving the outer belts and spiralling inwards under Poynting-Robertson drag. We present an empirical model fitted to N-body simulations that allows the fate of these dust particles when they encounter a planet to be rapidly calculated. High mass planets eject most particles, whilst dust passes low mass planets relatively unperturbed. Close-in, high mass planets (hot Jupiters) are best at accreting dust. The model predicts the accretion rate of dust onto planets interior to debris belts, with mass accretions rates of up to hundreds of kilograms per second predicted for hot Jupiters interior to outer debris belts, when collisional evolution is also taken into account. The model can be used to infer the presence and likely masses of as yet undetected planets in systems with outer belts. The non-detection of warm dust with the Large Binocular Telescope Interferometer (LBTI) around Vega could be explained by the presence of a single Saturn mass planet, or a chain of lower mass planets. Similarly, the detection of warm dust in such systems implies the absence of planets above a quantifiable level, which can be lower than similar limits from direct imaging. The level of dust detected with LBTI around $\beta$ Leo can be used to rule out the presence of planets more massive than a few Saturn masses outside of $\sim$5au.

\end{abstract}


\section{Introduction}

Many nearby stars have dusty analogues to our Solar System's asteroid and Kuiper belt, observed in the infrared (see review by \citealt{wyattreview, Matthews2016}). We
observe small dust, which we know must be continuously replenished by collisions
between larger parent bodies, as it has a short lifetime against collisions and radiative forces. Radiative forces from the star are strong
enough to place the smallest grains on unbound or weakly bound orbits, as observed by the large halos of debris systems
such as Vega \citep{vegasu05,Sibthorpe2010}, HR 4796 \citep{Schneider2018} or HR 8799 \citep{Matthews2014}. Small dust grains that are large
enough not to be blown out of the system can have their orbital
velocities reduced by radiative forces, such that they gradually
spiral inwards towards the star, under Poynting-Robertson drag
(PR-drag). This phenomena is well understood theoretically
\citep[\eg][]{burns}, and has long been considered critical to the evolution of dust grains in the inner Solar System \citep[\eg][]{Grun2001, Mann2006}. 

% review Mann2006 Grun 2001

Dust spirals inwards from all outer debris belts due to PR-drag, but is generally depleted by mutual collisions before migrating far from its source \citep{wyattnopr, vanlieshout2014, KennedyPiette2015}. Such a
dust population has been suggested as an explanation for the
mid-infrared excesses, resulting from warm dust, typically at $\sim1$au around sun-like stars, found around some
stars that also have far-infrared excesses from cold outer dust belts, typically at tens of au
\citep{Eps_hotdust, Mennesson2014, KennedyPiette2015}. Although suffering
from a small number of detections, there is already a statistically
significant link between mid and far-infrared excesses
\citep{Mennesson2014, Ertel2018}. Observations with the Large Binocular Telescope Interferometer (LBTI) find a 60\% occurrence rate for mid-infrared excesses in systems with cold, outer dust belts, compared to 8\% in systems without far-infrared detections \citep{Ertel2018}.  Even hotter dust is also observed closer in around some
main-sequence stars in the near-infrared using interferometry, with instruments such as
VLTI/PIONIER or CHARA/FLUOR \citep{Ertel2014, Absil2013}.  The link between such
hot dust and cold, outer debris belts is less clear \citep{Ertel2014} and an
explanation for this hot dust remains elusive \citep{Kral_exozodireview}, although a mixture of scattering by comets \citep{bonsor_exozodi, bonsor_pdm}, a coupling of PR-drag and pile-up at the sublimation radius \citep{Kobayashi2009, vanlieshout2014} and/or trapping in magnetic fields has been suggested \citep{Rieke2016}.

 A growing number of planets are known to orbit interior to cold,
 outer debris belts \citep[\eg][]{Marshall2014}. These planets can have a significant influence on the population of dust in the inner planetary systems. Planets can eject or accrete dust. The Earth receives a significant flux of meteoroids, many of which originate in the asteroid belt and have spiralled inwards under PR-drag \citep{Mann2004}. Planets interior to debris discs may receive a similar flow of material, and their influence on the atmospheric dynamics of these planets is unknown. Characterisation of the
 atmospheres of many close-in, massive planets have revealed the presence of dust or haze, which most likely is linked to internal atmospheric evolution, but could potentially have an external origin \citep{Madhusudhan2016}. 

%%%%%% ADD MORE MOTIVATION FOR ejection lack of warm dust etc (see Mark's comments) 

Following the evolution of dust particles from an outer debris belt to the inner regions of a planetary system is a complex problem, particularly in multi-planet systems. It is possible to make detailed models for our Solar System. These track the dynamical evolution of dust grains leaving the Kuiper belt \citep[\eg][]{LiouZookDermott1996}, or known comets \citep[\eg][]{Yang2018}, using N-body simulations to track their interactions with the planets, taking into account the influence of non-gravitational forces, including radiative forces or stellar wind drag. Such simulations are computationally intensive, particularly for the massive grains that migrate the slowest, but contain the most mass. Collisional evolution is even harder to account for and, generally, is only considered using a statistical approach, which does not allow for consideration of interactions with planets \citep[\eg][]{Eps_hotdust}. Models that couple dynamics and collisions are computationally expensive to run \citep[\eg][]{colgrooming09, Kral2013}. Whilst it may be feasible to simulate individual systems, N-body simulations for the wide range of parameter space available to exoplanets would take a prohibitively long time. Instead this work aims to provide an alternate, fast to calculate, empirical means of calculating the fate of dust
particles leaving a debris belt due to PR-drag. This enables it to be applied to the vast range of parameter space
probed by exo-planetary systems.

The empirical fate of dust will be assessed using a simple analytic model, compared to the results of more computationally intensive N-body simulations. This will be used to calculate how much dust is present in inner planetary systems, including the Solar System and how much dust is accreted by planets. \cite{Moromartin2005} performed
N-body simulations for a similar problem, but focussing on the dust
leaving the system, mainly ejected by planets. Their simulations had insufficient particle
numbers to trace accretion onto planets. In this paper, we perform
N-body simulations including sufficient particles to trace accretion
by planets interior to debris belts, as well as ejection, as described in \S\ref{sec:simulations}. We compare
the results of these simulations to a simple analytic model, and
present an empirical method to predict the fraction of particles that
approach a planet migrating due to PR-drag that are accreted or
ejected by the planet in \S\ref{sec:model} and \S\ref{sec:applications}. This model is then used to  make predictions for the mass accretion rates onto planets interior to outer debris belts (\S\ref{sec:macc}), to the Solar System (\S\ref{sec:ss}) and to predict the levels of dust in the inner regions of debris disc systems (\S\ref{sec:inner}), in relation to any planets that may orbit in these systems. We focus on two systems (Vega and $\beta$ Leo) where LBTI observations provide important constraints on any planetary companions.  Our conclusions are summarised in \S\ref{sec:conclusions}.

\section{Numerical Simulations}

\label{sec:simulations}

\begin{figure}
\includegraphics[width=0.48\textwidth]{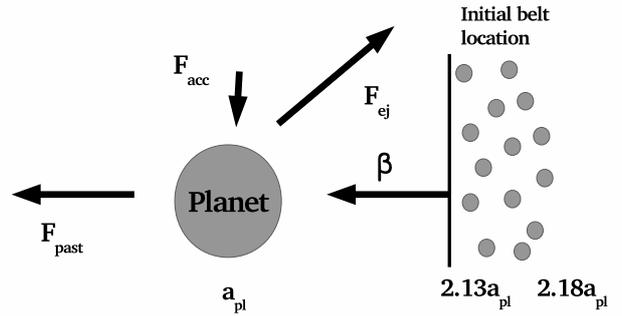}

\caption{A schematic diagram of our approach, indicating the location of the initial belt in the N-body simulations which track the fraction of particles ejected by the planet ($F_{\rm ej}$), the fraction accreted by the planet ($F_{\rm acc} $) and the fraction that migrate past the planet, going on to hit the star ($F_{\rm past}$), as described in \S\ref{sec:simulations}.   }
\label{fig:di}
\end{figure}

Numerical simulations are used to track the fate of particles leaving a debris belt and migrating inwards under Poynting-Robertson drag (PR-drag). We use the N-body code {\it Mercury}
\citep{Chambers99}, with the addition of migration due to PR-drag and radiation pressure \citep{Shannon2015}. We use the {\it hybrid} integrator, which switches between the {\it sympletic} and {\it Burlisch-Stoer} integrators for close encounters. A single planet
orbits interior to the dust belt, on a circular orbit, with semi-major
axis, $a_{\rm pl}$ and mass, $M_{\rm pl}$, around a star of mass
$M_\star=1M_\odot$, as shown on the cartoon in Fig.~\ref{fig:di}. A belt of test particles start exterior to the planet. In order to speed up the simulations, we do not need to track the particle's evolution from the outer belt all the way in to the planet, instead we start the particles at between $2.13a_{\rm pl}$ and $2.18a_{\rm pl}$, outside the planet's 3:1 resonance, where most dynamical interactions with the planet start. The particles migrate inwards at a rate specified by the ratio of the forces due to stellar radiation pressure to gravity, $\beta$, where: 
\begin{equation}
\beta=\frac{3 L_\star Q_{PR} }{8 \pi G M_\star \rho c D}
\label{eq:beta}
\end{equation}
is the ratio of the force due to radiation pressure to the
gravitational force on a particle of diameter $D$, density, $\rho$,
around a star of luminosity, $L_\star$, and mass, $M_\star$. $Q_{PR}$ is
the radiation pressure efficiency factor, assumed to be 1 in this work, a valid approximation resulting from geometric optics for grains larger than $\sim0.1\mu$m, $G$ and $c$ are the gravitational constant and the speed
of light, respectively. All variables are listed in Table~\ref{tab:params}.

The test particles all have low initial eccentricities of $e_i=0.01$, either {\it low} or {\it moderate} initial inclinations, with $I_i=0.03^\circ$ and $I_i=17^\circ$, and initial arguments of pericentre, longitudes of
ascending node and free, true anomalies that are randomly selected.  Low eccentricities when particles interact with the planet are likely given that significant migration under PR-drag will circularise orbits, although we note here that for the highest values of $\beta$ (fastest migration) there may not be sufficient time for orbits to be circularised. A timestep of $8 \left(\frac{a_{\rm pl}}{au}\right)^{3/2}$ days is used. 

The aim of the simulations is to track particles accreted by the planet. We, therefore, require that we have sufficient particles to resolve an accretion fraction of 0.3\% to $3\sigma$, which assuming Poisson statistics and $\sigma=\sqrt{N}/{N}$, requires at least 3,100 particles.
The simulations were run until all particles have either been ejected, accreted by the planet or hit the central star. For each simulation we track the fraction ejected ($F_{\rm ej}$), the fraction accreted by the planet ($F_{\rm acc} $) and the fraction that migrate past the planet and go on to hit the star ($F_{\rm past}$). The inner radius down to which the orbits of dust grains are followed is fixed at $a_{\rm pl}/10$ in order to speed up the simulations. This is sufficient that particles are no longer under the influence of the planet, and unlikely to change their fate. The planet density is set to $\rho_{pl}=5.52$g cm$^{-3}$ (Earth) for
$M_{pl}<30$M$_\oplus$, \ie rocky planets and $\rho_{pl}=1.33$g cm$^{-3}$
(Jupiter) for $M_{pl} \ge 30$M$_\oplus$ \ie gas giants. The influence of
changing the planet density is small.

\begin{figure}
\includegraphics[width=0.48\textwidth]{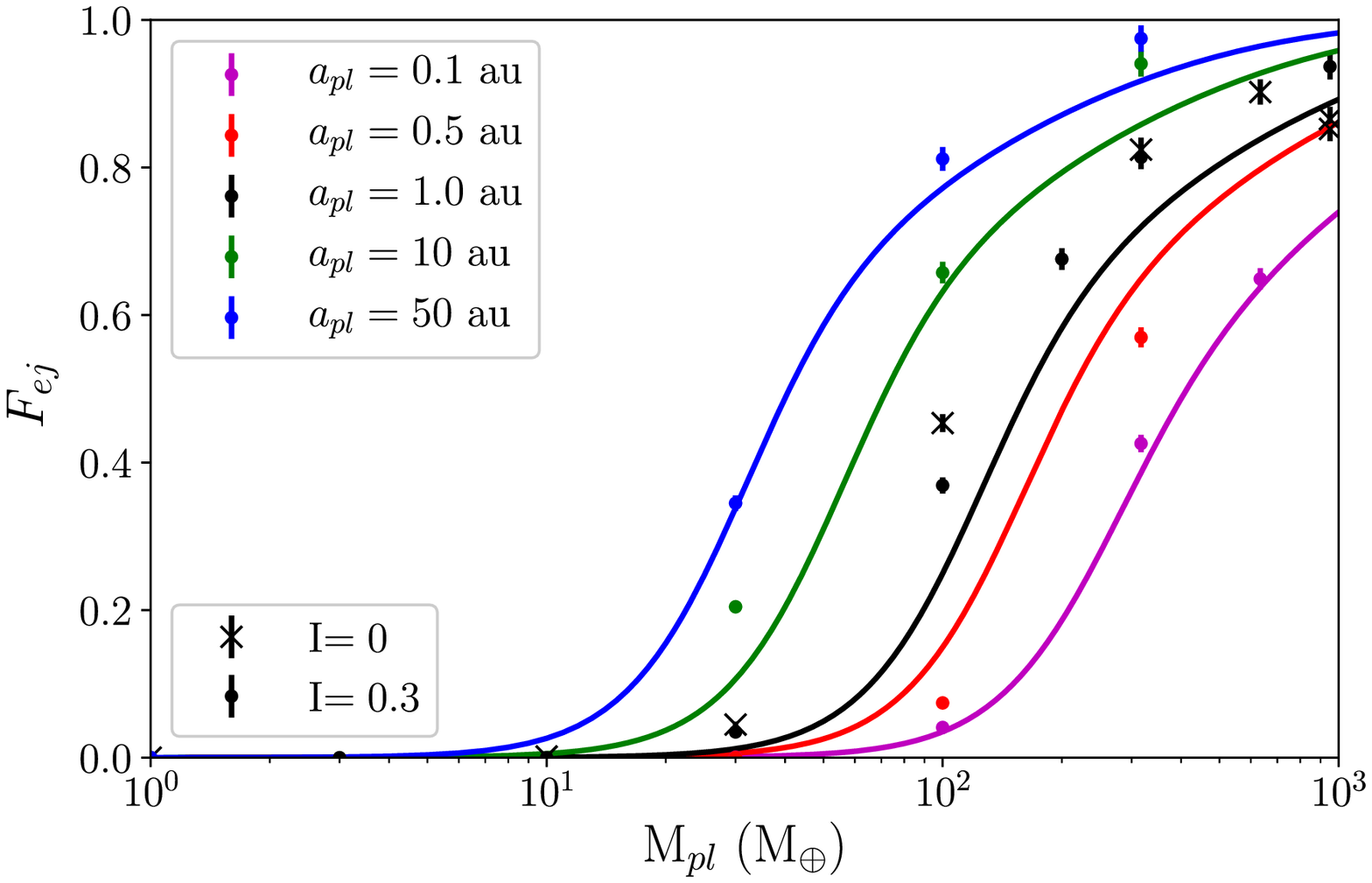}
\includegraphics[width=0.48\textwidth]{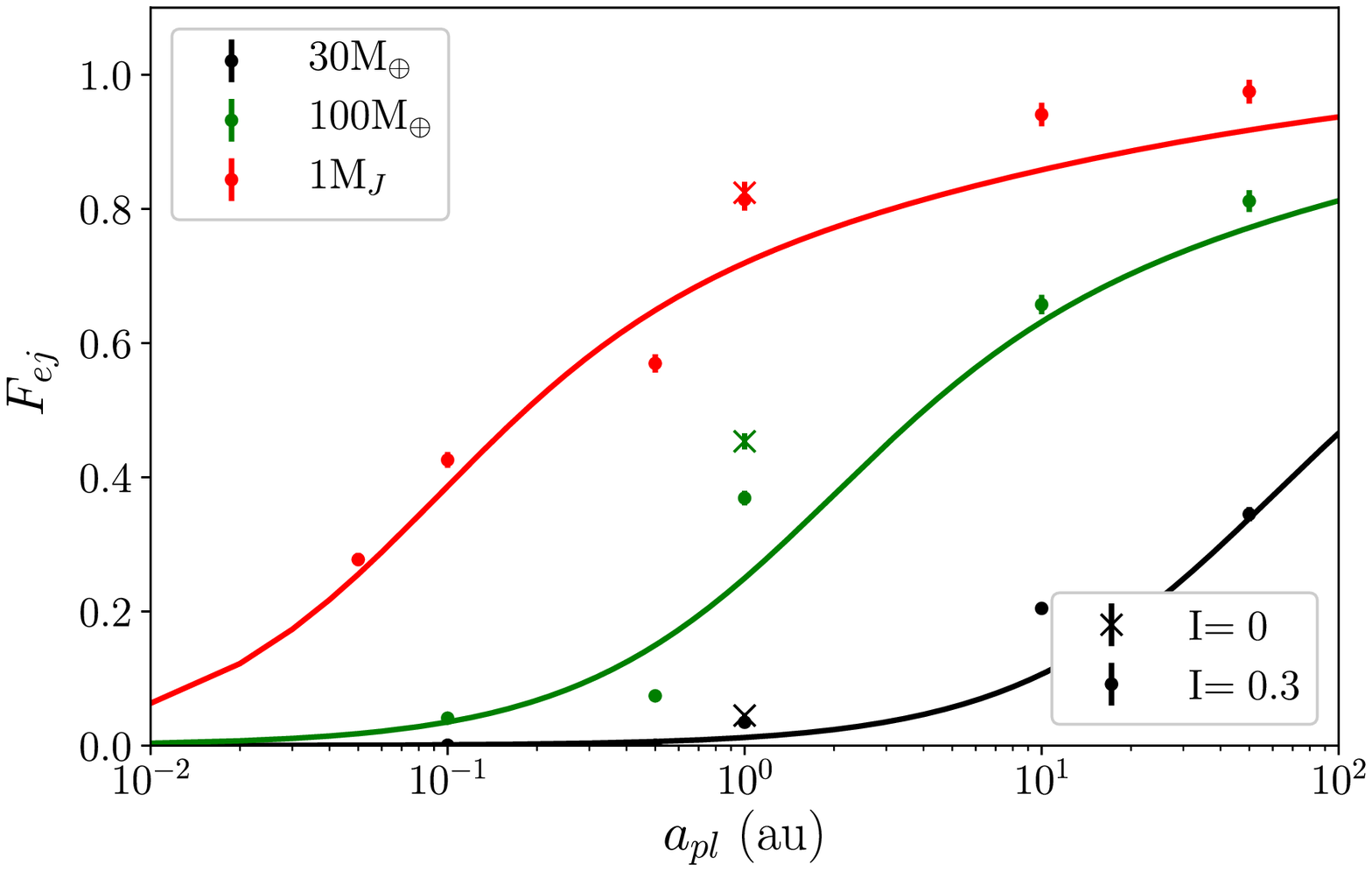}
\includegraphics[width=0.48\textwidth]{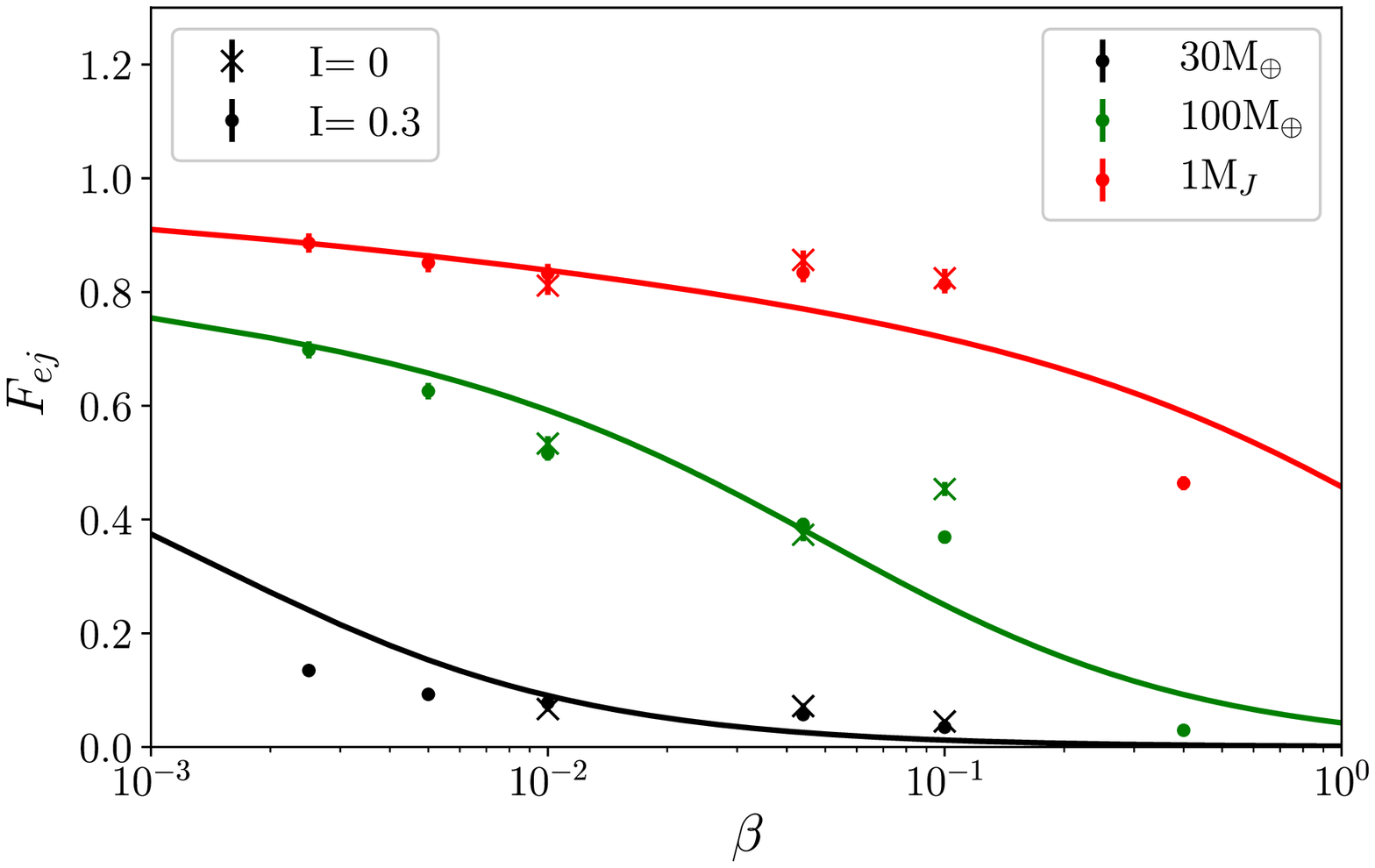}
\caption{The results of the numerical simulations showing the fraction of particles ejected as a function of planet mass, for $\beta=0.1$ (top panel), planet semi-major axis, for $\beta=0.1$ (middle panel) and particle size, for $a_{\rm pl}=1$au ($\beta$ Eq.~\ref{eq:beta} bottom panel). The solid lines show a fit to the results of the form Eq. ~\ref{eq:rej}~\ref{eq:fej_epsilon}, using the best-fit parameters in Table~\ref{tab:bestfit}. Error bars are $1\sigma$ , where $\sigma= \sqrt{N_{\rm ej}}/N$. }
\label{fig:bestfit}
\end{figure}

\begin{figure}
\includegraphics[width=0.48\textwidth]{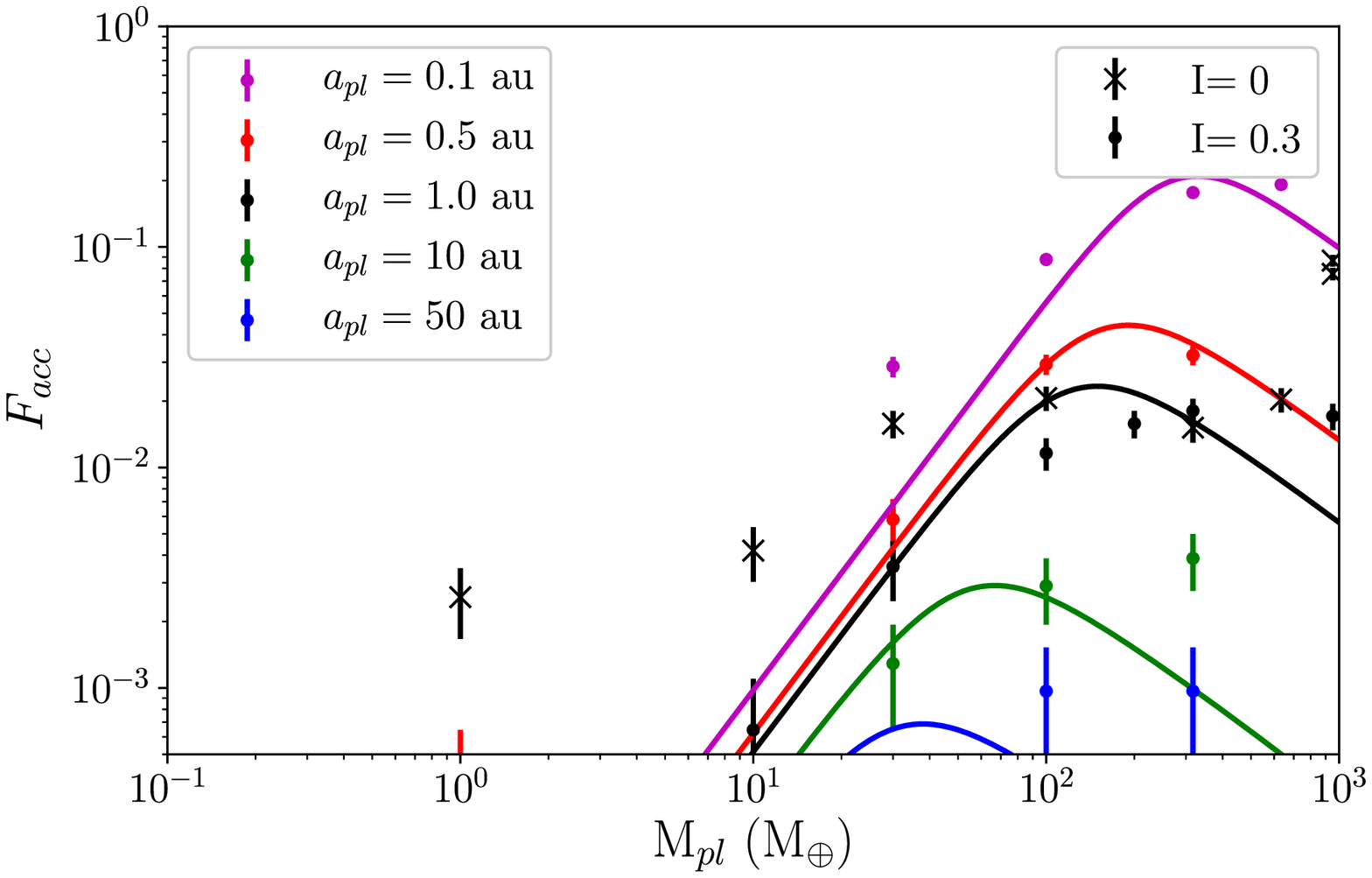}
\includegraphics[width=0.48\textwidth]{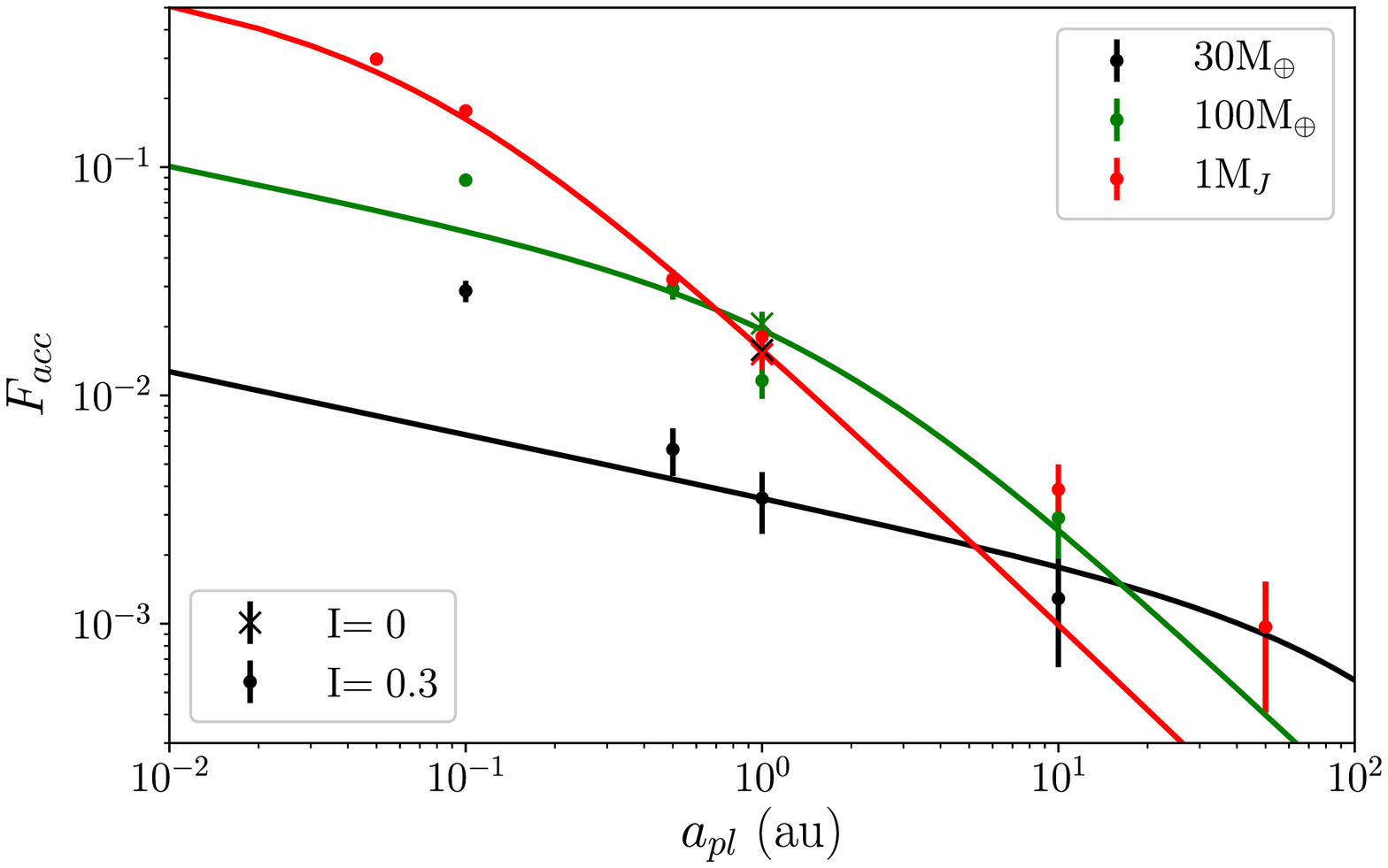}
\includegraphics[width=0.48\textwidth]{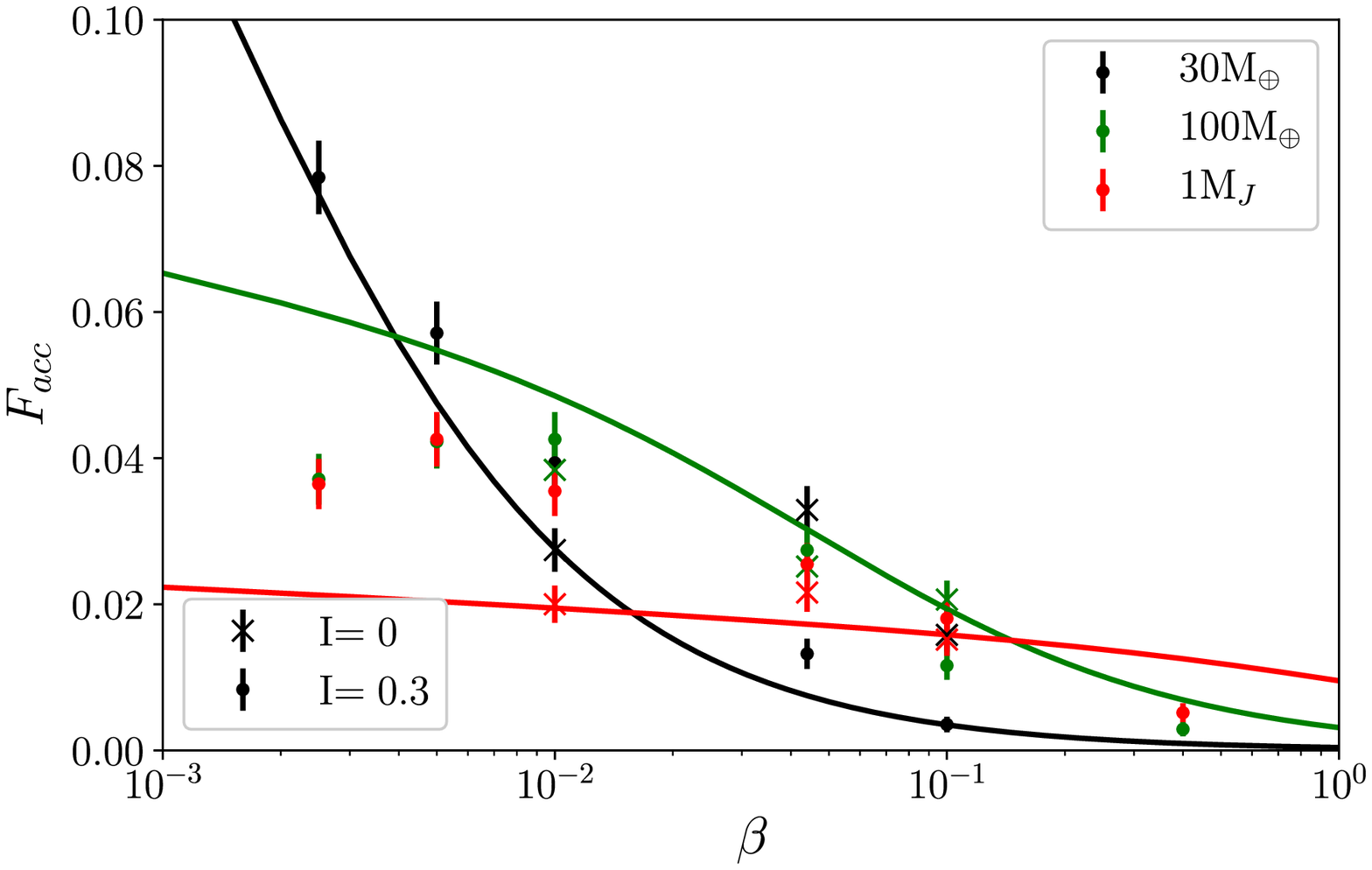}
\caption{The results of the numerical simulations showing the fraction of particles accreted as a function of planet mass, for $\beta=0.1$  (top panel), planet semi-major axis, for $\beta=0.1$  (middle panel) and particle size, for $a_{\rm pl}=1$au ($\beta$ Eq.~\ref{eq:beta} bottom panel). The solid lines show a fit to the results of the form Eq.~\ref{eq:racc} or Eq.~\ref{eq:facc_epsilon} using the best-fit parameters in Table~\ref{tab:bestfit}. Error bars are $1\sigma$ , where $\sigma= \sqrt{N_{\rm acc}}/N$. }%Error bars show Poisson distributed upper and lower limits for small particl ~\ref{eq:fej_epsilon} e numbers, taken from \citet{Gehrels1986}. }
\label{fig:bestfit_acc}
\end{figure}

A range of simulations were run varying the planet properties ($a_{\rm pl}$ and $M_{\rm pl}$) and the migration rate ($\beta$). For a sub-set of the simulations, the stellar mass, $M_\star$ and inclination, $I_i$, were also changed. The results of all simulations are summarised in Tables~\ref{tab:results}, ~\ref{tab:results_mstar}, and Figs~\ref{fig:bestfit}, ~\ref{fig:bestfit_acc}. The ejection rate is seen to increase steeply with planet mass, as seen by \cite{Moromartin2005}, ranging from no ejections to almost all particles ejected, for example for Earth mass to Jupiter mass planets at 10au with $\beta=0.1$. The same range in ejection rate is seen when varying semi-major axis at fixed planet mass, \eg for $100M_\oplus$ no particles are ejected at 0.1au and almost all particles are ejected at 100au. \cite{Moromartin2005} only saw an almost flat trend with semi-major axis, as they focussed on higher planet masses and higher semi-major axes, where the ejection rate remains close to 1. The ejection rate falls off weakly for smaller particles (higher $\beta$), in a similar manner to that seen by \cite{Moromartin2005}.  Fig~\ref{fig:bestfit_acc} shows that the accretion rate is almost always lower than the ejection rate, increasing only up to a maximum of about 20\% in these simulations. Accretion rates are highest for the highest mass planets, that are closest to the star, and accretion rates decrease with increasing $\beta$.

%{\bf Mention that particle initial inclinations are not as important as other parameters, and particle initial eccentricities are wiped out by resonant trapping}

%\firsttable

\section{The fate of particles that encounter a planet}

\label{sec:model}
\subsection{The Model}
The aim of this work is to provide an empirical means to predict the fraction of particles that spiral inwards under PR-drag towards a planet that are accreted, ejected and pass the planet, going on to hit the star, if no further planets are present. We base these predictions on the following simple analytic model. %Initial tests, however, found that the simple analytic model alone does not reproduce the simulation results well. Instead, we, therefore, choose to parameterise the analytic model in terms of its dependence on the key properties of the planet, mass ($M_{\rm pl}$), semi-major axis ($a_{\rm pl}$) and the size of the particles. 
As in the numerical simulations, we only consider planets on circular orbits.

Consider the number of particles passing the planet, $N(t)$, to be reduced by both ejections and accretions at rates $R_{\rm ej}$ and $R_{\rm acc}$ per particle, respectively, \ie where $1/R_{\rm acc}$ is the mean time for any given particle to be accreted if it remained on its current orbit. If the initial number of particles is $N_0$, and both rates are constant throughout the time the particles interact with the planet, and there are no further loss mechanisms, then: 
\begin{eqnarray}
\dot{N} &=& -R_{\rm acc} N - R_{\rm ej} N \\
N(t) &=& N_0\, e^{-(R_{\rm acc}+ R_{\rm ej})t}.
\end{eqnarray}

The total number of particles ejected ($N_{\rm ej}$) can then be calculated by integrating the rate of ejections over the time that the particle remains interacting with the planet, $\Delta t$, such that $N_{\rm ej} = \int^{\Delta t}_0 R_{ej}\, N(t)\, dt$, and the fraction ejected is: 
\begin{equation}
F_{\rm ej} = \frac{N_{\rm ej}}{N_0}=  \frac{R_{\rm ej}}{(R_{\rm acc} + R_{\rm ej})} \left( 1- e^{-(R_{\rm ej} + R_{\rm acc}) \Delta t} \right).
\label{eq:fej}
\end{equation}
In a similar manner, the number accreted, $N_{\rm acc} = \int^{\Delta t}_0 R_{\rm acc}\, N(t) \, dt$, and the fraction accreted: 
\begin{equation}
F_{\rm acc}  = \frac{N_{\rm acc}}{N_0}=  \frac{R_{\rm acc}}{(R_{\rm acc} + R_{\rm ej})} \left( 1- e^{-(R_{\rm ej} + R_{\rm acc}) \Delta t} \right).
\label{eq:facc}
\end{equation}

The time that the particle remains under the potential influence of with the planet, $\Delta t$, is taken to be the time to traverse the planet-crossing region, migration from $a = a_{\rm pl}/(1-e)$ to $a=a_{\rm pl}/(1+e)$, where $a$, $e$ are the orbital parameters of the particles, and $a_{\rm pl}$ is the semi-major axis of the planet.  We now make the assumption that the particles are on almost circular orbits whilst migrating past the planet, which is a reasonable assumption as in general their eccentricity will have decayed following migration by PR-drag, except where $\beta$ is high or the particles have migrated insufficient distance. We note here that the validity of this approximation may break down due to resonant interactions (see later).  For almost circular orbits, the particle's semi-major axis decays under PR-drag as  \citep{Wyatt1950, burns, Shannon2015}:
\begin{equation}
\dot{a_{\rm PR}}\approx  -1.25 \,  \left(\frac{M_\star}{M_\odot}\right)\left( \frac{a}{{\rm au}}\right)^{-1}\,  \beta \; + \mathcal{O}(e^2) \; \;  {\rm au \; kyr}^{-1}.
\label{eq:apr}
\end{equation}

Thus, the time to traverse the planet's orbit is given by: 
\begin{equation}
\label{eq:deltat}
 \Delta t \approx  \frac{1602}{\beta} \left(\frac{M_\star}{M_\odot}\right)^{-1}
\left( \frac{a_{pl}}{\rm au }\right)^2 e + O(e^3) \; \; \; \; {\rm yr}.
\end{equation}

The rate at which a particle interacts with the planet with a sufficiently small impact parameter such that it is accreted is given by: 
\begin{equation}
R_{\rm acc} = n \, \pi b_{\rm acc}^2 \, v_{\rm rel},
\end{equation} 
where $b_{\rm acc}$ is the impact parameter required for accretion, $v_{\rm rel}$ the relative velocity between the particle and the planet, and $n$ the density of colliders, given by $1/V_{\rm tor}$, where the volume of the torus ($V_{\rm tor}$) occupied by particles with semi-major axis, $a$, eccentricity, $e$, inclination, $I$ and randomly distributed orbital elements is given by \citep{Sykes1990}
\begin{equation}
n= \frac{1}{V_{ \rm tor}}=\frac{1}{8 \pi a^3 e \sin I} + \mathcal{O}(e^2).
\label{eq:vtor}
\end{equation}

The relative velocity between the planet and the particle, assuming that the particle is on an approximately circular orbit with $r=a_{\rm pl}$, can be calculated by considering the velocity of a planet on a circular orbit, $v_{\rm K}^2 = \frac{\mu }{a_{\rm pl}}$, where $\mu = G\, M_\star$, and the velocity of a particle on a circular orbit that experiences gravity reduced by a factor $(1-\beta)$: 
\begin{equation}
v_{\rm pp}^2 = v_{\rm K}^2(1-\beta), 
\end{equation}
where the particle has approximately the planet's semi-major axis at the point of interaction. The planet's orbit is inclined by $I$ relative to the particle, such that the planet's velocity is given by: 
\begin{equation}
{\bf v_{\rm K} } =  \begin{pmatrix} 0 \\v_{{\rm pl}}\cos I  \\  v_{{\rm pl}}\sin I  \end{pmatrix}
\end{equation}
and the particle's velocity: 
\begin{equation}
{\bf v_{\rm pp} } =   \begin{pmatrix} 0 \\  v_{{\rm pp}} \\ 0 \end{pmatrix},\end{equation}
then,
\begin{equation} v_{\rm rel}^2 = v_{\rm K}^2 ( 2- \beta -2 \sqrt{1-\beta} \cos I ).
\label{eq:vrel_appendix}
\end{equation}

If $\beta=0$ this reduces to the standard expression in terms of the Tisserand parameter, $T$, $\frac{v_{\rm rel}^2}{v_{\rm K}^2}=3-T = 2 (1-\cos I)$.

The impact parameter for accretion is given by the planet's radius ($R_{\rm pl}$) multiplied by a gravitational focussing factor, such that  
\begin{equation}
b_{\rm acc}=R_{\rm pl} \sqrt{1+ \frac{v_{\rm esc}^2}{v_{\rm rel}^2}},
\label{eq:bacc}
\end{equation}
where $v_{\rm esc}= \sqrt{\frac{2GM_{\rm pl}}{R_{\rm pl}}}$ is the planet's escape velocity. For  particles on nearly circular orbits, $v_{\rm rel} \ll v_{\rm  esc}$, such that:
\begin{eqnarray}
\label{eq:bacc1}
&b_{\rm acc} \approx  \frac{R_{\rm pl} \, v_{\rm esc}}{v_{\rm rel}}
=1.5\times 10^{10}m \times \\\nonumber
&\left(\frac{M_{\rm pl}}{M_\odot}\right) ^{2/3}\left(\frac{M_\star}{M_\odot}\right)^{-1/2}  \left(\frac{\rho_J}{\rho_{\rm pl}}\right)^{1/6} \left( \frac{a_{\rm pl}}{\rm au}\right)^{1/2}\left(\frac{v_{\rm K}}{v_{\rm rel} }\right). 
\end{eqnarray} 
%%&=&\left (\frac{6 }{\pi}\right)^{1/6} \left(\frac{ G\,au^{1/2} M_\odot^{7/6}}{\rho_j^{1/6}}\right)  \\
 All these expressions together lead to an accretion rate proportional to: 
\begin{equation}
R_{\rm acc} \Delta t =\propto K_{\rm acc} M_{\rm pl}^{4/3} a_{\rm pl} ^{-1/2} M_\star^{-3/2} \beta^{-1},
\label{eq:racc_analytic}
\end{equation}
where the constant of proportionality, $K_{\rm acc}$, depends only on the particle's orbital parameters when it interacts with the planet, the planet's density and is approximately independent of particle size ($\beta$). This simple model suggests a weak dependence on $\beta$ resulting from the $v_{\rm rel}$ term in Eqs.~\ref{eq:vrel_appendix}~\ref{eq:bacc1}, which we ignore. In order to fit the simulation results we assume the form: 
\begin{equation}
R_{\rm acc} \Delta t =K_{\rm acc} M_{\rm pl}^{\alpha_a} a_{\rm pl} ^{\gamma_a} M_\star^{\delta_a} \beta^{\eta_a}.
\label{eq:racc}
\end{equation}

The dependence of $R_{\rm acc}$ on planet mass, semi-major axis, particle size ($\beta$) and stellar mass are parametrised in terms of four parameters $\alpha_a$, $\gamma_a$, $\eta_a$ and $\delta_a$, which will be determined empirically from fitting the simulation results. Eq.~\ref{eq:racc_analytic} shows analytic predictions for their values, which are used to fix $\delta_a = -3/2$ as insufficient simulations were made to explore this parameter fully.

The rate of ejections can be determined in a similar manner. In this calculation, the cross-sectional area for ejections is given by $\pi (b_{\rm ej}^2- b_{\rm acc}^2)$, where $b_{\rm ej}$ is the impact parameter for ejections. Whilst the impact parameter for ejection depends on the orientation of the interaction, here we assume that ejection occurs if the change in the particle's velocity due to the interaction $\Delta v > (\sqrt{2} + 1) v_{K}$.  

Using Rutherford scattering to estimate $\Delta v$ gives 
\begin{equation}
\left(\frac{b_{\rm ej}}{a_{\rm pl}} \right)= \left(\frac{v_{\rm K, pl}}{v_{\rm rel}}\right)^2\left(\frac{M_{\rm pl}}{M_\star}\right)\left(\frac{4 v_{\rm rel}^2}{\Delta v^2}-1\right)^{1/2}, 
\label{eq:bej}
\end{equation}
which is independent of $\beta$ for $\beta\ll 1$, but again a weak dependence on $\beta$ occurs due to the $v_{\rm rel}$ term, which becomes important for large $\beta$.

Combining these gives: 
\begin{eqnarray}
R_{\rm ej} \Delta t &=& n \, \pi (b_{\rm ej}^2 -b_{\rm acc}^2) v_{\rm rel}  \Delta t \nonumber \\
&\approx&K_{\rm ej} a_{\rm pl}^{1/2} M_{\rm pl}^2 M_\star^{-5/2} \beta^{-1}\nonumber \\ &&-K_{\rm acc}M_{\rm pl}^{4/3} a_{\rm pl}^{-1/2} M_\star^{-3/2} \beta^{-1}, 
\label{eq:rej_analytic}
\end{eqnarray}
which we force to be always positive and model as:
\begin{equation}
R_{\rm ej} \Delta t =K_{\rm ej} M_{\rm pl}^{\alpha_e} a_{\rm pl} ^{\gamma_e} M_\star^{\delta_e} \beta^{\eta_e} - K_{\rm acc} M_{\rm pl}^{\alpha_a} a_{\rm pl} ^{\gamma_a} M_\star^{\delta_a} \beta^{\eta_a}.
\label{eq:rej}
\end{equation}
Again, the four constants, $K_{\rm ej}$, $\alpha_e$, $\gamma_e$,  $\eta_e$ will be determined empirically from fitting the simulation results, whilst the stellar mass dependence, $\delta_e$, is taken to be $-5/2$ from the analytics.

There are a number of reasons why this simple analytic model may not give a perfect match to the simulation results and an empirical model is required. For example, particles are scattered multiple times by the planet, and particles may become trapped in resonance prior to interacting with the planet, both of which lead to higher particle eccentricities and inclinations at interaction. In fact, analytic predictions suggest that trapping in the exterior 2:1 mean motion resonance is almost 100\% efficient for planets more massive than a few Earth masses \citep{Shannon2015} for $\beta = 0.1$. Particles trapped in the 2:1 resonance evolve to eccentricities of around $(\frac{1}{5})^{1/2}$ before leaving the resonance. These factors are accounted for by allowing $\alpha$, $\gamma$ and $\eta$ to vary from the analytic predictions. 

Another factor to consider is the migration of particles scattered interior to the planet, which can quickly leave the influence of the planet, resulting in shorter interaction times than stated in Eq.~\ref{eq:deltat}. A significant number of particles are scattered inwards and migrate out of reach of the planet when the ejection rate is high. This tends to occur at high planet masses. Thus, to incorporate this in the empirical model, we add an additional parameter, $\epsilon$, which reduces the interaction timescale, such that Eq.~\ref{eq:fej} and Eq.~\ref{eq:facc} become: 

\begin{eqnarray}
\label{eq:fej_epsilon}
F_{\rm ej} &=& \frac{R_{\rm ej}}{(R_{\rm acc} + R_{\rm ej})} \left( 1- e^{-\frac{(R_{\rm ej} + R_{\rm acc})\Delta t}{(1+R_{\rm ej}\Delta t)^\epsilon}} \right),\\
F_{\rm acc} & =& \frac{R_{\rm acc}}{(R_{\rm acc} + R_{\rm ej})} \left( 1- e^{-\frac{(R_{\rm ej} + R_{\rm acc}) \Delta t}{(1+R_{\rm ej}\Delta t)^\epsilon}} \right).
\label{eq:facc_epsilon}
\end{eqnarray}

In order to determine the values of the free parameters, we use a Markov chain Monte Carlo method (MCMC) to maximise the likelihood, using the {\it emcee} package of \cite{emcee}, assuming a normal distribution, with errors on the number of particles ejected or accreted, given by $\sigma_{\rm ej}({\bf k})=\frac{N_{\rm ej}({\bf k})^{1/2}}{N_{\rm 0}}$ and $\sigma_{\rm acc}({\bf k})=\frac{N_{\rm acc}({\bf k})^{1/2}}{N_{\rm 0}}$, where $N_{\rm ej}$ and $N_{\rm acc}$ are the total number of particles ejected or accreted during the simulation and ${\bf k}$ labels the set of simulation parameters ($M_{\rm pl}$, $a_{\rm pl}$ and $\beta$) used. The likelihood function is given by  
\begin{eqnarray}
ln \;  \mathcal{L} = -2\Sigma_{{\bf k}}\left(\frac{( F_{\rm ej}({\bf k})^{\rm model}-F_{\rm ej}({\bf k})^{\rm sims})^2}{\sigma_{\rm ej}({\bf k})^2}\right) \nonumber \\ -2\Sigma_{{\bf k}}\left( \frac{( F_{\rm acc}({\bf k})^{\rm model}-F_{\rm acc}({\bf k})^{\rm sims})^2}{\sigma_{\rm acc}({\bf k})^2}\right),
\label{eq:likelihood}
\end{eqnarray}
where $F_{\rm ej}^{\rm model}$ and $F_{\rm acc}^{\rm model}$ are the fraction of particles accreted and ejected in the model, derived from Eq.~\ref{eq:fej_epsilon},~\ref{eq:facc_epsilon}, Eq.~\ref{eq:facc} and Eq.~\ref{eq:fej}, and depend on the 9 free parameters, $K_{\rm acc}$, $\alpha_a$, $\gamma_a$,  $\eta_a$, $K_{\rm ej}$, $\alpha_e$, $\gamma_e$,  $\eta_e$ and $\epsilon$. $F_{\rm ej}^{\rm sim}$ and $F_{\rm acc}^{\rm sim}$ are the number of particles ejected and accreted in the N-body simulations. Uniform priors are assumed for all free parameters. % range?! 

Both the particle's initial eccentricity ($e_i$) and initial inclination ($I_i$) have the potential to influence the ability of planets to eject or accrete particles. Our best-fit solution is determined based on a set of fiducial simulations in which $I_i=0.3$ and $e_i=0.01$, however, we made a few simple tests to show that these results are actually valid over a range of initial inclinations and initial eccentricities. This is because most particles are influenced by outer resonances with the planet before interacting and in fact, many particles are influenced by either the 2:1 mean-motion resonance or eccentricity-inclination resonances exterior to the planet, such that
their inclinations and eccentricities evolve to similar values, irrespective of the
initial values, before they interact with the planet. Simulations with $I_i=0$ were
performed for a sub-set of simulations with $a_{\rm pl}=1$ au and $\beta =0.1$ (see
Table~\ref{tab:results} or Fig.~\ref{fig:bestfit} and Fig.~\ref{fig:bestfit_acc}), and the difference between the
fraction of particles ejected or accreted in the simulations compared to the empirical model was always less than
10\%. In a similar manner, simulations with $a_{\rm pl}=1$ au, $\beta =0.1$ and $I_i=0.3$ were performed for $e_i=0.01$, $e_i=0.1$ and $e_i=0.4$ (see Table~\ref{tab:highecc} and Fig.~\ref{fig:highecc}), and the difference between the
fraction of particles ejected or accreted between the models was always less than
5\%. Very different behaviour was seen if eccentricities were increased above $e_i>0.4$, which given this limited set of simulations suggests that the model may be valid up to eccentricities of around 0.4 as for such high eccentricities, \ie above the maximum found in the 2:1 resonance, trapping probabilities and the ability of outer resonances to influence the particle's behaviour can be significantly different, and we deem that the empirical model presented here is no longer valid in this regime.

\begin{figure*}
\includegraphics[width=0.8\textwidth]{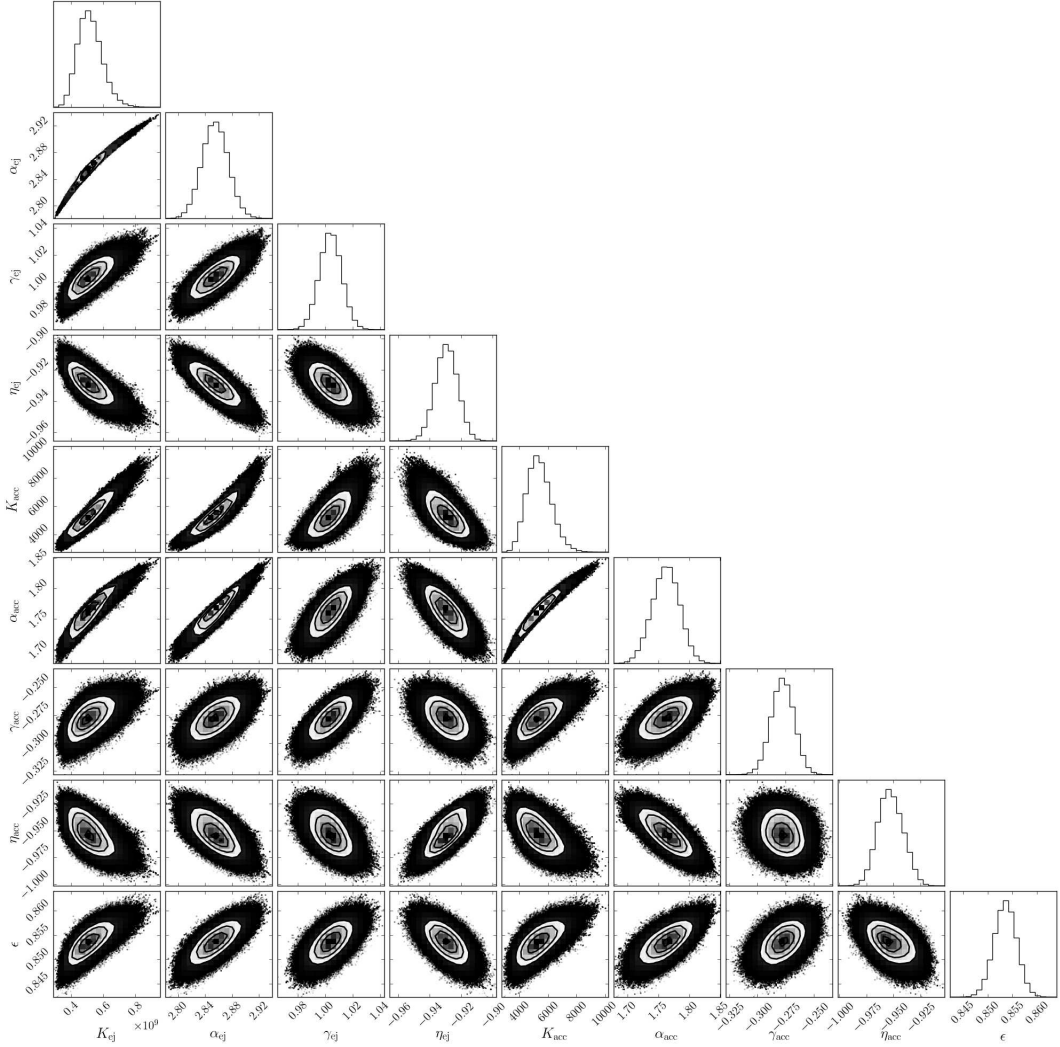}

\caption{The posterior probability distributions of each parameter in the empirical fit to the ejections and accretions seen in the N-body simulations (Eq.~\ref{eq:fej}, ~\ref{eq:rej}, ~\ref{eq:racc})), calculated by maximising the likelihood (Eq.~\ref{eq:likelihood}). Plot created using {\it corner} from \citet{corner}.  }
\label{fig:fej_corner}
\end{figure*}

The posterior probability distribution of each parameter in the fit is shown in Fig.~\ref{fig:fej_corner}. Almost all walkers converge to a single best-fit solution, although we note that as the model is limited and unable to fit the data perfectly, alternative solutions may be equally valid. Our best fit parameters are listed in Table~\ref{tab:bestfit}, and the best-fit solutions are plotted as a function of planet mass, semi-major axis and particle size ($\beta$) in Figs~\ref{fig:bestfit} and ~\ref{fig:bestfit_acc}.

\subsection{Comparison between the model and the simulation results}

Fig.~\ref{fig:bestfit} shows the fraction of particles ejected as a function of planet mass (top), semi-major axis (middle) and particle size or $\beta$ (bottom). Solid lines show the best-fit model, with parameters listed in Table~\ref{tab:bestfit}, whilst the individual data points show simulation results. As discussed briefly in \S\ref{sec:simulations}, the fraction of particles ejected increases with planet mass, which is explained by the analytics as being because larger planets can more readily impart a sufficiently large kick to eject particles. The model does a good job of reproducing the form of this behaviour, with the best-fit exponent, $\alpha_{\rm ej}$ varying by a small amount from the analytic prediction in order to achieve this (see Table~\ref{tab:bestfit}). The parameter $\epsilon$ is critical in achieving the fit at large planet masses, where the fraction of particles ejected would otherwise tend to one. This is because some particles are scattered by long range interactions that are not quite sufficient to eject them, but can place them on orbits from where they quickly migrate inwards, out of reach of further interactions with the planet. Thus, the fraction of particles ejected is reduced due to the inclusion of the $\epsilon$ parameter for large ejection rates $R_{\rm ej}$.

The fraction of particles ejected increases with semi-major axis, which the analytics show is predominantly because the timescale over which the particles interact with the planet increases (Eq.~\ref{eq:deltat}). Again the model produces a good fit to the observations, with the best-fit exponent, $\gamma_{\rm ej}$ differing from the analytic prediction by a factor of $\sim 2$ (see Table~\ref{tab:bestfit}). The fraction of particles ejected decreases for small particles (large $\beta$) that migrate quickly out of the region where they can interact with the planet. Our empirical model deviates slightly from the N-body simulations for large values of $\beta$. This can partly be attributed to an oversimplification in the model, which ignores a significant $\beta$ dependence that is more complex than a power-law, and contributes at high $\beta$.

Fig.~\ref{fig:bestfit_acc} shows the fraction of particles accreted by a planet as a function of planet mass (top), semi-major axis (middle) and particle size or $\beta$ (bottom). In general, as noted in \S\ref{sec:simulations}, higher mass planets are better at accreting particles. However, for the largest planets, ejection becomes the dominant outcome (see top panel Fig.~\ref{fig:bestfit}) and particles do not survive long enough to be accreted. Our model exhibits this behaviour due to the competition between the $R_{\rm acc}$ and $R_{\rm ej}$ terms in Eq.~\ref{eq:fej_epsilon},~\ref{eq:facc_epsilon}. At low planet mass, the model reduces to the behaviour demonstrated in Eq. B2 of \cite{Wyatt99}, where accretion increases strongly with planet mass. However, for the lowest mass planets, \eg Earth mass planets at 1au, insufficient particles were included to follow accretion rates in detail. Such planets have lower probabilities to trap particles in outer resonances \citep{Shannon2015} and thus the fate of particles is influenced more strongly by their initial parameters, as can be seen in Fig.~\ref{fig:bestfit_acc} by the difference in accretion rates for the $I=0$ and $I=0.3$ simulations for low mass planets.

%% The empirical fit suffers from the small number statistics and the low number of particles accreted by planets, particularly at low planet masses or large semi-major axes. The N-body simulations indicate that the initial particle inclination may play a role, particularly with low initial inclinations leading to higher accretion for low mass planets. 

The fraction of particles accreted by the planet decreases with the planet's semi-major axis (middle panel Fig.~\ref{fig:bestfit_acc}), which is explained by the analytics as the volume of the torus occupied by the particles (Eq.~\ref{eq:vtor}) increases faster with semi-major axis than the interaction timescale (Eq.~\ref{eq:deltat}) and impact parameters (Eq.~\ref{eq:bacc}).  The best-fit model describes this behaviour successfully. The fraction of particles accreted by the planet decreases with $\beta$, as smaller particles migrate faster past the planet (bottom panel Fig.~\ref{fig:bestfit_acc}). The model does a reasonable job of fitting the dependence on $\beta$ (particle size), although it is clear that this dependence is not strong and displays complexity beyond this simple model. This is expected, as the model (Eq.~\ref{eq:rej}, ~\ref{eq:racc},~\ref{eq:fej_epsilon},~\ref{eq:facc_epsilon}) misses out the complex dependence on $\beta$ of the $\frac{v_{\rm rel}}{v_{\rm K}}$ term. Nonetheless, we deem that the model can make satisfactory predictions regarding the fraction of particles accreted. For ejections the difference between the model predictions and the simulation results is always always less than a factor 2, for accretions it is always less than a factor of 3, for planet masses higher than $10M_\oplus$.

\begin{table}
\begin{tabular}{cc cc}
\hline
Parameter & Dependence & Analytic & Numerical \\
\hline
$K_{\rm ej}$ && &$5.14 \times 10^7$  \\

$\alpha_{\rm ej}$ &$M_{\rm pl}$& 2&$2.85$ \\
$\gamma_{\rm ej}$ &$a_{\rm pl}$&1/2 & $1.00$\\

$ \eta_{\rm ej}$ &$\beta$& -1&$-0.93$ \\

$ \delta_{\rm ej}$ &$M_\star$& -5/2&\\

\hline

$K_{\rm acc}$ && &$5350^{+900}_{-780}$  \\

$\alpha_{\rm acc}$& $M_{\rm pl}$& 4/3&$1.76$ \\

$\gamma_{\rm acc}$ &$a_{\rm pl}$&-1/2 & $-0.28$\\

$ \eta_{\rm acc}$ &$\beta$& -1&$-0.95$ \\

$ \delta_{\rm acc}$ &$M_\star$& -3/2&\\

\hline
$\epsilon$& & 0  & $0.85$ \\
\hline
\end{tabular}
\caption{ The analytic and numerical values of the constants in predicting the fraction of particles ejected of the form Eq.~\ref{eq:fej_epsilon},~\ref{eq:facc_epsilon}, Eq.~\ref{eq:racc} and Eq.~\ref{eq:rej}.  }

\label{tab:bestfit}
\end{table}

\begin{figure*}
\includegraphics[width=0.48\textwidth]{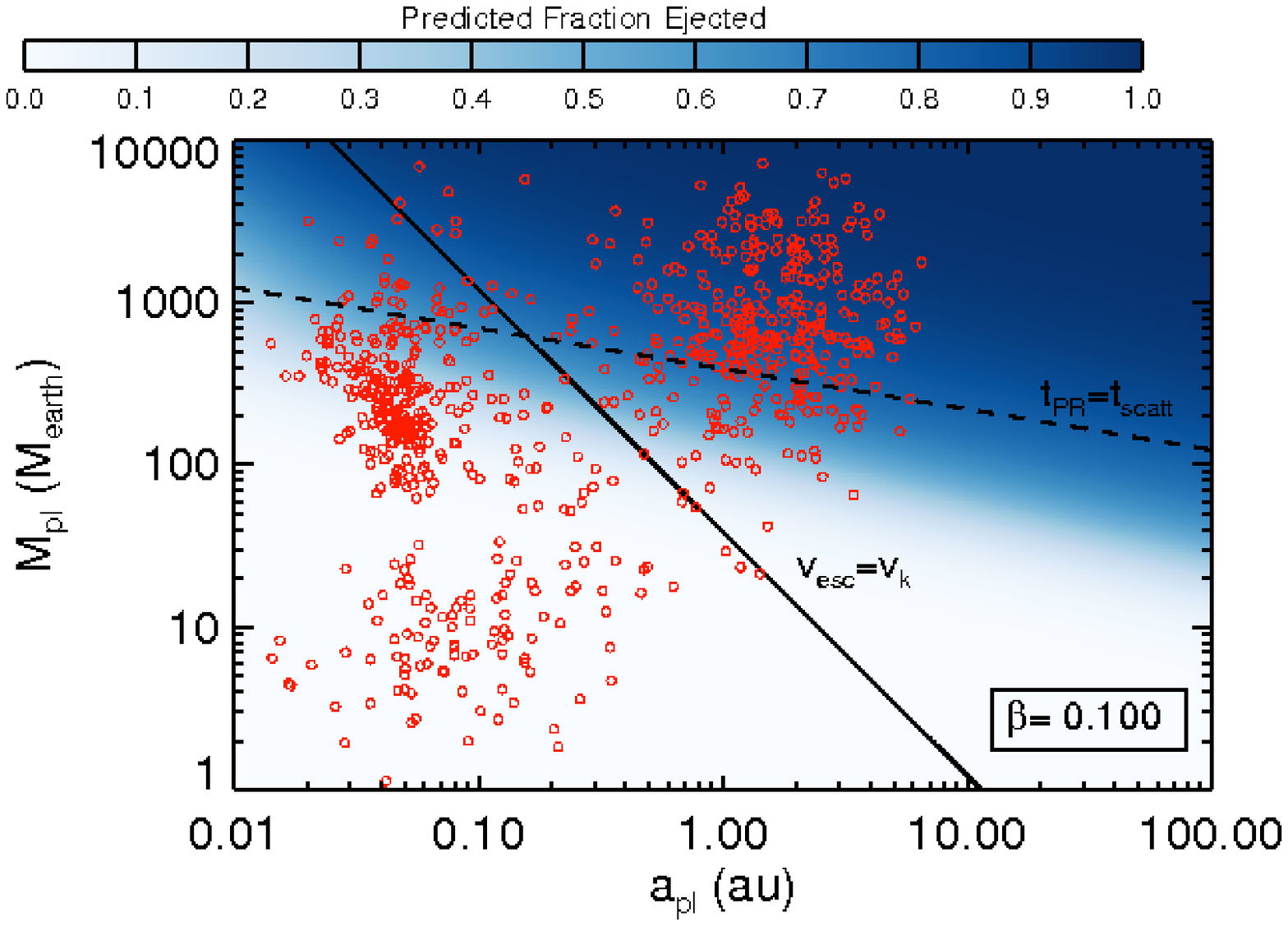}

\includegraphics[width=0.48\textwidth]{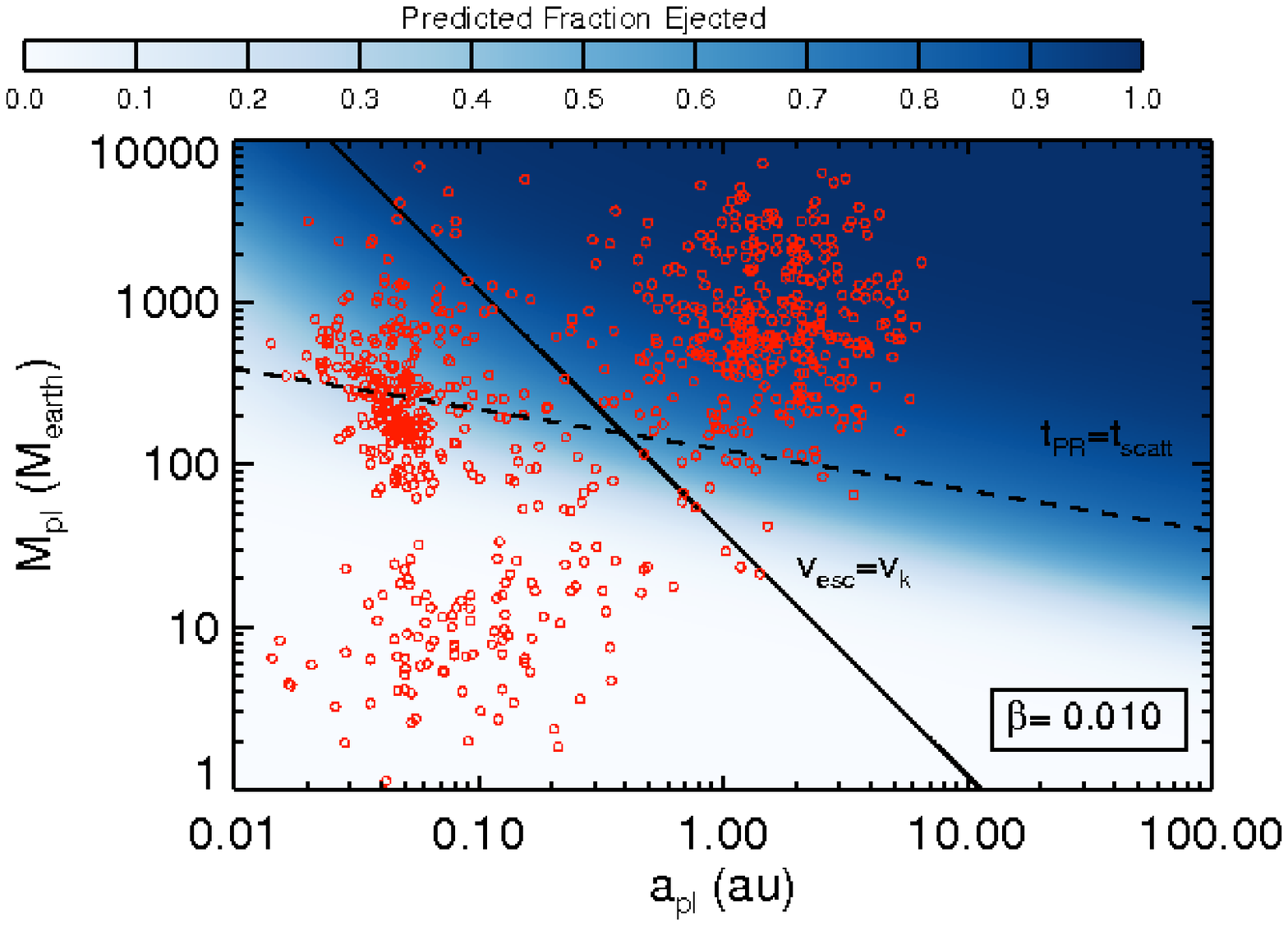}
\includegraphics[width=0.48\textwidth]{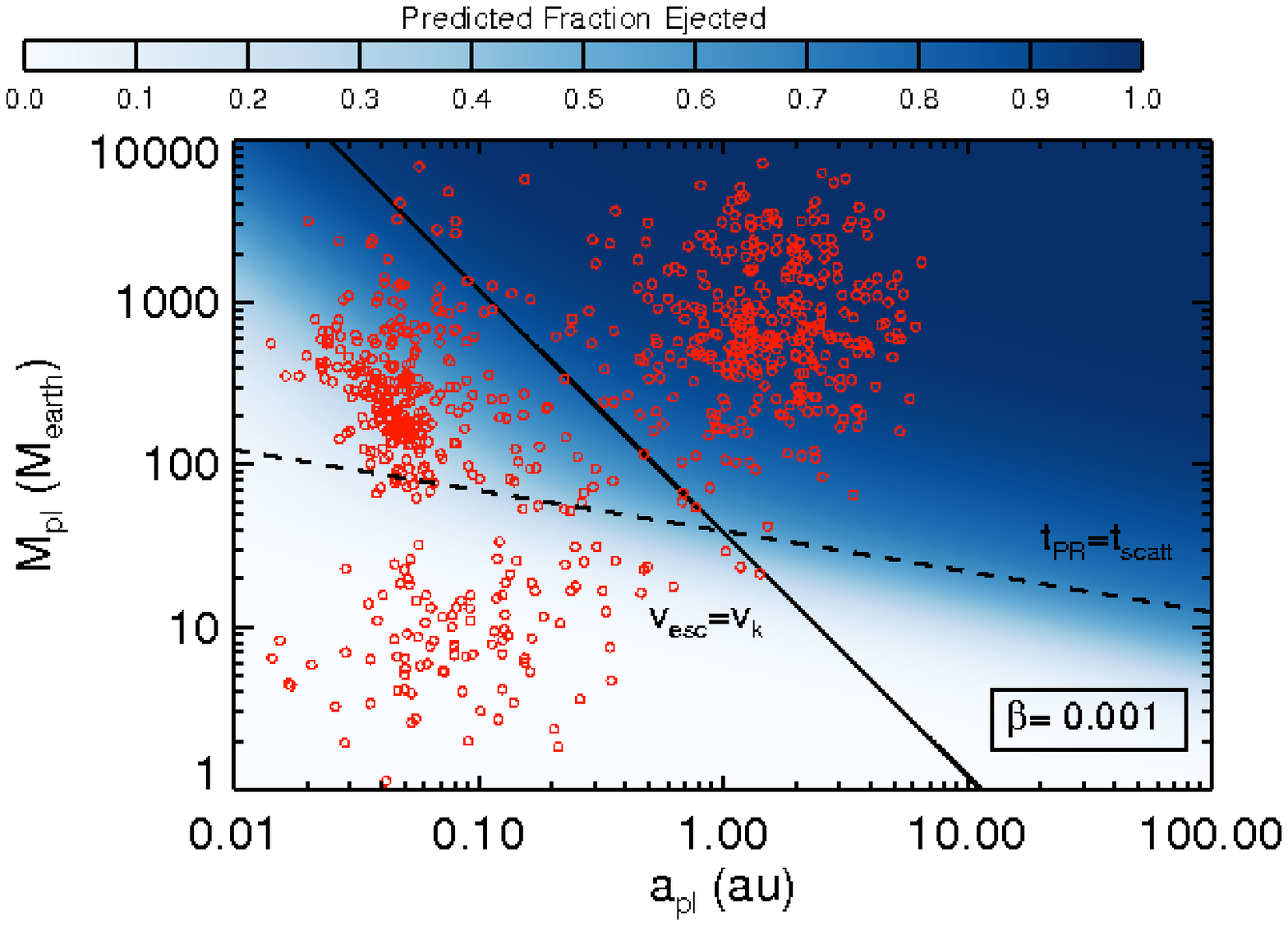}

\caption{Predictions for the fraction of particles ejected by planets, as a function of the planet mass and semi-major axis, for $\beta=0.1,0.01,0.001$, based on the best-fit empirical model, calculated using Eqs.~\ref{eq:racc}~\ref{eq:rej}~\ref{eq:fej_epsilon},~\ref{eq:facc_epsilon} and parameters from Table~\ref{tab:bestfit}. The solid line shows $v_{K} = v_{\rm esc}$ and the dashed line shows where the time for the particles to migrate past the planet by PR-drag (Eq.~\ref{eq:deltat}) is equal to the time for particles to be ejected (Eq.~\ref{eq:mplequal}).   }
\label{fig:apl_mpl}
\end{figure*}

\begin{figure*}

\includegraphics[width=0.48\textwidth]{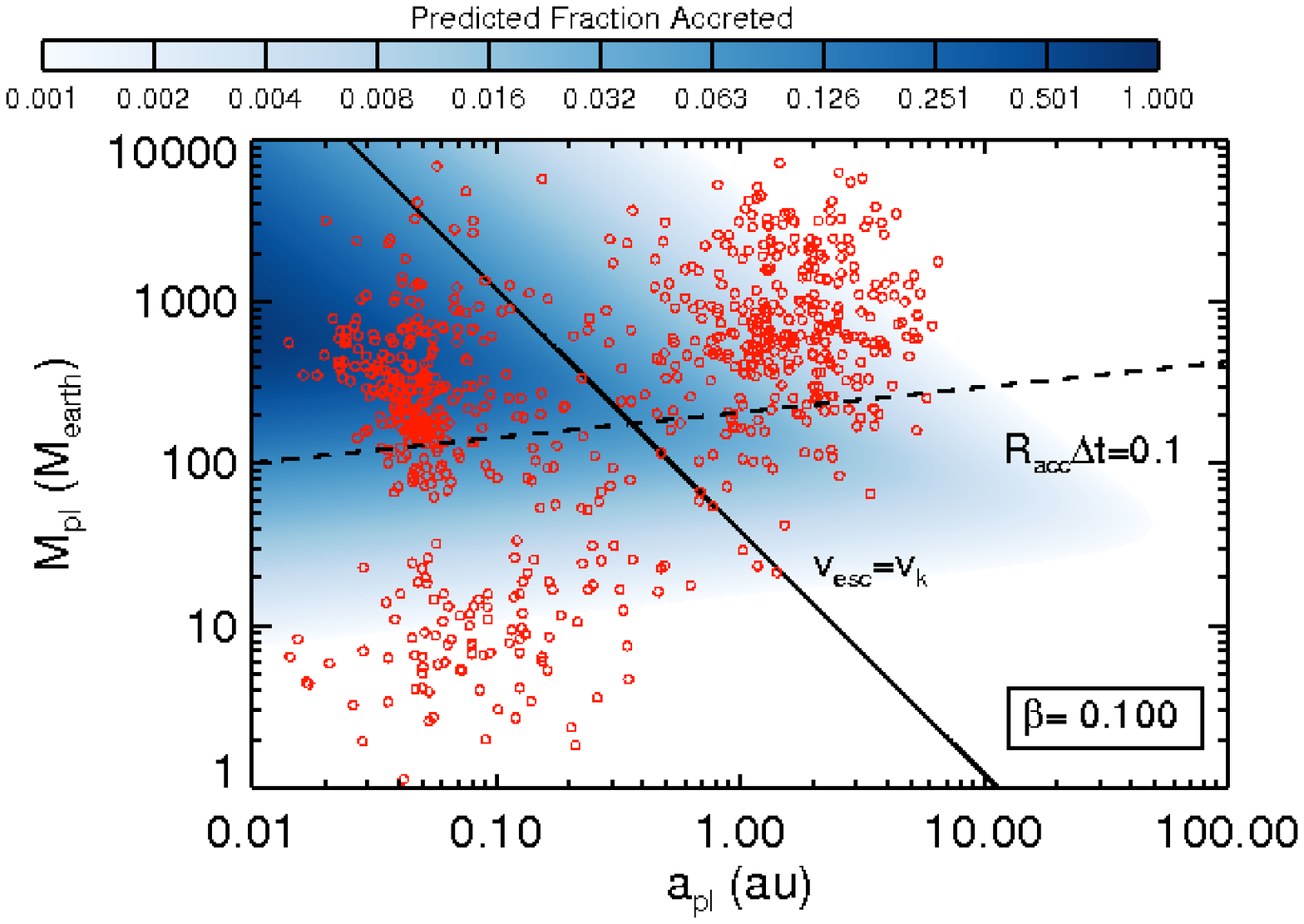}

\includegraphics[width=0.48\textwidth]{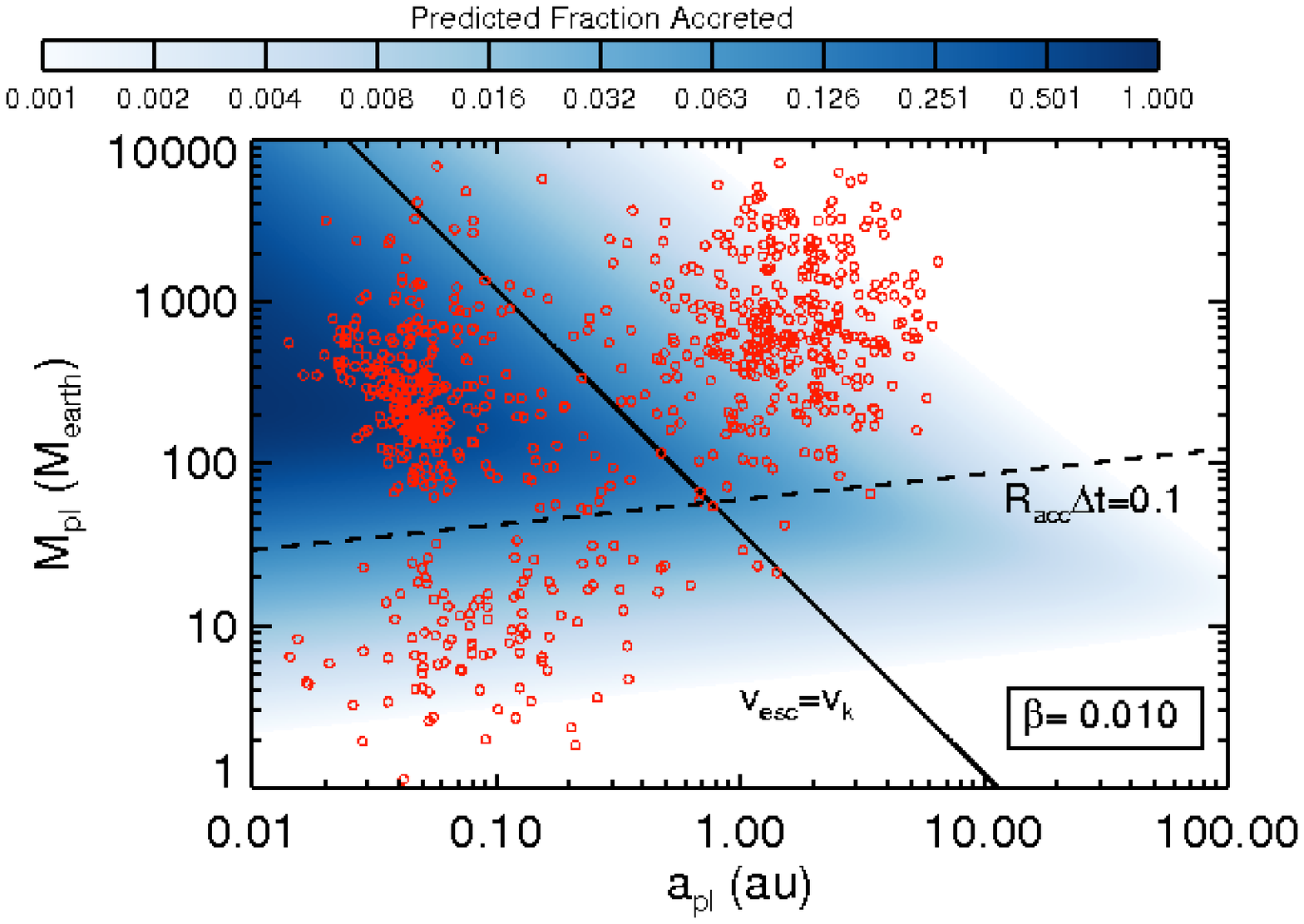}

\includegraphics[width=0.48\textwidth]{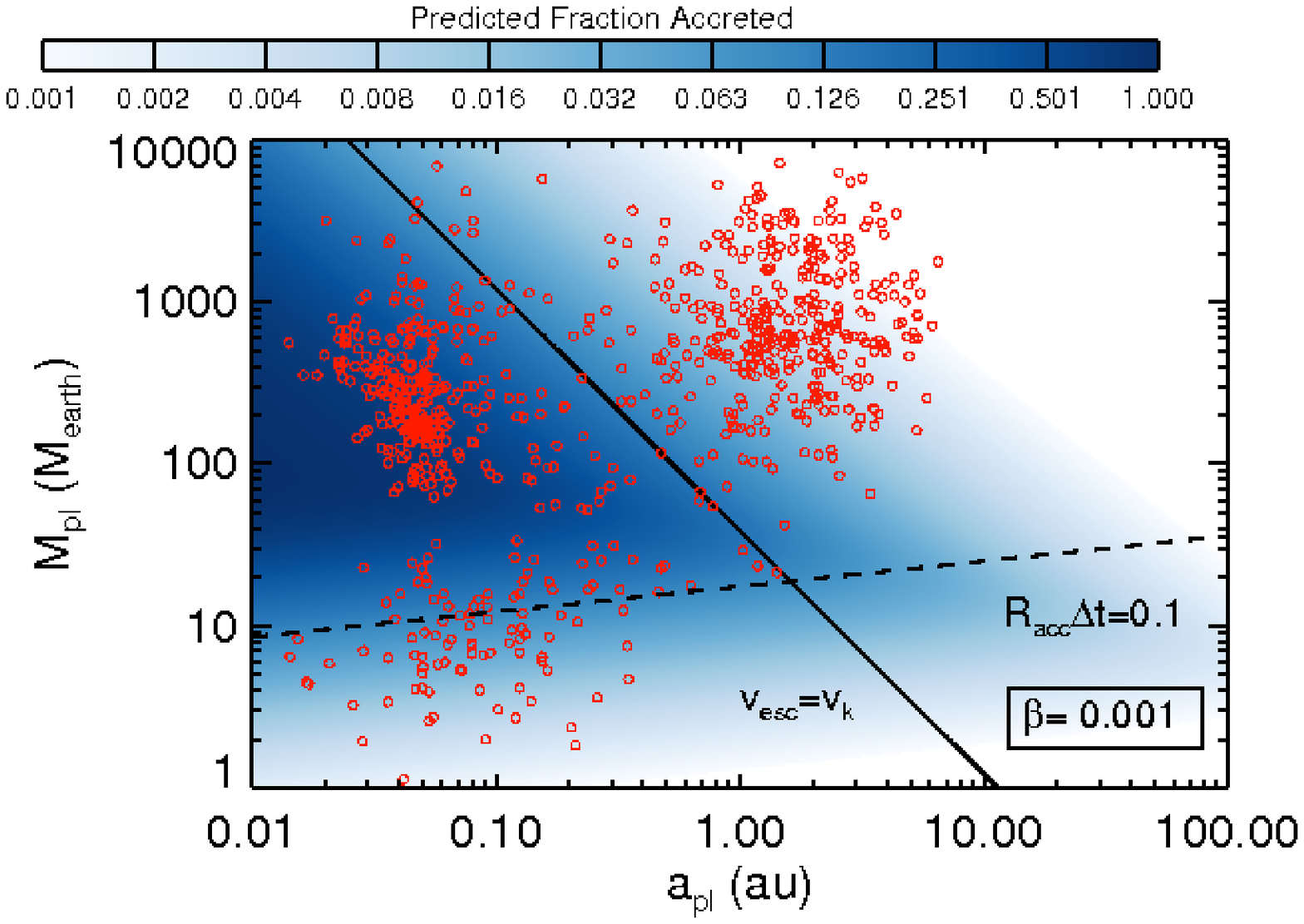}

\caption{The same as Fig.~\ref{fig:apl_mpl}, except for accretion. The dashed line shows where the timescale for accretion is equal to that for PR-drag, according to the empirical fit presented here, calculated by setting $R_{\rm acc} \Delta t = 1\%$.   }
\label{fig:apl_mpl_predict_facc}
\end{figure*}

\begin{figure*}

\includegraphics[width=0.48\textwidth]{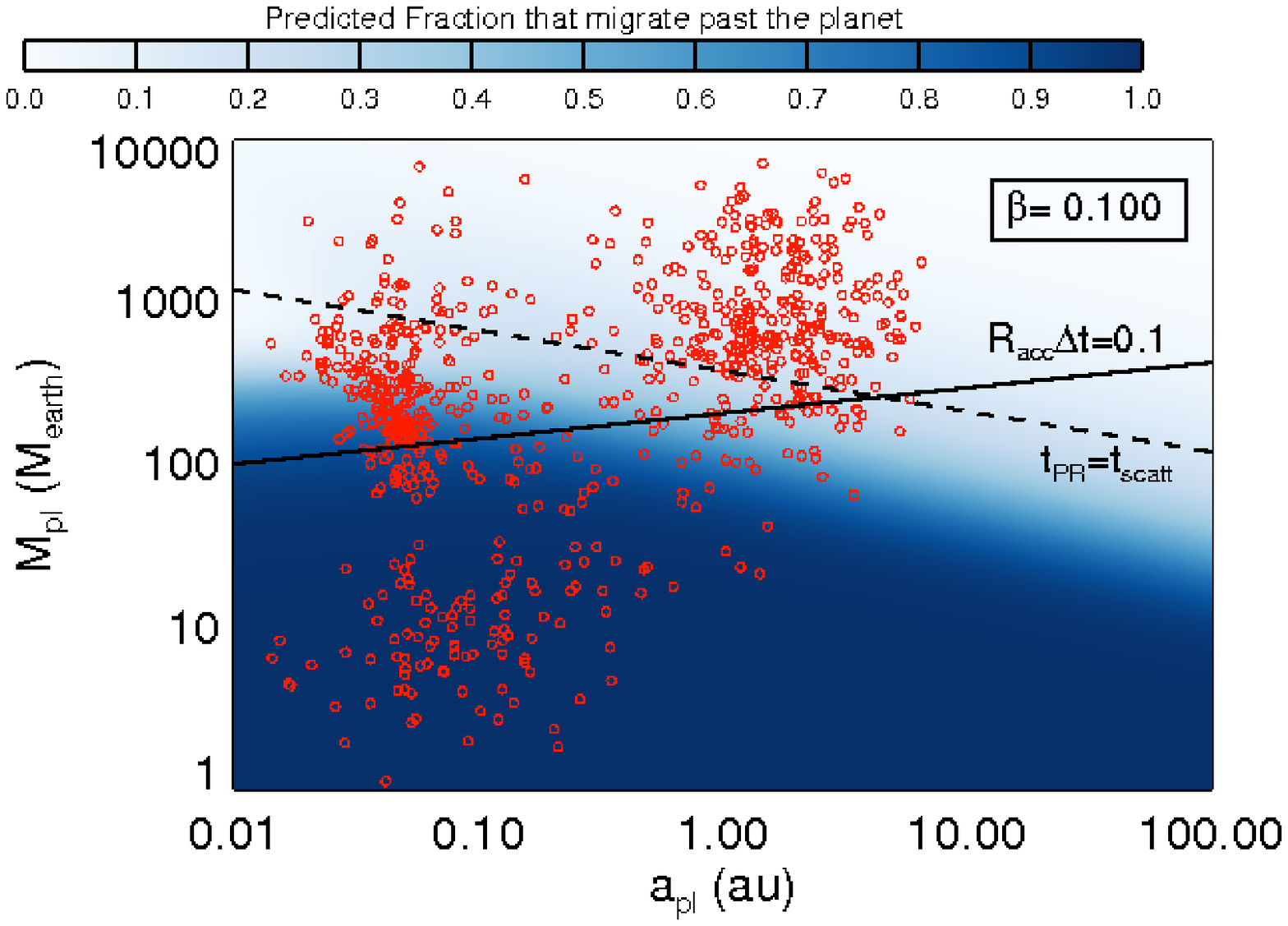}

\includegraphics[width=0.48\textwidth]{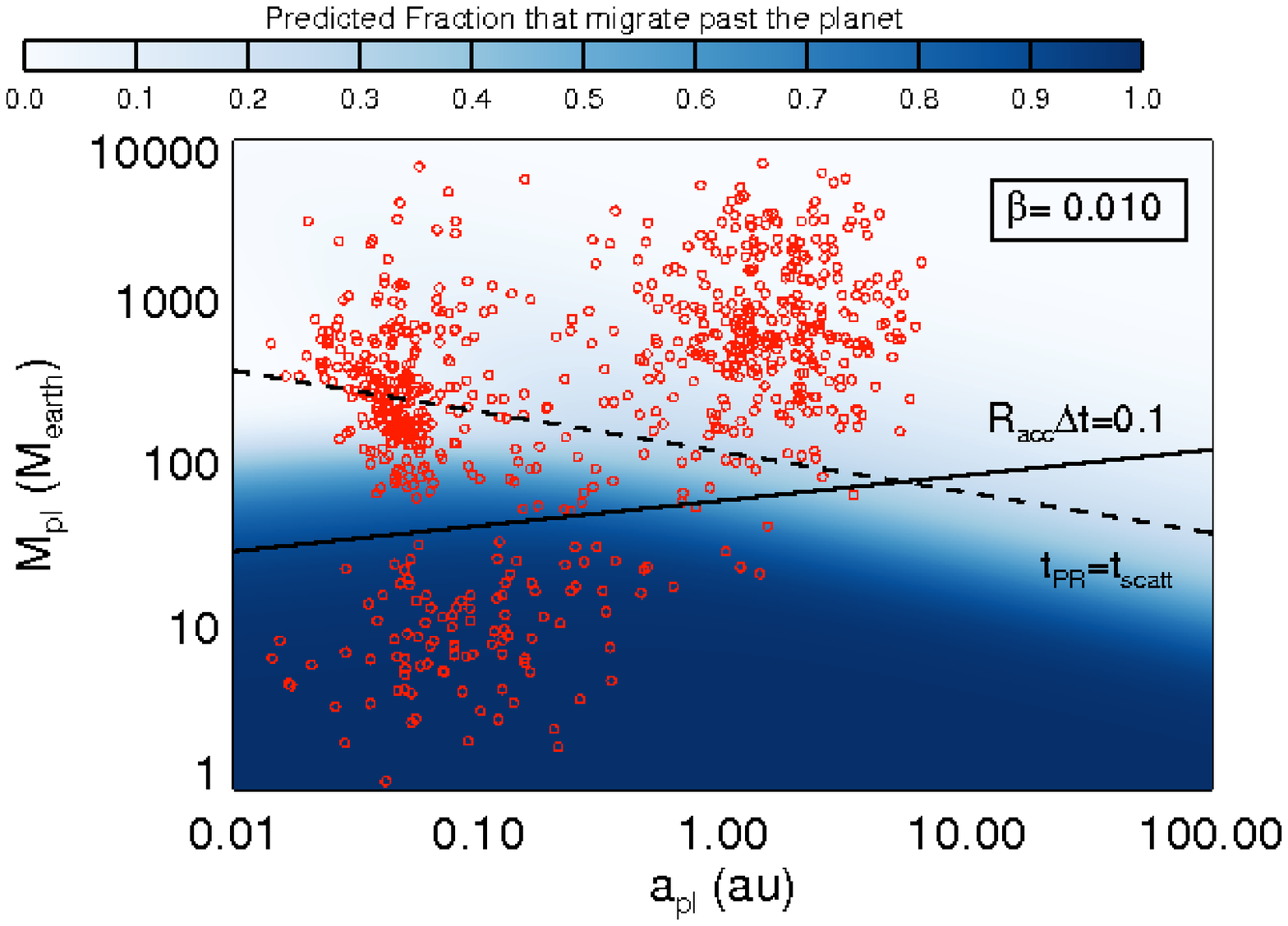}

\includegraphics[width=0.48\textwidth]{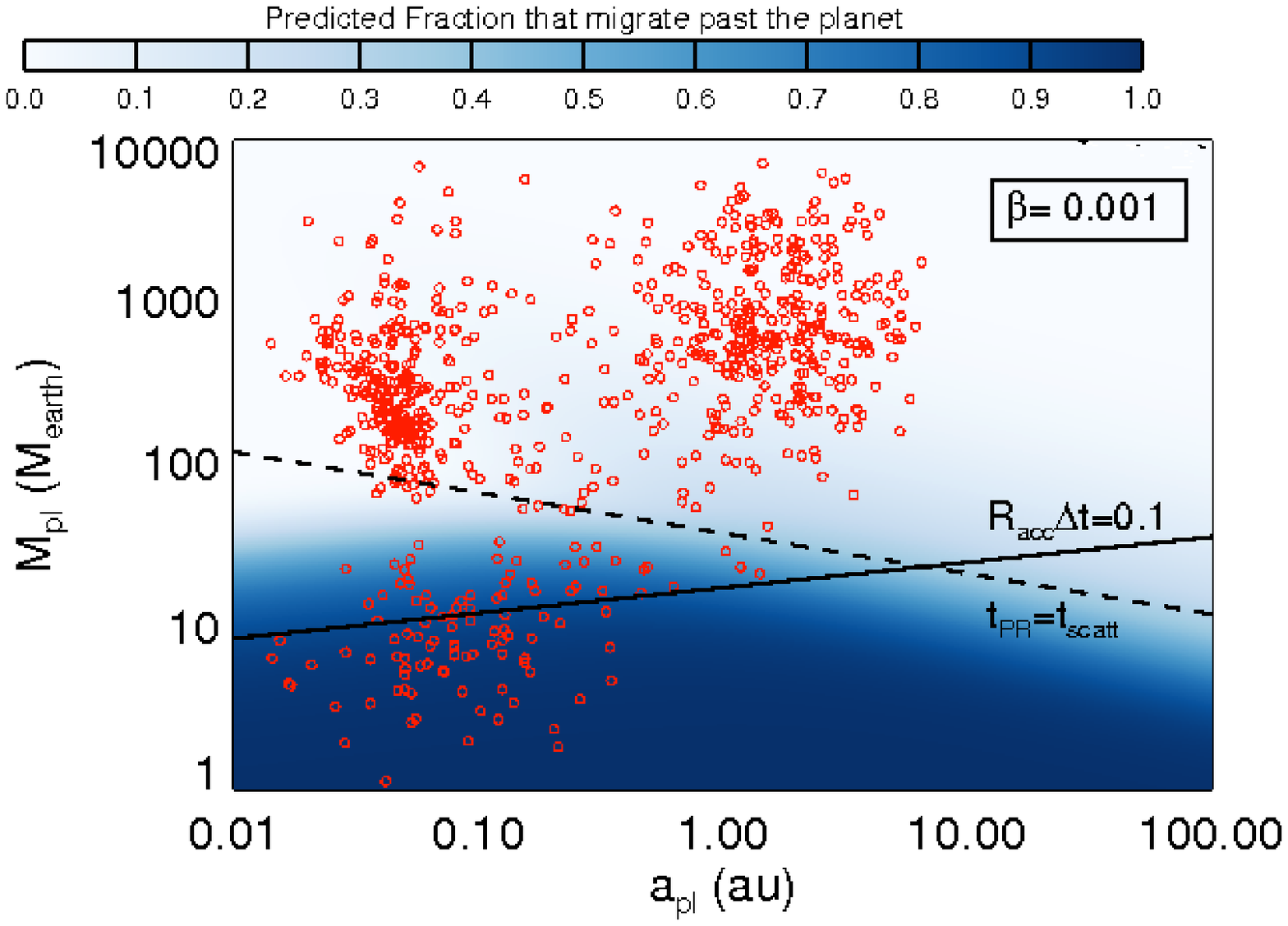}

\caption{The same as Fig.~\ref{fig:apl_mpl}, except for the fraction of particles that migrate past a planet, suffering neither accretion nor ejection. The dashed line shows where the time for the particles to migrate past the planet by PR-drag (Eq.~\ref{eq:deltat}) is equal to the time for particles to be ejected (Eq.~\ref{eq:mplequal}), and the solid line shows where the timescale for accretion is equal to that for PR-drag, according to the empirical fit presented here, calculated by setting $R_{\rm acc} \Delta t = 1\%$.}
\label{fig:apl_mpl_predict_hitstar}
\end{figure*}

\section{A model for the fate of particles that encounter a planet}
\label{sec:applications}

The model presented in \S\ref{sec:model} can be used to predict the
average fate of particles spiralling inwards under PR-drag and whether
they are accreted or ejected by any individual planet on a circular
orbit using Eq.~\ref{eq:racc},~\ref{eq:rej},~\ref{eq:fej_epsilon}, and
~\ref{eq:facc_epsilon} and the best-fit parameters from Table~\ref{tab:bestfit}.

The fraction of particles predicted to be ejected (accreted) by a planet can be summarised in terms of the planet's semi-major axis and mass, as shown on Fig.~\ref{fig:apl_mpl} (Fig.~\ref{fig:apl_mpl_predict_facc}). The red dots indicate all the known exo-planets. As expected, high mass planets eject almost all particles they encounter, whereas low mass planets eject almost no particles. A transition between ejection as the dominant outcome, compared to accretion as the dominant outcome is expected to occur for planets where the Keplerian velocity is approximately equal to the escape velocity \citep{Wyatt2017}, which is shown by the black solid line on Fig.~\ref{fig:apl_mpl} and Fig.~\ref{fig:apl_mpl_predict_facc}. 

However, there is another criterion required for planets to eject particles, as for some planets, particles migrate past too fast for them to be ejected. This occurs for planets at large semi-major axis. Analytically we can estimate when this transition occurs by comparing the timescale for particles to migrate past the planet due to PR-drag (Eq.~\ref{eq:deltat}) with the timescale for the planet to eject particles. This is estimated by considering cometary diffusion, and the timescale for this to lead to ejection, as derived in \cite{Tremaine93} (Eq. 3) and \cite{BrasserDuncan2008} (Appendix A). Setting these two timescales to be equal puts the transition from ejection to migration past the planet at: 
\begin{equation}
M_{\rm pl, equal} = 830 \, M_\oplus \;  \left(\frac{\beta}{e} \right)^{1/2} \left(\frac{M_\star}{M_\odot}\right)^{5/4} \left(\frac{a_{\rm pl}}{{\rm au}}\right)^{-1/4}, 
\label{eq:mplequal}
\end{equation}
where $e$ is the particle's eccentricity. The dependence on eccentricity is small, and given that this is unknown, we set the eccentricity to a plausible value of $e=(\frac{1}{5})^{1/2}$, the eccentricity at which particles leave the 2:1 resonance \citep{Shannon2015} to plot the dashed line on Fig.~\ref{fig:apl_mpl}. This line separates the two regions of parameter space between where the dominant outcome is ejection and where the rate of ejection is low or negligible. Planets that lie above both the solid and dashed lines on Fig.~\ref{fig:apl_mpl} are best at ejecting particles.

%Some high mass planets that lie beneath the black solid line on Fig.~\ref{fig:apl_mpl} can still eject some particles, despite the fact that accretion is the dominant outcome. This occurs when particles interact multiple times with the planet, eventually increasing their relative velocities above the escape velocity, such that they can be ejected. It is notable that the model predicts fewer ejections here, but nevertheless there is a non-negligible ejection rate above $M_{\rm pl, equal}$ and below $v_{K}=v_{\rm esc}$, as shown in the top left-hand corner of Fig.~\ref{fig:apl_mpl}. 

In terms of accretion, Fig.~\ref{fig:apl_mpl_predict_facc} shows that significant accretion only occurs for planet masses below $v_{K}=v_{\rm esc}$, noting the log-scale. However, for the lowest mass planets, particles migrate past the planet before they have time to be either accreted or ejected. The dotted line on Fig.~\ref{fig:apl_mpl_predict_facc} shows where the timescale for accretion is equal to that for PR-drag, according to the empirical fit presented here, calculated by setting $R_{\rm acc} \Delta t = 0.1$, using the model parameters shown in Table~\ref{tab:bestfit}. Planets that are good at accreting particles lie above the dashed line and below the solid line on Fig.~\ref{fig:apl_mpl_predict_facc}.

Fig.~\ref{fig:apl_mpl_predict_hitstar} shows the fraction of particles that are not lost in interactions with the planet. In general, whether or not particles are ejected dominates their fate and the fraction that hit the star, \ie migrate past the planet without interacting, is very similar to the fraction that are not ejected. Thus, the dashed line, $t_{\rm PR}=t_{\rm scatt}$ also explains this behaviour. The only exception to this is for planets where accretion is the dominant outcome which is at small semi-major axes and planet masses close to $M_{\rm pl, equal}$. Particles that migrate past the planet are of interest as in multiple planet systems they are the particles that can go on to interact with other planets and may be the particles that make it into the inner regions to replenish an exozodiacal cloud.

%We can use the analytic approximation of \cite{Wyatt99} to predict when particles are accreted. The blue dotted line on Fig.~\ref{fig:apl_mpl_predict_facc} shows when $P_{\rm acc}=0.5$, according to Eq.~\ref{eq:pacc99}. Accretion by planets is, however, limited by ejection, once the Keplerian velocity of the planet (approximately the Keplerian velocity of the particle interacting with the planet) is greater than the escape velocity for the planet, shown by the black solid lines on Fig.~\ref{fig:apl_mpl_predict_facc}. Particles hit the star when they are neither ejected, nor accreted. This is the region that lies below the green dashed line ($t_{PR}= t_{\rm scatt}$) and below the analytic approximation of \cite{Wyatt99} shown by the blue dotted line. Noting that \cite{Wyatt99} predicts a much stronger dependence on $\beta$ than the analytic predictions made in this paper. {\bf \cite{Wyatt99} has $1/\beta$, which is much stronger than the $\beta$ dependence that we see in the sims, and comes out of the analytic prediction. I don't fully understand this yet??}. 

\subsection{Low mass stars}
\label{sec:lowmass}

\begin{figure}

\includegraphics[width=0.48\textwidth]{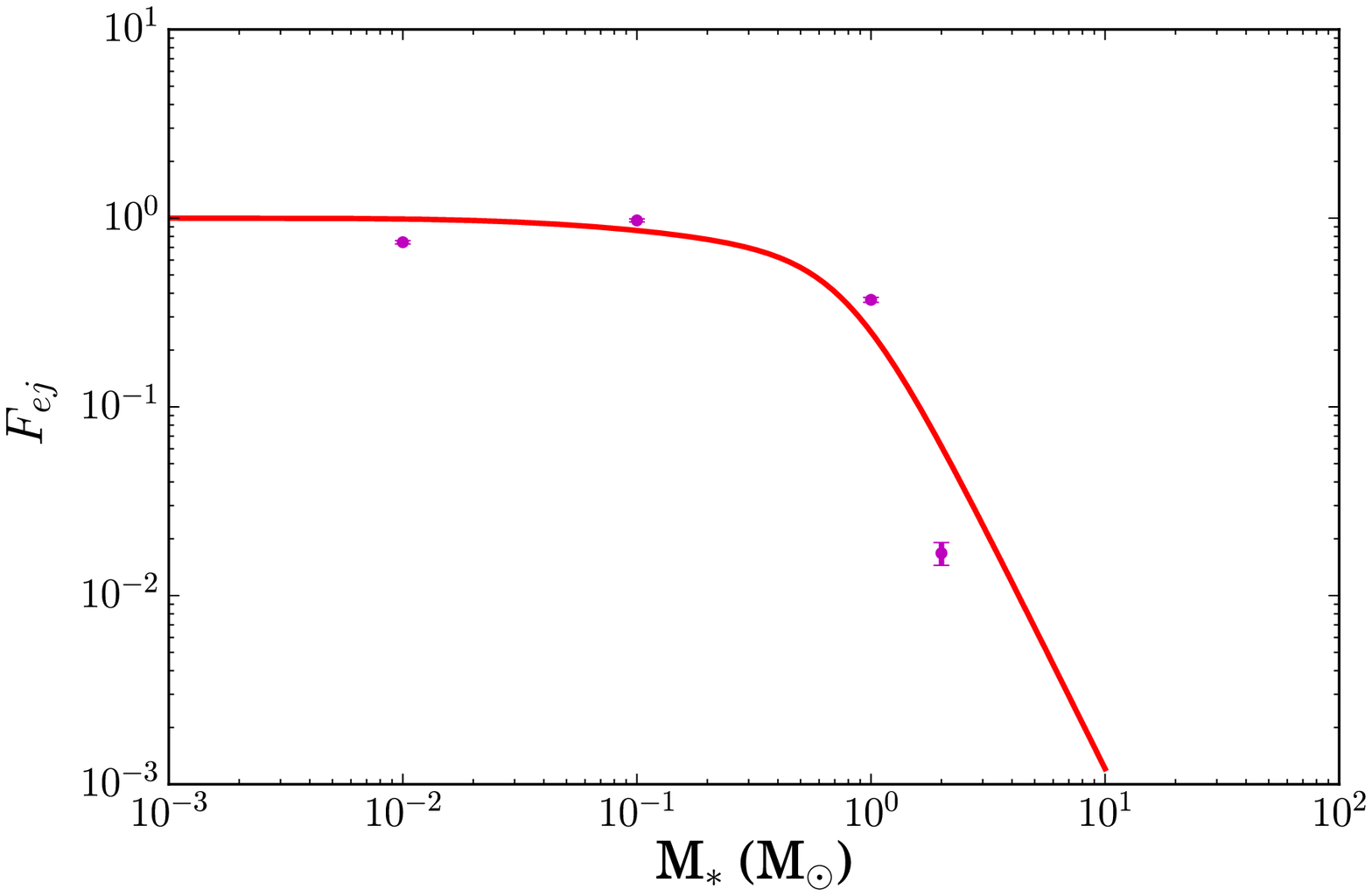}
\includegraphics[width=0.48\textwidth]{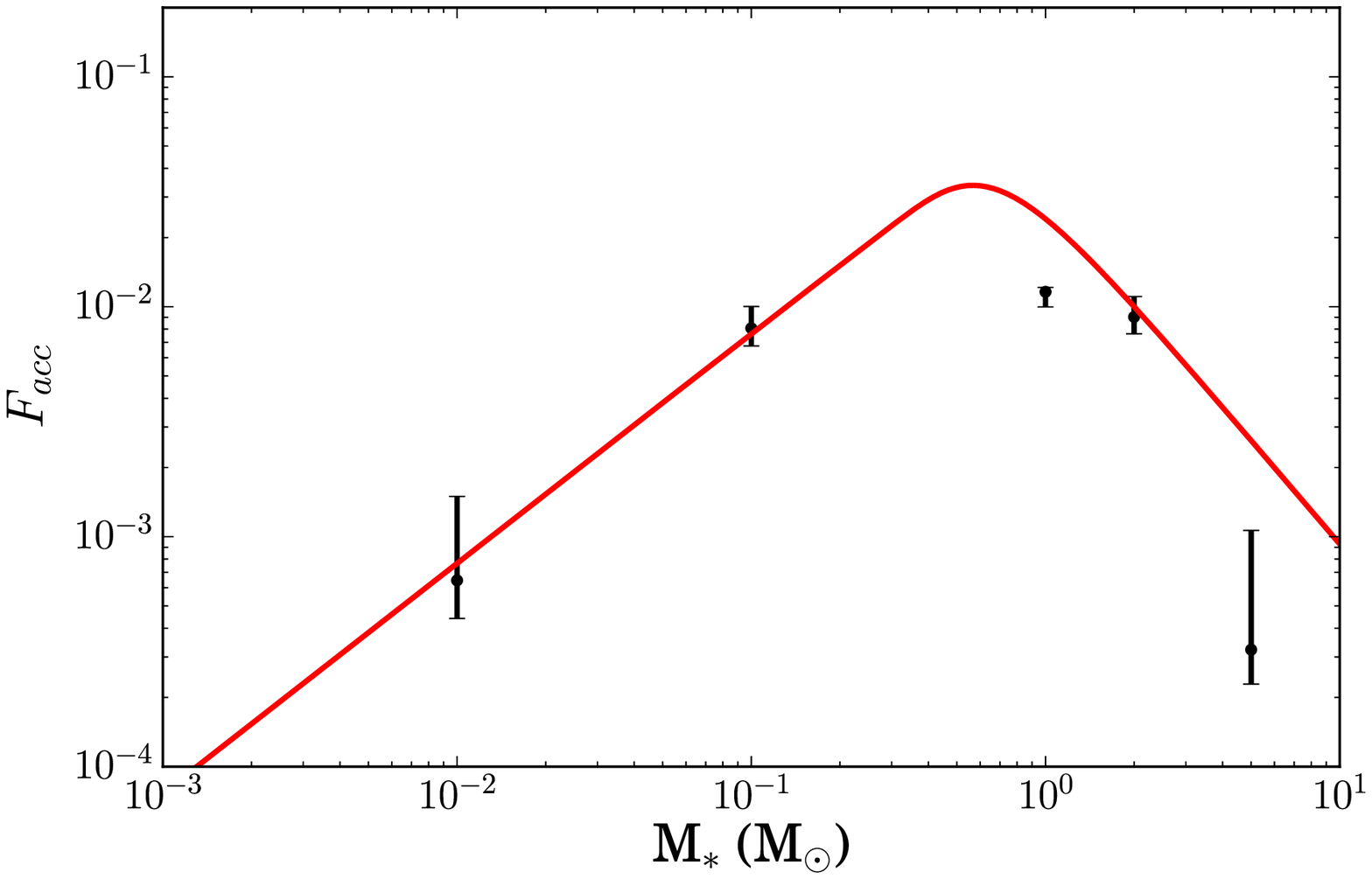}
\includegraphics[width=0.48\textwidth]{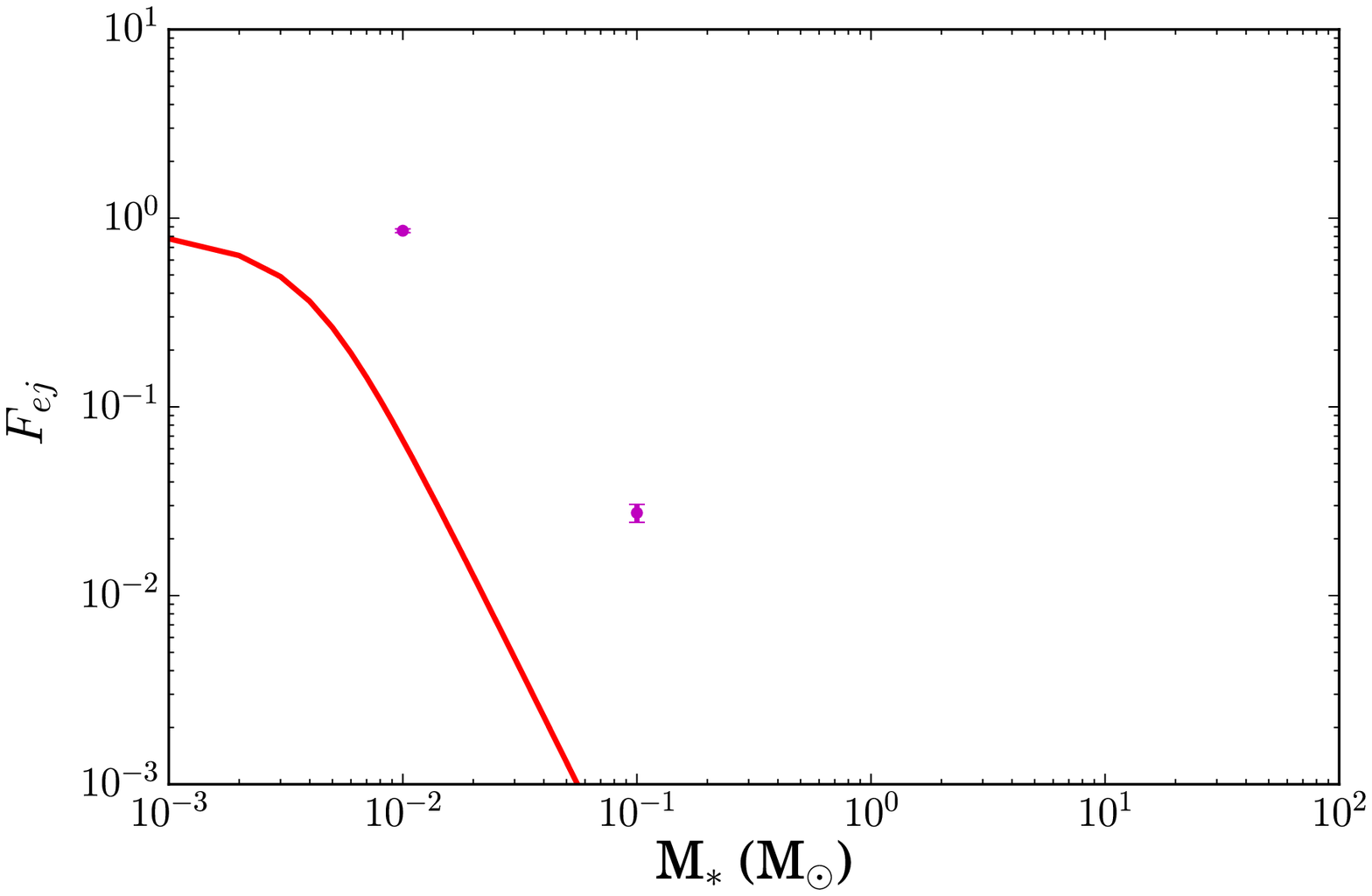}
\includegraphics[width=0.48\textwidth]{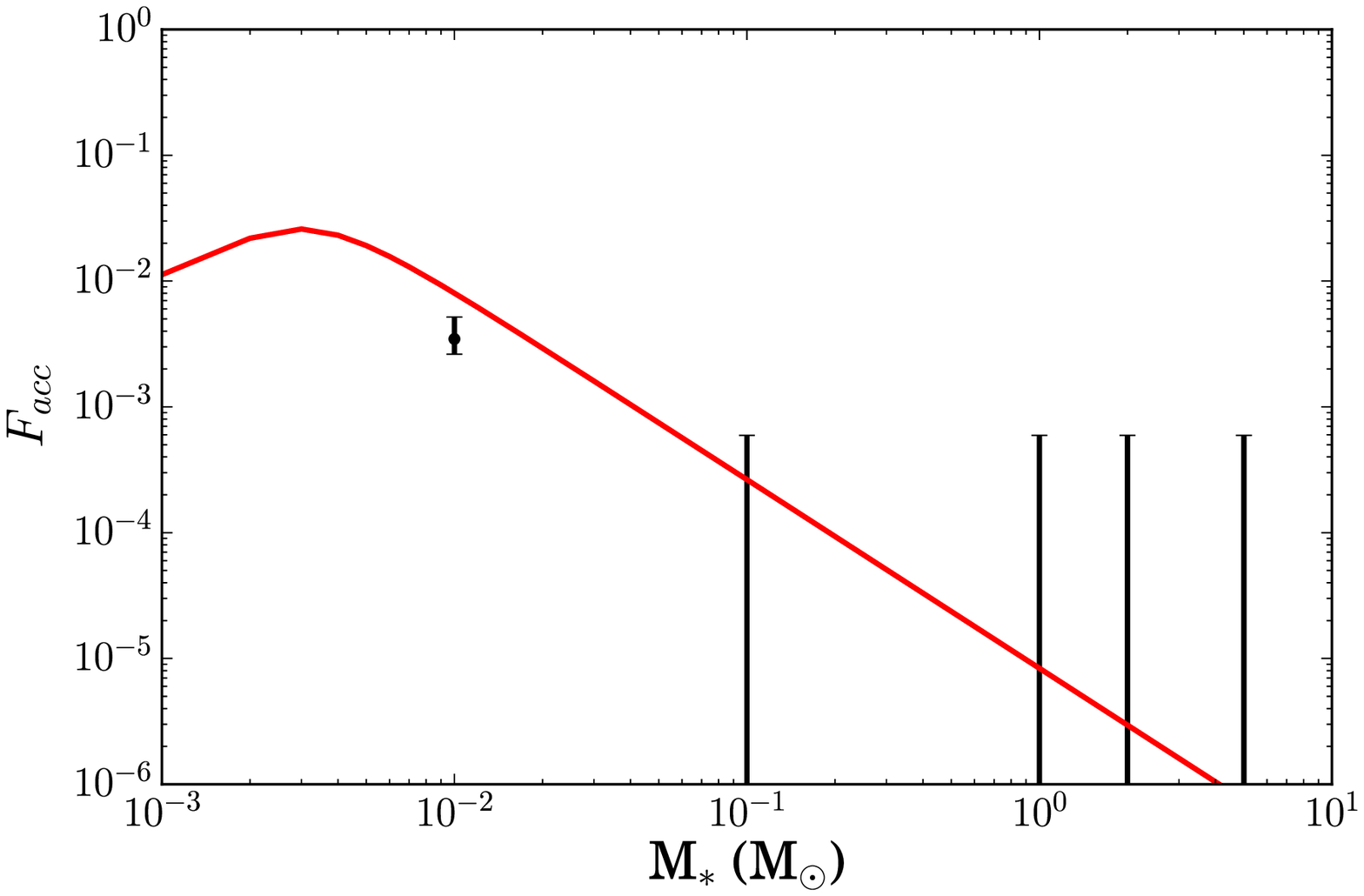}

\caption{The fraction of particles ejected and accreted in the numerical simulations for $M_{\rm pl}=100M_\oplus$ (top two plots), $M_{\rm pl}=1M_\oplus$ (bottom two plots), $\beta=0.1$ and $a_{\rm pl} = 1$au, varying stellar mass ($M_\star$), compared to predictions, based on Eq.~\ref{eq:rej}, Eq.~\ref{eq:racc} using $\delta_a=1/2$ and $\delta_e=-5/2$, as described in \S\ref{sec:lowmass}. }
\label{fig:lowmass}
\end{figure}

Our numerical simulations focussed on solar mass stars. However, we can use the analytic approximation to make a prediction for the dependence on stellar mass (Eq.~\ref{eq:fej}, Eq.~\ref{eq:facc}). The analytic approximation calculates the fraction of particles ejected or accreted as a function of $\beta$, which for radiation pressure corresponds to the particle size. For such low mass stars, however, it is questionable whether radiation pressure is sufficiently strong to lead to high values of $\beta$, and in fact, it has been suggested that forces due to the stellar wind may dominate \citep[\eg][]{AugereauAUmic}. Nonetheless, we test the extension of the analytic model by a handful of numerical simulations, noting that care should be taken in applying this model to low mass stars, particularly for the high values of $\beta$ considered and the planet masses that may be a significant fraction of the stellar mass. For this suite of simulations the stellar mass was varied, for $M_{\rm pl}=100M_\oplus$, $\beta=0.1$ and $a_{\rm pl} = 1$au and $M_{\rm pl}=1M_\oplus$, $\beta=0.1$ and $a_{\rm pl} = 1$au. The results of the numerical simulations, compared to the analytic predictions are shown in Fig.~\ref{fig:lowmass}, based on Eq.~\ref{eq:rej}, Eq.~\ref{eq:racc} using $\delta_a=-3/2$ and $\delta_e=-5/2$. The simulations are by no means comprehensive and they indicate, as in a similar manner to the other parameters, an empirical fit to the numerical simulations might lead to slightly different values of $\delta_a$ and $\delta_e$, however, the analytically predicted values do a reasonably good job of predicting general trends as the stellar mass changes. The fraction of particles ejected increases for lower stellar masses, as does accretion, until the stellar mass is sufficiently low that ejection becomes the dominant outcome and planets are no longer as good at accreting, as seen on Fig.~\ref{fig:lowmass}. 

\begin{figure}

\includegraphics[width=0.48\textwidth]{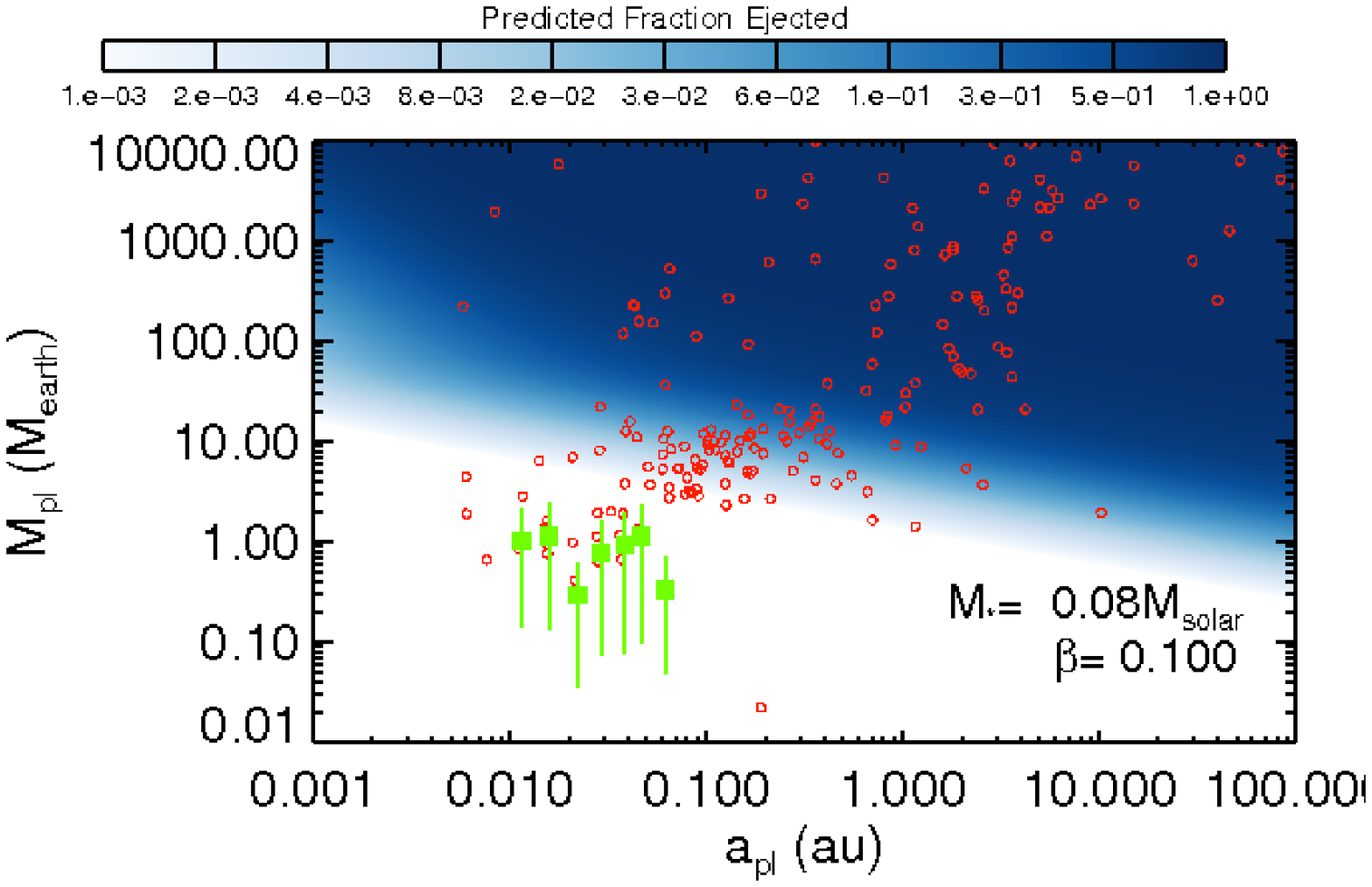}

\includegraphics[width=0.48\textwidth]{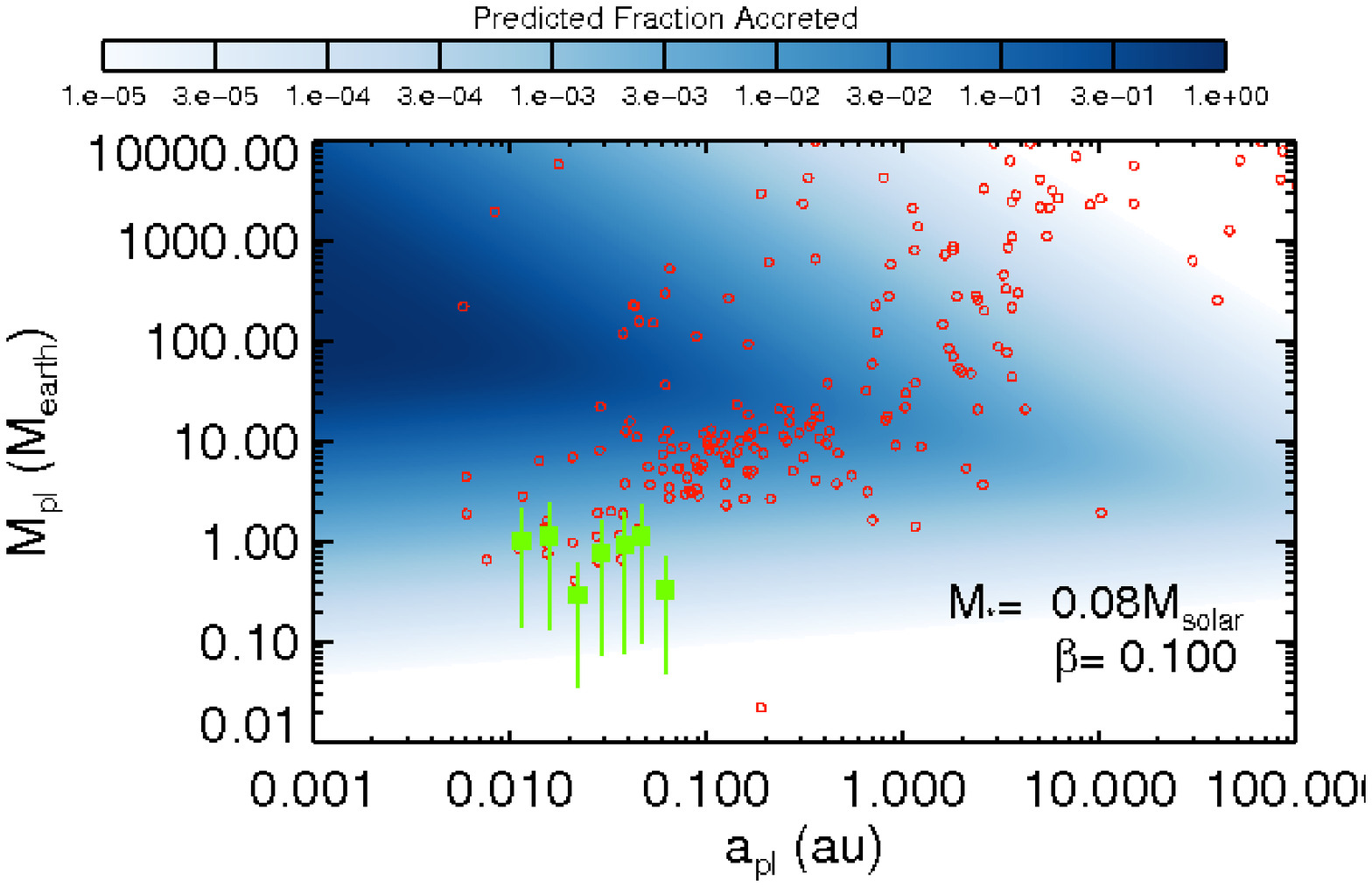}

\caption{ Predictions for the fraction of particles accreted and ejected by planets orbiting low mass stars ($M_\star=0.08M_\odot$) as a function of planet semi-major axis, $a_{\rm pl}$ and planet mass, $M_{\rm pl}$. Over-plotted in red are exoplanets orbiting stars with $M_\star<0.5M_\odot$ and as the green points, the TRAPPIST-1 planets \citep{Gillon2017}.}
\label{fig:lowmass2}
\end{figure}

 Fig.~\ref{fig:lowmass2} shows predictions for the fraction of
 particles ejected and accreted, as a function of the planet's
 semi-major axis and mass, for lower mass stars
 ($M_\star=0.08M_\odot$). A famous example of a multi-planet system
 around a low mass star is the TRAPPIST-1, planetary system
 \citep{Gillon2017}. Most particles migrate past the TRAPPIST-1 planets. Our model predicts that for large grains, $\beta=0.001$, $F_{\rm ej} <0.2\%$, and $F_{\rm acc} < 10$\%, whereas for small grains e.g. $\beta=0.1$, $F_{\rm ej} <10^{-3}\%$, and $F_{\rm acc} \sim 0.1$\%. The TRAPPIST-1 planets are better at accreting than ejecting particles spiralling inwards due to PR-drag. However, a caveat is that test simulations for a TRAPPIST-1-like planet ($M_{\rm pl}=1M_\oplus$ at 0.01au around a $0.08M_\odot$ star) find accretion rates that are higher than predicted by the model (5\% for $\beta=0.1$). This is likely due to a limitation in the model that does not always provide a good fit for low mass planets, particularly relevant at small semi-major axis (see the middle panel of Fig.~\ref{fig:bestfit_acc}), as noted in \S\ref{sec:limitations}.

%Fig.~\ref{fig:sigma} shows that the surface density of dust is only slightly reduced as it migrates past the TRAPPIST-1 planets. It also indicates that the TRAPPIST-1 planets are unlikely to accrete significant quantities of dust leaving any outer debris disc and migrating past the planet due to PR-drag. 

%{\bf I run a N-body simulation with 1mearth at 0.01 au around a $0.08M_\odot$ star $\beta=0.1$ and find that 5\% of particles are accreted by the planet!}

\subsection{Limitations of the Model}

\label{sec:limitations}

The model presented here is designed to be a fast alternative to detailed simulations, for predicting the fate of particles leaving a debris belt due to PR-drag. It does a reasonable job of reproducing the results of those simulations, within the limited parameter space tested. Clearly there are details of such simulations that the simple model misses. In particular, it does not do as good a job of reproducing the behaviour seen in simulations for large values of $\beta$ (small dust grains). This is related to changes in the inclination and eccentricity distributions of particles at the point of interaction with the planet. The analytic model is derived assuming that eccentricities are low when particles interact with the planet, an approximation which may not be strictly valid following resonant interactions. Multiple interactions with the planet can increase eccentricities and inclinations in a manner not fully accounted for in the model. Another limitation regards accretion rates for low mass planets, where in general the simulations were limited by the number of particles included and the fate of particles depends more strongly on their initial parameters. This is because resonant trapping is less efficient for low mass planets. Clearly there are further subtleties related to the inclination/eccentricity distribution of particles as they interact with the planet, that this simple model misses. For example, changes to the initial inclinations or eccentricities of particles at the start of simulations can change the fraction of particles ejected or accreted. In addition to which, the model only includes planets on circular orbits and the behaviour for planets on eccentric orbits may differ significantly. Another point to note is that the scaling with stellar mass presented here has not been robustly tested by simulations and is based purely on the analytic model.

\section{Applications}

The model presented here can be applied in multiple contexts, including predicting the mass accretion rates onto planets interior to outer debris belts, predicting the levels of dust in inner planetary systems, based on the architecture of the outer planetary system, and using observed levels of dust in the inner regions to predict the presence of outer planets. In this section we apply this model to a sub-set of observed systems in order to make predictions regarding the dust levels in their inner regions. The properties of these systems are listed in Table~\ref{tab:planet}, which is limited to those with known outer debris belts and multiple planets orbiting interior to these belts listed in \citep{Marshall2014}, or those with known outer belts and LBTI observations that characterise the levels of dust in the inner planetary system \citep{Ertel2018}.

The model presented here makes predictions regarding how the presence of planets changes the levels of dust leaving an outer belt due to PR-drag that reach the inner planetary system. Mutual collisions between these dust particles also play a critical role, but unfortunately are harder to model. For the purposes of this work, we rely on the simple model of \cite{wyattnopr} that traces the collisional evolution of a population of single size dust grains, to predict the depletion of dust due to collisions. We note, however, that collision rates could be higher \citep{vanlieshout2014, KennedyPiette2015} and that whether or not PR-drag is indeed the dominant transport mechanism to explain dust observed in the inner regions of planetary systems remains an open question \citep{Kral_exozodireview}.

In this simple model the effective optical depth (equivalent to the surface density of cross-sectional area) of the outer belt, $\tau(r_0)$ is depleted at a distance r from the star as (Eq. 4 of \cite{wyattnopr}): 

\begin{equation}
\tau (r) = \frac{\tau(r_0)}{1+ 4 \eta_0(1-\sqrt{\frac{r}{r_0}})}
\label{eq:sigma_r}
\end{equation}
where $r_0$ is the radius of the outer belt and $\eta_0= \frac{5000}{\beta}\tau(r_0)\sqrt{\left(\frac{r_0}{\rm au}\right)\left(\frac{M_\odot}{M_*}\right)}$. The effective optical depth of the outer belt, $\tau(r_0)$, can be related to the observed properties of the outer belt, including its fractional luminosity, $f$, radius, $r$ and width, $dr$, assuming that all of the grains emit efficiently as black-bodies
\begin{equation}
\tau(r_0)= \frac{2 \; f\;  r_0}{dr}.
\label{eq:sigma0}
\end{equation}

For the radius of the outer belt, $r_0$, we take the inner edge of the belt as determined from resolved imaging, where available and otherwise use the black-body radius determined from a black-body fit to the SED \citep{Marshall2014}. The belt width is generally undetermined or poorly constrained and therefore, $dr=0.1r_0$ is assumed for all systems. This assumption does not affect the conclusions signitificanly, since if the belt is broader, the assumption of a narrow belt supposes that the emission (fractional luminosity) comes from a narrow region, which, therefore, has a higher initial collision rate, such that the dust is ground down faster and the evolution, therefore, tends towards the same evolution as would have resulted from a broader belt. Morever, changing $dr$ from $0.1r_0$ to $r_0$ results in a $<10\%$ change in the effective optical depth at 1au (for $r_0=200$au). This is because the profile of $\tau(r)$ tends to a constant value for small radii and a change in the belt width only changes this constant value slightly. On the other hand, a significant change in the location of the belt can mean that we are no longer in the regime where $\tau(r)$ tends to a constant value, rather closer to the outer belt, where $\tau(r)$ can decrease steeply with $r$, such that for example changing $r_0=100$au to $r_0=200$au can produce changes in $\tau(1{\mathrm au})$ of $>50\%$.

In addition to this, this simple model may underestimate the rate of collisions based on the observed fractional luminosity, as in many cases the emission is dominated by small grains that are inefficient in their emission at the relevant infrared wavelengths. The model is limited by the assumption of a single grain size and the lack of calibration against observations. The model of \cite{wyattnopr} would benefit from future updates to include multiple grain sizes \citep{vanlieshout2014}, and to allow grain size dependent sink terms, such as the ejection or accretion by planets presented here, as well as effects like resonant trapping \citep{Shannon2015}. The predictions made here can be readily updated to include any improved collision model, as available and would greatly benefit from any improvements.

%%%%%%%%%%%%%%%%%%%%%%%%%%%%%%%%%%%%%%%%%%%%%%%%%%%%%%%%%%%%%%%%%%%%%%%%%%%%%%%%%%%%%%%%%%%%%%%%%%%%%%%%%%%%%%%%%%%%

\begin{table*}

\begin{tabular}{cccccccc}
%n & lstar & mstar & rbb & lir & apl & mpl \\
\hline\hline
Name & $L_*$   & $M_*$ & Radius & $f$ & $a_{\rm pl}$ & $M_{\rm pl}$ & $e_{\rm pl}$ \\
&$L_\odot$ & $M_\odot$&  au   &$10^{-6}$ & au & $M_J$ &\\
\hline\hline
q$^{1}$ Eri & 1.52 & 1.11&  85$^a$ & 405 & 2.022  & 0.93 & 0.16\\ %\citep{Liseau2010}
$\tau$ Ceti & 0.53 &0.78 &5$^b$ & 7.8 & 0.105,  0.195,  0.374,  0.552,  1.35 & 0.0063,   0.0098,   0.011, 0.014 ,  0.0208 & 0.16,0.03,0.08,0.05  \\ %\citep{Lawler2014} palnet properties out of date
HD 19994 & 3.84 & 1.3 &90 & 5.4 & 1.306  & 1.33  & 0.266\\% not resolved
HD 20794 & 0.66 &0.7  &24$^c$ & 2.4 & 0.1207 ,  0.2036,  0.3498  & 0.0085, 0.0074,0.015 & 0,0, 0.25 \\%\citep{Kennedy2015}
$\epsilon$ Eri & 0.43 &0.82 & 11$^d$ &  108 & 3.38 & 1.05 & 0.25\\ %\citep{Booth2017}
HD 40307 & 0.25  &0.75& 24 & 4.3 & 0.047,  0.08,  0.13 & 0.01291,0.0211,0.0281 & 0,0,0\\
61 Vir & 0.84 & 0.93 &30$^e$& 28 & 0.05006,  0.2169,  0.4745& 0.0161,0.0334,0.0716 &0.12,0.14,0.35 \\  % 30-150 au Marino2017
70 Vir & 2.9 & 1.1 &50 & 4.8 & 0.4836 & 7.46 & 0.4 \\
GJ 581 & 0.012 & 0.31 &25$^f$& 91 & 0.04061,  0.0729,  0.2177,  0.02846 & 0.05,0.017,0.019,0.0061& 0.031,0.07,0.25,0.32\\  % 25-60au \citep{Lestrade2012}
HD 210277 & 1.0  &1.09& 155  &5.1 & 1.131  & 1.273 & 0.476 \\ % cold debris discs?! 

HR 8799 & 5.4 & 1.47 &145$^g$ &49$^h$ & 14.5,27,42.9,68 & 9,10,10,7 & -,0.1,0,0 \\ % \citet{Booth2016}

HD 82943 & 1.0 & 1.14 &67$^i$& 100$^i$ & 0.746,1.19,2.145 & 14.4,14,0.29& 0.425,0.203,0 \\ % 67-300au 
HD 69830 & 0.62 & 0.86 &1$^j$ & 190 & 0.0186,0.079,0.63 &0.165,0.143,0.253  & 0.1,0.13,0.07,0.31,0.33\\ %\citep{resolveHD69830}
%HD 38858 & 0.58 & 0.89 30 &  88 & & \\ %\citep{Kennedy2015}
\hline
$\beta$ Leo & 13.3 &2.3 &30$^k$&20 & &\\ % nb this isn't the belt inner edge from Churcher et al ! 
Vega & 57 & 2.9 & 85$^l$& 19 &   & \\
\hline
\hline
\end{tabular}

\caption{The properties of a sample of planet-hosting debris discs, with the addition of Vega and $\beta$ Leo. Planet data from \citet{Marshall2014} or exoplanet.eu. Fractional luminosities ($f$) of the outer belt are taken from \citet{Marshall2014} unless otherwise referenced. Radii are the inner edge of a resolved disc, where resolved imaging exists, otherwise, black-body radii from \citet{Marshall2014}. $^a$\citet{Liseau2010} $^b$\citet{Lawler2014} $^c$\citep{Kennedy2015} $^d$\citet{Booth2017} $^e$\citet{Marino2017} $^f$\citet{Lestrade2012} $^g$\citet{Booth2016} $^h$\citet{wyatt07} $^i$\citet{ Kennedy2013b} $^j$\citep{resolveHD69830} $^k$\citet{Churcher11} $^l$\citet{Sibthorpe2010} }
\label{tab:planet}
\end{table*}

%%%%%%%%%%%%%%%%%%%%%%%%%%%%%%%%%%%%%%%%%%%%%%%%%%%%%%%%%%%%%%%%%%%%%%%%%%%%%%%%%%%%%%%%%%%%%%%%%%%%%%%%%%%%%%%%%%%%

%%%%%%%%%%%%%%%%%%%%%%%%%%%%%%%%%%%%%%%%%%%%%%%%%%%%%%%%%%%%%%%%%%%%%%%%%%%%%%%%%%%%%%%%%%%%%%%%%%%%%%%%%%%%%%%%%%%%
\begin{figure}
\includegraphics[width=0.48\textwidth]{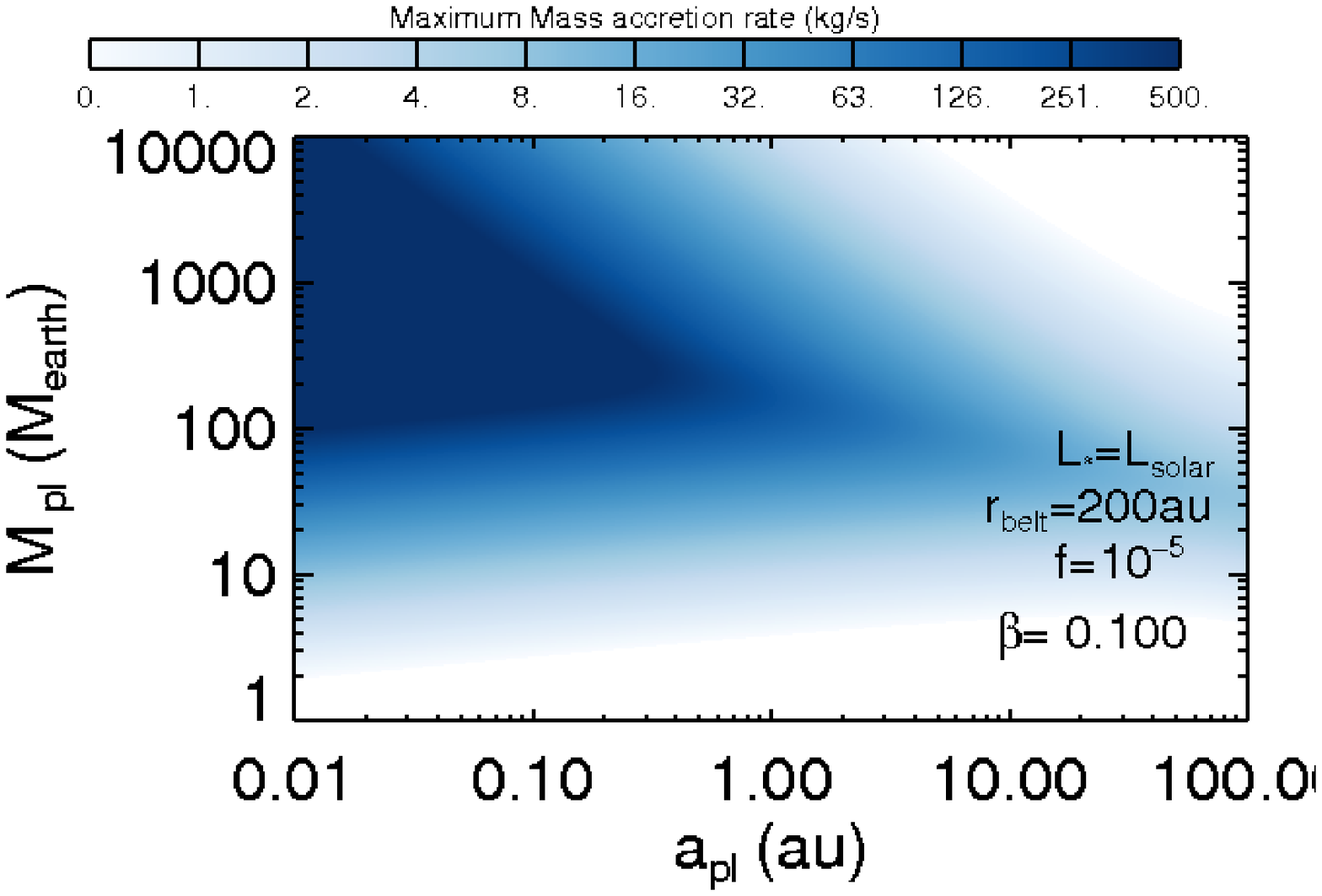}
\includegraphics[width=0.48\textwidth]{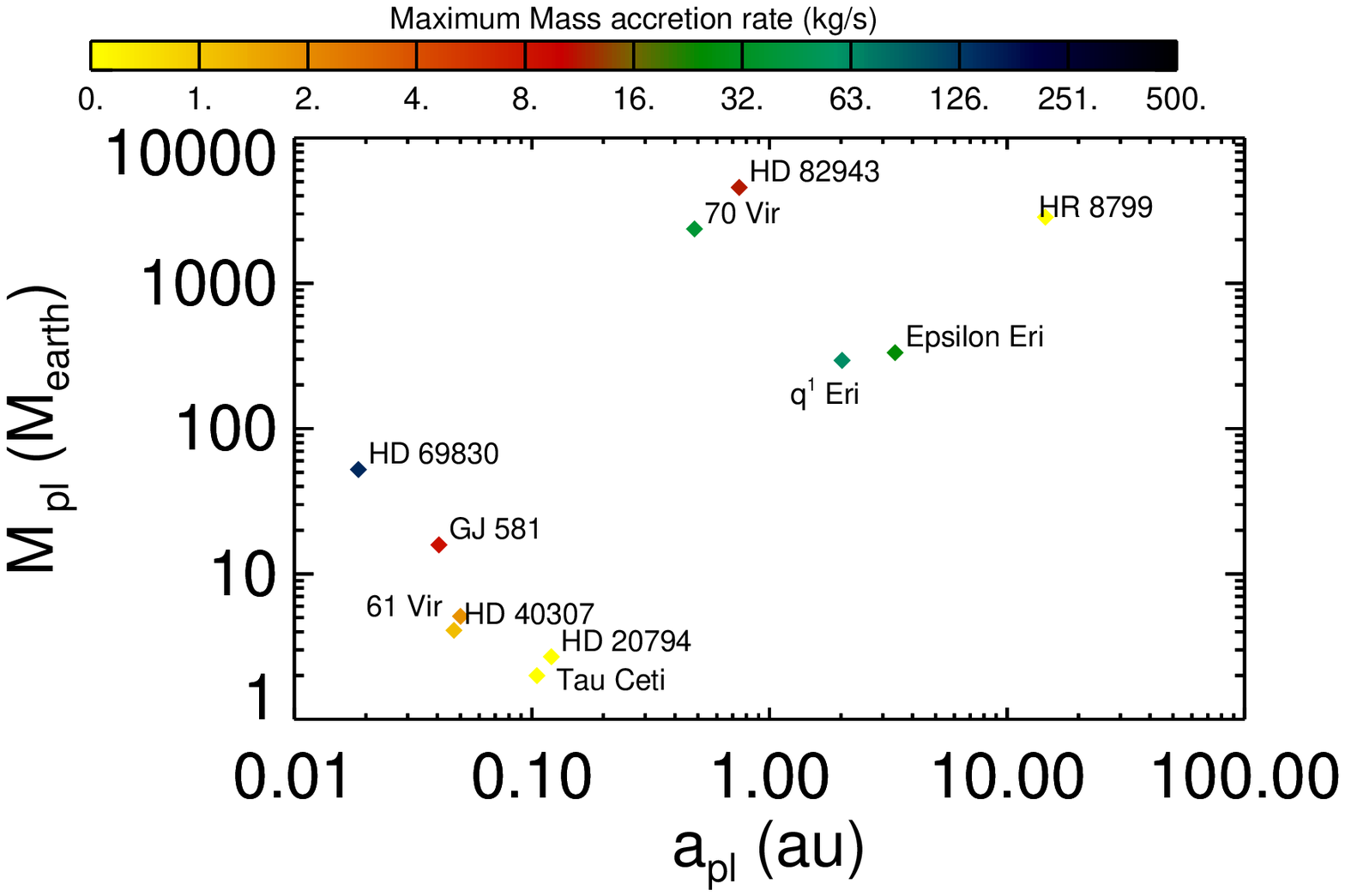}
\caption{Predictions for the mass accretion rate of $\beta=0.1$ dust grains onto planets interior to debris belts, calculated using Eq.~\ref{eq:macc}. Top panel: For a typical outer debris belt with $r_{\rm belt}=200$au and fractional luminosity $f=10^{-5}$ around a solar mass star. Bottom panel: Accretion rates onto the outer planet interior to those debris belt systems listed in Table~\ref{tab:planet}.  }
\label{fig:macc}
\end{figure}

%%%%%%%%%%%%%%%%%%%%%%%%%%%%%%%%%%%%%%%%%%%%%%%%%%%%%%%%%%%%%%%%%%%%%%%%%%%%%%%%%%%%%%%%%%%%%%%%%%%%%%%%%%%%%%%%%%%%

\subsection{Accretion onto planets interior to debris belts}
\label{sec:macc}

Planets interior to outer debris discs can accrete dust that migrates inwards under PR-drag from the outer belt. The mass accretion rate due to PR-drag at a radius $r$, interior to a belt at $r_0$,  can be calculated by considering the density of particles migrating inwards and their velocity \citep{vanlieshout2014} and is given by :
\begin{equation}
\dot{M}_{PR}(r)= F_{\rm acc} \; \frac{Q_{PR}\;  \tau(r) L_*}{c^2},
\label{eq:macc}
\end{equation}
where $\tau(r)$ is taken from Eq.~\ref{eq:sigma_r} and $F_{\rm acc}$ is the fraction of material that passes a given planet that is accreted calculated using Eq.~\ref{eq:facc}. The planets that are best at accreting PR dust are essentially hot Jupiters, with small semi-major axes and large planet masses, as shown by Fig.~\ref{fig:apl_mpl_predict_facc}. Before looking at the accretion predicted for the planets in the systems with known debris from Table~\ref{tab:planet}, we will first use Fig.~\ref{fig:macc} (top) to show how planet properties affect the predicted accretion rate. This shows the predicted mass accretion rates onto these planets, assuming that they orbit interior to a debris belt that lies at  $r_{\rm belt}=200$au with a fractional luminosity of $f=10^{-5}$, around a solar luminosity star, the particles have $\beta=0.1$ and any other planets that may exist in the system are ignored. This reinforces the expectation from Fig.~\ref{fig:apl_mpl_predict_facc} that close-in, high mass planets accrete at the highest rates. Typical mass accretion rates for Jupiter mass planets orbiting at 0.01au can be as high as hundreds of kilograms per second. The lower panel of Fig.~\ref{fig:macc} shows predictions for those systems with known debris belts and planets listed in Table~\ref{tab:planet}. The highest accretion rates are predicted for systems like HD 69830 or HD 210277, which can be as high as hundreds of kilograms per second. 

%% see Mark's comments here - make everything consistent! 

One of the aims of this work is to predict the amount of dust potentially accreted by planets interior to debris belts. The influence of this dust accretion on the atmospheres of these planets depends in a complex manner on the exact depth and temperature profile of the atmosphere, as well as how quickly the material sinks, how long the
system has been accreting for and the primordial budget of similar
species in the upper atmosphere.

It is interesting to note that in order to explain CO detections in the atmosphere of Saturn, a
steady-state accretion of CO at a rate of $\sim 35$kgs$^{-1}$ is
required \citep{Cavalie2010}, showing that the postulated levels of accretion can have an observable consequence. However, for Saturn, this CO is, instead, postulated to originate from
the recent accretion of a cometary body \citep{Cavalie2010}. Indeed, Fig.~\ref{fig:macc} shows
that such high mass accretion rates (even assuming that the dust
grains accreted had a generous CO mass fraction of \eg 10\%) would not
occur for planets like Saturn, instead, only for planets much closer to the star. The high temperatures of such planets result in a higher diffusivity of CO \citep{Zahnle2014} and therefore, much
higher abundances of CO in the upper atmosphere would be expected
naturally. Thus, for exoplanets, we do not necessarily expect that the accretion of
material spiralling inwards from an outer debris belt via PR-drag could be detected in atmospheres with current instrumentation, but it may nonetheless play a role in the evolution
of these planetary atmospheres that can be explored now that we are able to quantify the level of accretion expected.

% However, there is potential to interpret the presence of dusty hazes in the atmospheres of such planets interior to debris belts as resulting from PR-drag dust, and as an indicator of a planetary system empty of massive planets interior to the debris belt that would remove such material. 

%- explain how mass accretion rate is calculated.. 

%- can these accretion rates be detected? Depends on how long they continue? 1e3kg/s for 100Myr = 1e18kg (1e-6Mearth)

%\subsection{Dust interior to multi-planet chains}
%\label{sec:multi}
%Many planetary systems have chains of multiple planets, and in many instances these planets orbit interior to a debris belt. Dust that leaves the belt due to PR-drag and spirals inwards will encounter each planet in turn. The model presented in \S\ref{sec:model} can be used to predict the fraction of the dust that is ejected or accreted by each planet and the fraction that passes the planet and continues to migrate inwards, and encounter the next planet. %The presdictions made in this section consider the dynamical depletion of PR-drag dust as it migrates past a chain of planets, ignoring any collisional depletion that is likely to occur. 
\subsection{Dust dragged in by PR-drag in the Solar System}
\label{sec:ss}
\begin{figure}
\includegraphics[width=0.48\textwidth]{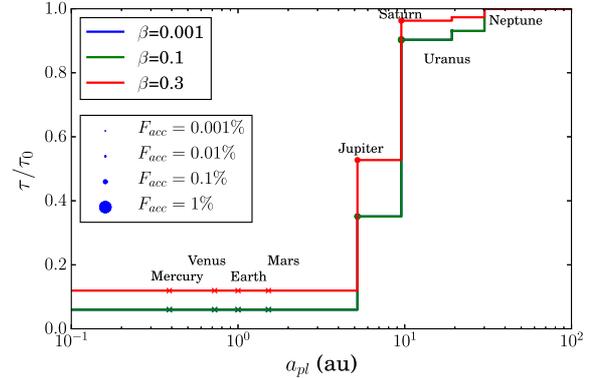}
\caption{The change in effective optical depth of dust migrating inwards via PR-drag from the Kuiper belt, as it encounters the Solar System planets. The empirical model (Eq.~\ref{eq:racc}~\ref{eq:rej}~\ref{eq:fej_epsilon}~\ref{eq:facc_epsilon}) with best-fit parameters from Table~\ref{tab:bestfit} is used to calculate the fraction of dust ejected or accreted by each planet in turn. Collisional depletion is ignored. The fraction of the dust that approaches each planet that is accreted is shown by the size of each marker at the position of each planet. }
\label{fig:sigmass}
\end{figure}

In Fig.~\ref{fig:sigmass} we apply our model to the Solar System and calculate the fraction of dust leaving the Kuiper-belt due to PR-drag that is accreted, ejected and migrates past each planet from Neptune to Mercury. This ignores any collisional evolution in the dust population, as this will play a neglible role in low density debris discs, like that in the Solar System \citep{Vitense2012} and enables direct comparison with previous work. Both Neptune and Uranus lie in a regime where they eject a small fraction of the dust that migrates past them, depending on the speed at which it migrates ($\beta$). Saturn is better still at ejecting dust, and Jupiter is extremely efficient and ejects almost all the Kuiper belt dust that reaches it. None of the planets are very efficient at accreting dust, but the larger outermost planets can accrete on the order of a percent of the dust that approaches them. The model predicts that $<10\%$ of the dust leaving the Kuiper belt due to PR-drag reaches the inner Solar System and the terrestrial planets, with the highest fraction reaching the inner regions for the smallest particles (largest $\beta$).

The Solar System provides a good test case to compare the model presented here to other more detailed simulations. Based on N-body simulations considering a single planet, \cite{Vitense2012} predict similar levels of particles migrating past the planet, although clearly some differences exist, \eg for $\beta=0.259$, \cite{Vitense2012} find that for Neptune, Uranus, Saturn, 93\%, 95\% and 66\% of particles survive, compared to our model which finds 97\% 99\% and 52\%. For $\beta=0.106$, \cite{Vitense2012} find 82\%, 80\% and 50\% which can be compared to 93\%, 97\% and 40\% from our model predictions. We note particularly that, as discussed in \S\ref{sec:limitations} the model is less valid for lower mass planets. 
The model presented here under-predicts the Kuiper belt dust grains that reach the inner Solar System compared to more detailed models of \cite{LiouZookDermott1996} that use N-body simulations that consider both solar radiation pressure, solar wind drag and gravitational interactions with the planets to find that 20\% of Kuiper belt dust grains (1-9$\mu$m or $\beta=0.6-0.1$) evolve all the way to the Sun. This is to be compared with the $<10\%$ of dust grains predicted by the empirical model presented here. The higher ejection rate by the giant planets that we predict could be related to the fact that our simulations did not include solar wind drag, or the importance of gravitational scattering by multiple planets, but is also partly expected from Fig.~\ref{fig:bestfit} which shows that we over-predict the ejection rate for high $\beta$ ($\beta>0.4$ small dust grains). This is because such small grains can be scattered inwards by an initial kick that is insufficient to eject them, but enables them to migrate quickly out of reach of the planet before a subsequent kick strong enough to eject them can occur. However, as noted earlier our predictions are accurate to within a factor of 3, even in the limit of small particles (large $\beta$), and we highlight again here an important advantage of this model is that it rapidly predicts the fate of PR-drag particles, even for large grains (small $\beta$), which are computationally expensive to simulate using N-body simulations.

%\cite{Yang2018} fraction that arrive a<1au in the Solar System could be the same order of magnitude as in our models? 

%\cite{Vitense2012} use numerical simulations including PR-drag to monitor the fate of particles leaving the Kuiper belt and encountering a single planet. They find that Saturn ejects about 50\% of all dust particles, for $\beta=0.1$, which is a similar level to our empirical model (58\%). 

%vitense - numerical simulations with PR-drag up to 50\% ejected by Saturn - for $\beta>0.1$ I see similar levels...  

%This is an underestimation compared to numerical simulations \cite[\eg][]{?}, who predict that ? \% reaches the inner Solar System. One reason that our model underpredicts the dust that reaches the inner Solar System is that it is based on particles encountering the planets with low initial eccentricities or inclinations, whereas in the Solar System by the time particles reach Jupiter they will have high eccentricities or inclinations and many will be scattered past to Jupiter, to locations where they rapidly migrate out of Jupiter's reach. Also, in our simulations with one planet, almost all particles with trapped in the outer 2:1 resonance with the planet. This trapping behaviour is greatly more complex in such multi-planet systems. Nonetheless our simple model can reasonably reproduce the results of numerical simulations with long computational times. 

%%%%%%%%%%%%%%%%%%%%%%%%%%%%%%%%%%%%%%%%%%%%%%%%%%%%%%%%%%%%%%%%%%%%%%%%%%%%%%%%%%%%%%%%%%%%%%%%%%%%%%%5
%%%%%%%%%%%%%%%%%%%%%%%%%%%%%%%%%%%%%%%%%%%%%%%%%%%%%%%%%%%%%%%%%%%%%%%%%%%%%%%%%%%%%%%%%%%%%%%%%%%%%%%%%%%%%%%%%%%%%%%%%%%%%%%

\begin{figure}

\includegraphics[width=0.48\textwidth]{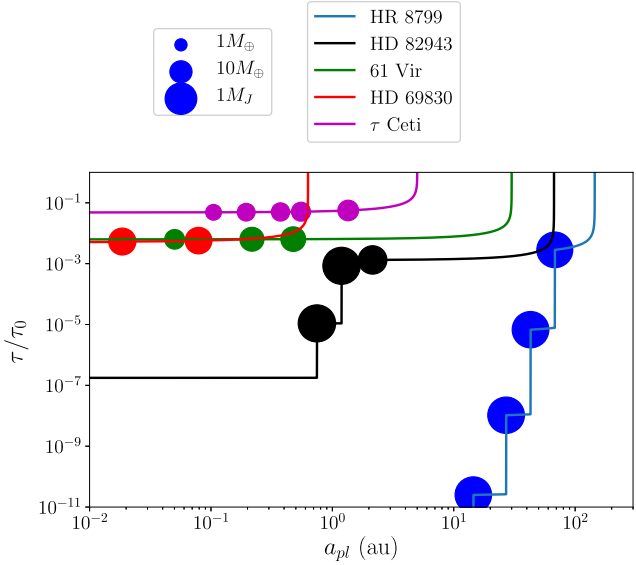}

\caption{The change in effective optical depth of dust migrating inwards via PR-drag from known debris discs exterior to four example multi-planet systems. Planet properties are listed in Table~\ref{tab:planet}. As the dust encounters each known planet in the system, both collisions (Eq.~\ref{eq:sigma_r}) and dynamical scattering (Eq.~\ref{eq:fej_epsilon}) are included, assuming $\beta=0.1$. }%Data from exoplanet.eu and \citet{wyatt07, Kennedy2013b, Sibthorpe2018, Booth2016}. }
\label{fig:sigma}
\end{figure}

%%%%%%%%%%%%%%%%%%%%%%%%%%%%%%%%%%%%%%%%%%%%%%%%%%%%%%%%%%%%%%%%%%%%%%%%%%%%%%%%%%%%%%%%%%%%%%%%%%%%%%%%%%%%5
\begin{figure}
\includegraphics[width=0.48\textwidth]{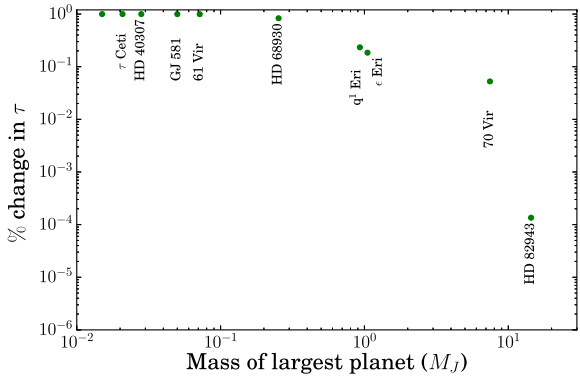}
\caption{Predictions for the change in dust effective optical depth (surface density) at 0.01au for the planet-hosting debris discs listed in Table~\ref{tab:planet} as a function of the highest mass planet in the system. Systems with high mass planets \ie HR 8799 predict a change in $\tau< 10^{-11}$ which falls below the axis limits of this plot.   }
\label{fig:mpl_max}
\end{figure}
%%%%%%%%%%%%%%%%%%%%%%%%%%%%%%%%%%%%%%%%%%%%%%%%%%%%%%%%%%%%%%%%%%%%%%%%%%%%%%%%%%%%%%%%%%%%%%%%%%%%%%%%%%%%%%%%%%

%% List of references for DD properties

%% beta Uma Booth et al, 2013

\subsection{Dust in inner planetary systems}
\label{sec:inner}
The level of dust dragged in by PR-drag from an outer belt that reaches an inner planetary system depends on the collisional and dynamical evolution of the dust as it moves through the planetary system. The dynamical evolution of the dust depends strongly on the presence and orbits of planets interior to the outer belt. In particular in this work we have shown that massive planets, particularly those orbiting at small semi-major axis, can severely deplete the levels of dust dragged in by PR-drag that reach the inner planetary system (\eg Fig.~\ref{fig:apl_mpl_predict_hitstar}).

Using the simple model for collisional depletion (Eq.~\ref{eq:sigma_r}), alongside the model for dynamical depletion (Eq.~\ref{eq:fej_epsilon}~\ref{eq:facc_epsilon}), Fig.~\ref{fig:sigma} shows predictions for the levels of depletion in the optical depth (surface density) interior to four example systems with known outer debris belts and known multiple planets, whose properties are listed in Table~\ref{tab:planet}. The effective optical depth of dust is reduced by ejections and accretions as each planet is encountered. For systems with high mass outer planets, such as HR 8799, most dust is ejected by the planets and the effective optical depth drops significantly before the inner regions, whereas for systems with close-in, lower mass planets such as 61 Vir or $\tau$ Ceti, dust levels remain close to those due to the depletion from mutual collisions.

Fig.~\ref{fig:mpl_max} shows predictions for the level of depletion in the optical depth (surface density) of dust interior to those systems with known planets and debris discs listed in Table~\ref{tab:planet}. This additional depletion due to the planets is plotted as a fraction of the optical depth in dust that would have been predicted at 0.01au due to collisions alone (Eq.~\ref{eq:sigma_r}) and is plotted on Fig.~\ref{fig:mpl_max} as a function of the mass of the largest known planet in each system. This shows that depletion is highest for those systems with the highest mass planets, although there is also a dependence on the location of those planets. Comparison of these model predictions with observations of dust in the inner regions of planetary systems with known outer belts can, therefore, be used to make predictions regarding the presence (absence) of further planets in these systems.

\subsection{Using mid-infrared observations to predict planets}
\label{sec:lbti}
Both the Large Binocular Telescope Interferometer (LBTI) and the Keck Interferometer Nuller (KIN) have been used to search for emission from dust in the inner regions, close to the habitable zones, around nearby stars \citep{Mennesson2014,Defrere2015, Weinberger2015,  Ertel2018}. Many stars exhibit high levels of dust in the mid-infrared \citep{Mennesson2014}. A definitive explanation for this emission is not as yet clear, however, there is evidence that points towards the importance of dust leaving outer debris belts by PR-drag \citep{Mennesson2014, Ertel2018}, which could potentially be detectable in the mid-infrared \citep{KennedyPiette2015}. There is a significantly higher incidence of mid-infrared excesses around stars with far-infrared excesses (cold, outer belts) \citep{Mennesson2014, Ertel2018}. High levels of dust in the near-infrared are also observed for many main-sequence stars using CHARA/FLUOR or VLTI/PIONIER \citep{Absil2013, Ertel2014}. A good explanation for this dust, which is at levels much higher than expected due to PR-drag, is missing from the literature \citep{Kral_exozodireview}. We, therefore, focus the discussion here on warm dust, observed in the mid-infrared.

\subsubsection{The absence of planets when exozodiacal dust is detected with LBTI} 

Dust in the inner regions of planetary systems with outer debris discs is inevitable as dust will always leak inwards due to PR-drag, and for many systems this dust will be detectable in the mid-infrared, even accounting for the collisional evolution. If massive planets orbit interior to the outer belt, these can significantly reduce the levels of dust reaching the inner regions. Thus, where dust is detected, if PR-drag is responsible for its presence, the model presented here can be used to rule out the presence of planets.

Specifically, we compare our models to LBTI observations at 11$\mu$m, which probe regions at roughly 100-500mas stellocentric separations. LBTI observes bright nearby main-sequence stars, so the angular scale corresponds to a few au, similar to the terrestrial planet region in the Solar system. To produce observables, we follow \cite{Kennedy_lbti, KennedyPiette2015} and take the absolute optical depth $\tau$ from the model for a given set of source belt and planet parameters, and assume $\beta=0.1$ and a blackbody temperature law, to create a model of the disc surface brightness as a function of stellocentric radial distance. The disc model extends from radii interior to the LBTI inner working angle out to the source belt, though this extent does not influence the results because the inner disc is nulled by the observing technique, and the outer disc is too cool to contribute significant flux at 11$\mu$m. This model is then attenuated by the LBTI transmission pattern to produce the disc flux observed when the star is `nulled', and this flux is divided by the stellar flux to obtain the observable, the null depth\footnote{The null depth measured by LBTI is analogous to the disc/star flux ratio at 11$\mu$m, with the difference that the disc flux is that transmitted through the LBTI transmission pattern, see \cite{Kennedy_lbti} for a full description}

\begin{figure}
\includegraphics[width=0.48\textwidth]{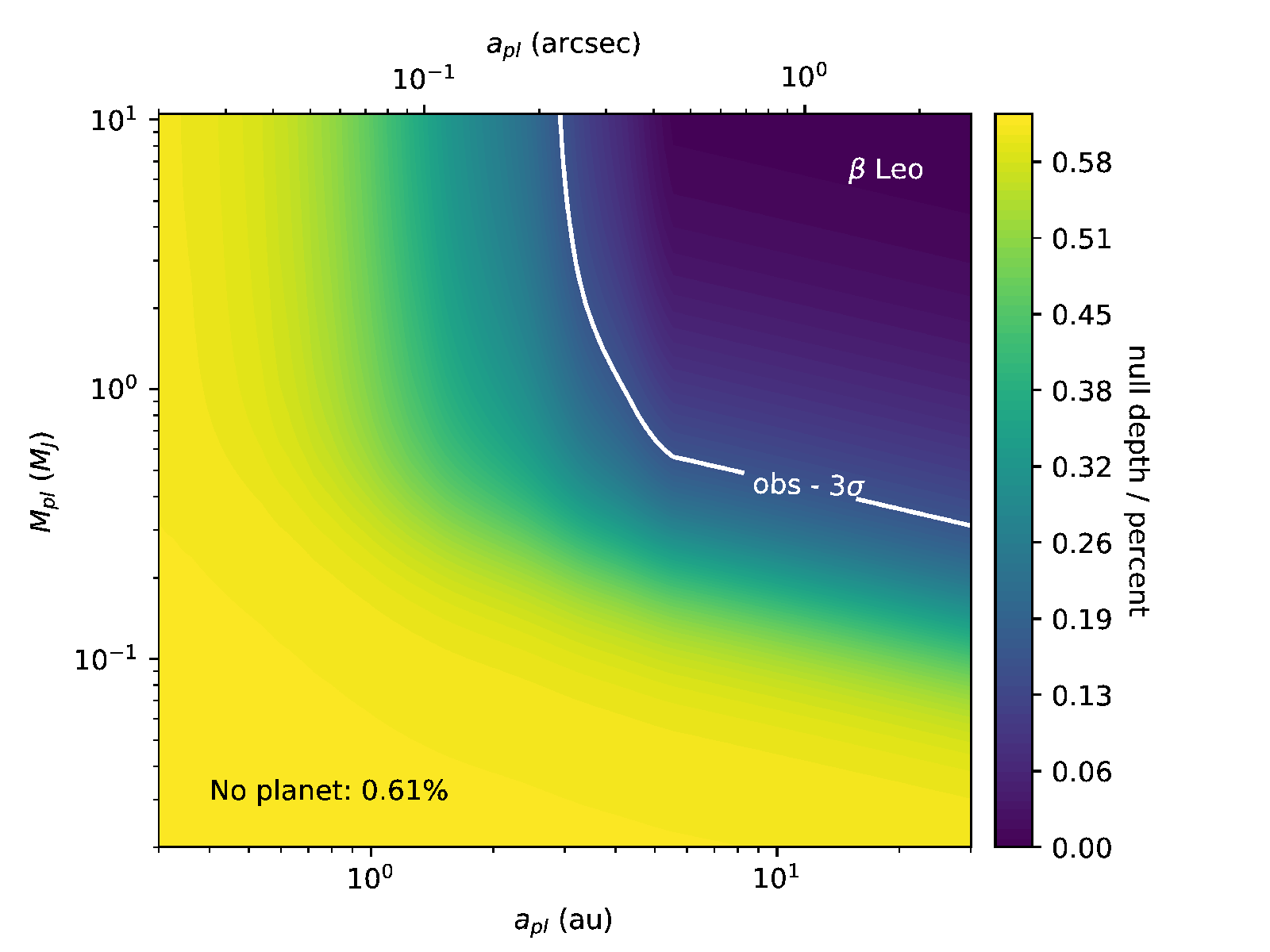}
\caption{ Contours showing the predicted null depths for LBTI observations of $\beta$ Leo, based on the collisional evolution of dust spiralling inwards from the observed outer belt due to PR-drag and the presence of a single planet of given semi-major axis and mass. If the P-R
drag model is correct and no planet orbits $\beta$ Leo, the predicted null depth
is 0.61\%. Planets above the solid lines would remove sufficient dust spiralling inwards due to PR-drag that the predicted null depth with LBTI would be $3\sigma$ below those observed \citep{Ertel2018} (see discussion in \S\ref{sec:lbti}). } 
\label{fig:lbti_det}
\end{figure}

LBTI observations of $\beta$ Leo detect warm dust in the inner regions with a null depth of $1.16 \times 10^{-2}\pm 3.3\times10^{-3}$ \citep{Ertel2018}. Given its outer belt with an inner edge at 30au with a fractional luminosity of $2 \times 10^{-5}$ \citep{Churcher11}, Eq.~\ref{eq:sigma_r} can be used to predict the level of dust in the inner regions due to PR-drag. This dust would produce a predicted null depth that is $3\sigma$ below that observed if a planet more massive than the solid line on Fig.~\ref{fig:lbti_det} orbited interior to the outer belt. %Thus, whilst the PR-drag model is not yet calibrated with sufficient accuracy to draw a firm conclusion, in its current application the presence of planets with a few Saturn masses at tens of au can be ruled out for $\beta$ Leo. 
Thus, if we assume that the P-R drag model is correct (i.e. the predicted null depth of 0.61\% in the no-planet case is correct), then planets more massive than Saturn between a few au and the outer belt can be ruled out. While the model in the no-planet case is consistent with the data at 2$\sigma$, further observations are needed to calibrate the P-R drag models so that future assertions about planet absence or presence can be made with confidence.

%%%%%%%%%%%%%%%%%%%%%%%%%%%%%%%%%%%%%%%%%%%%%%%%%%%%%%%%%%%%%%%%%%%%%%%%%%%%%%%%%%%%%%%%%%%%%%%%%%%%%%%5
%%%%%%%%%%%%%%%%%%%%%%%%%%%%%%%%%%%%%%%%%%%%%%%%%%%%%%%%%%%%%%%%%%%%%%%%%%%%%%%%%%%%%%%%%%%%%%%%%%%%%%%%%%%%%%%%%%%%%%%%%%%%%%%

\subsubsection{The planets that could explain the non-detection of exozodiacal dust with LBTI} 

For those systems with outer belts where no dust is detected in the inner regions with LBTI, it becomes relevant to ask how the dust levels were reduced to the observed levels. We postulate that the presence of planets that eject or accrete the dust before it reaches the inner regions could be responsible for the discrepancy and make predictions for the necessary properties of these planets.

%%%%%%%%%%%%%%%%%%%%%%%%%%%%%%%%%%%%%%%%%%%%%%%%%%%%%%%%%%%%%%%%%%%%%%%%%%%%%%%%%%%%%%%%%%%%%%%%%%%%%%%5
\begin{figure}
\includegraphics[width=0.48\textwidth]{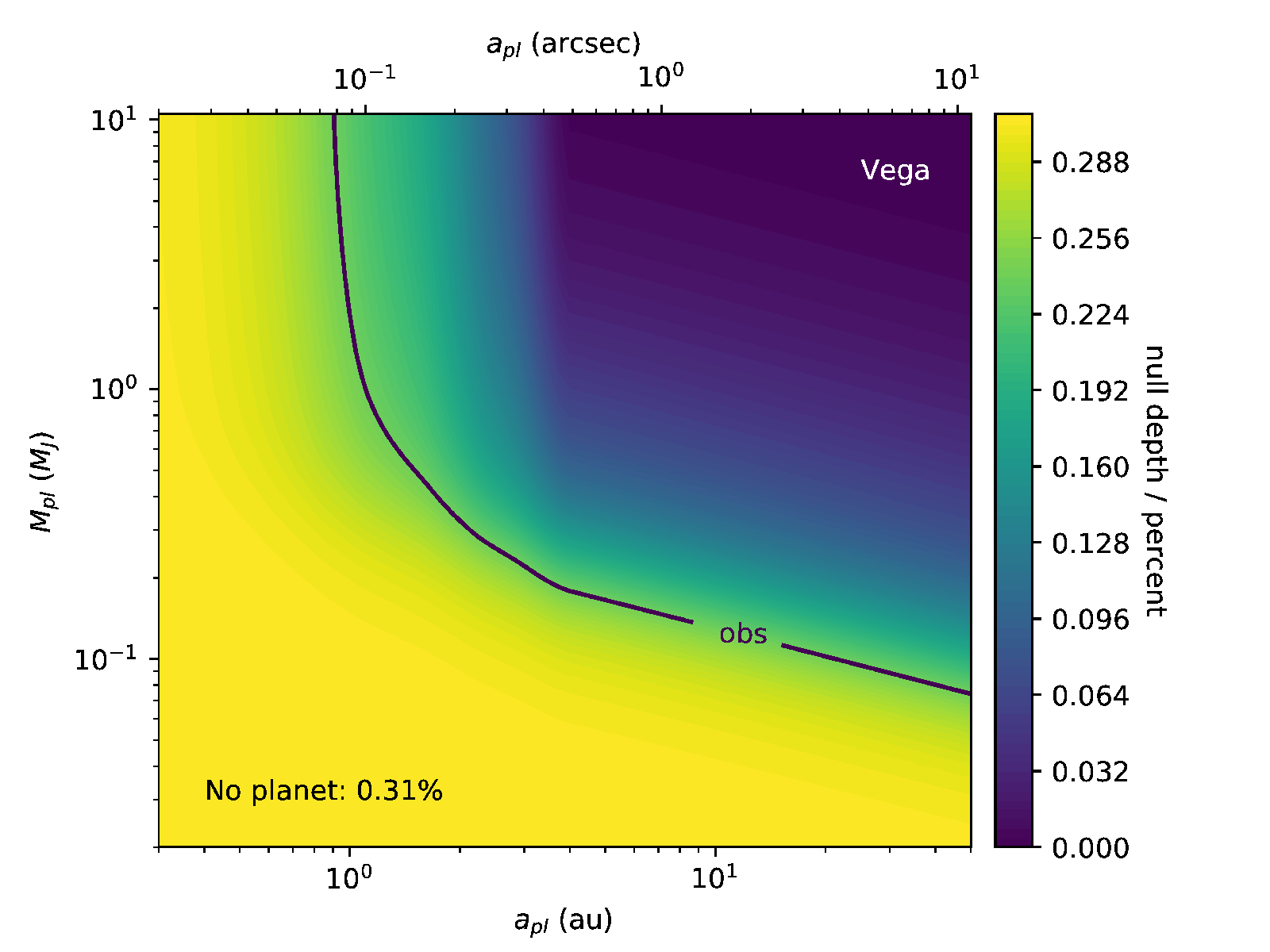}
\caption{ Contours showing the predicted null depths for LBTI observations of Vega, based on the collisional evolution of dust spiralling inwards from the observed outer belt due to PR-drag and the presence of a single planet of given semi-major axis and mass. If the P-R
drag model is correct and no planet orbits Vega, the predicted null depth
is 0.31\%. The observed null depth, including a $3\sigma$ error, is consistent with the predictions of the PR-drag model, without the need to invoke the presence of any planets. However, if we take the observed null depth at face value, the presence of a single planet above the solid line could reduce the predicted flux (0.31\%) arriving from the outer belt to that observed (0.24\%) \citep{Ertel2018} (see discussion in \S\ref{sec:lbti}). } 
\label{fig:lbti_nondet}
\end{figure}
%%%%%%%%%%%%%%%%%%%%%%%%%%%%%%%%%%%%%%%%%%%%%%%%%%%%%%%%%%%%%%%%%%%%%%%%%%%%%%%%%%%%%%%%%%%%%%%%%%%%%%%%%%%%%%%%%%%%%%%%%%%%%%%

To take Vega as an example, the presence of a bright, cold, outer belt would lead to dust in the inner planetary system. However, LBTI observations do not detect any dust \citep{Ertel2018}, giving an observed null depth of $2.4\times 10^{-3} \pm 1.5\times10^{-3}$\footnote{New LBTI observations for Vega indicate that an update is required to this model, which will be included in future work, but were not available in sufficient time to include in this work.}. Given the observed outer belt at 85 au, with a fractional luminosity of $1.9\times 10^{-5}$ \citep{wyatt07, Sibthorpe2010},  reduced by collisions using Eq.~\ref{eq:sigma_r} and accretion and ejection by a single planet, Fig.~\ref{fig:lbti_nondet} shows the predicted null depths as a function of the planet's mass and semi-major axis. The observed null depth, including a $3\sigma$ error, is consistent with the predictions of the PR-drag model, without the need to invoke the presence of any planets. However, if we take the observed null depth at face value (0.24\%), the presence of a single Saturn mass planet at around $\sim 10$au could reduce the predicted flux (0.31\%) arriving from the outer belt to the observed (0.24\%), assuming the PR-drag model is correct. While the model in the no-planet case is consistent with the data at 2$\sigma$, further observations are needed to calibrate the P-R drag models so that future assertions about planet absence or presence can be made with confidence.

%The orange dotted line on Fig.~\ref{fig:r_vega} indicates the upper limit on the mass of any planet orbiting Vega from angular differential imaging (ADI) using Gemini North (see Fig.5 of \cite{Marois2006}). If a single planet is to explain the reduced levels of exozodiacal emission due to ejecting particles, whilst being consistent with the non-detections in direct imaging, it must lie above the blue solid line and below the green dotted line on Fig.~\ref{fig:r_vega}, or in other words by a roughly $5M_J$ planet at 10-30au. Many such planets will be accessible to observations in the near future \eg {\emph JWST}. 

% NB Eps Eri detection is left out as this is likely significantly exterior to 300K and only with 'conservative aperture' Ertel 2018

% 110 Her outer debris disc is not real... 

%future targets for LBTi? 

%table of debris disc properties for the LBTI detection/non-detections 
% zeta lep : KIN 

% Vega Sibthorpe2010 85au 

% Sigma Boo - not resolved Sibthorpe2018 HD 128167 14e-6 8au

% beta uma booth   f=1.3e-5 43 au d=24.5pc

% beta leo   30-70 au churcher et al 

\section{Conclusions}

\label{sec:conclusions}

\begin{itemize}
\item{We present a simple empirical model for calculating the fate of dust leaving a debris disc and migrating inwards under PR-drag when it encounters a planet. }

\item{The model enables the fate of dust to be calculated rapidly, avoiding the need for computationally intensive simulations, in particular it predicts the fraction of particles accreted or ejected by a planet, as a function of the planet properties. }

\item{The model considers planets on circular orbits, and predicts the rate at which dust particles spiralling inwards under PR-drag are ejected and accreted (Eq.~\ref{eq:racc}, Eq.~\ref{eq:rej}):
\begin{eqnarray}
R_{\rm acc} \Delta t &=&K_{\rm acc} M_{\rm pl}^{\alpha_a} a_{\rm pl} ^{\gamma_a} M_\star^{\delta_a} \beta^{\eta_a}\nonumber \\
R_{\rm ej} \Delta t &=&K_{\rm ej} M_{\rm pl}^{\alpha_e} a_{\rm pl} ^{\gamma_e} M_\star^{\delta_e} \beta^{\eta_e} - K_{\rm acc} M_{\rm pl}^{\alpha_a} a_{\rm pl} ^{\gamma_a} M_\star^{\delta_a} \beta^{\eta_a}.\nonumber
\end{eqnarray}
with best-fit parameters listed in Table~\ref{tab:bestfit}, which are used to determine the fraction of particles accreted or ejected by a planet (Eq.~\ref{eq:fej_epsilon}~\ref{eq:facc_epsilon}):
\begin{eqnarray}
F_{\rm ej} &=& \frac{R_{\rm ej}}{(R_{\rm acc} + R_{\rm ej})} \left( 1- e^{-\frac{(R_{\rm ej} + R_{\rm acc})\Delta t}{(1+R_{\rm ej}\Delta t)^\epsilon}} \right)\nonumber\\
F_{\rm acc} & =& \frac{R_{\rm acc}}{(R_{\rm acc} + R_{\rm ej})} \left( 1- e^{-\frac{(R_{\rm ej} + R_{\rm acc}) \Delta t}{(1+R_{\rm ej}\Delta t)^\epsilon}} \right).\nonumber
\end{eqnarray}
}

\item{This model shows that most particles are ejected by high mass planets, particularly at large semi-major axis, where the timescale for ejection is shorter than the PR-drag timescale (Eq.~\ref{eq:deltat}) (see \S\ref{sec:applications}).}

\item{Ejection is the dominant outcome for planets where the Keplerian velocity is significantly smaller than the escape velocity ($v_{\rm K} \ll v_{\rm esc} $) and the timescale for particles to be scattered is shorter than the timescale for them to migrate past the planet (Eq.~\ref{eq:mplequal}).}

\item{This model shows that high mass, close-in planets, \ie hot Jupiters, are best at accreting dust dragged in by PR-drag and can be used to predict the rate at which such planets accrete dust.  }

\item{In multi-planet systems, the presence or absence of dust interior to a chain of planets with an outer debris disc provides clues as to the presence (or absence) of as yet undetected massive planets in the planetary system.}

\item{ LBTI detections rule out the presence companions with masses greater than a few Saturn mass outside of $\sim 5$au for $\beta$ Leo, whilst the non-detection of warm dust for Vega could be explained by the presence of a single Saturn mass planet, or a chain of lower mass planets, orbiting interior to the outer belt. }
\end{itemize}

\section{Acknowledgements}
Discussions with Jeremy Leconte, Alan Jackson, Sebastian Marino and the initial part III project of  B. A. Greenwood-Rogers were of great benefit to this work. AB acknowledges a Royal Society Dorothy Hodgkin Fellowship. AS is partially supported by funding from the Center for Exoplanets and Habitable Worlds. The Center for Exoplanets and Habitable Worlds is supported by the Pennsylvania State University, the Eberly College of Science, and the Pennsylvania Space Grant Consortium. GMK is supported by the Royal Society as a Royal Society University Research Fellow.

\bibliographystyle{mn}

\bibliography{ref}

\appendix 
\section{Appendix}

\begin{table*}
\begin{center}

\begin{tabular}{|cc p{10cm} |}

\hline
Symbol & Units & Description\\
\hline

$\alpha_a$ & & Parameter that describes dependence of accretion rate on planet mass \\
$\alpha_e$& & Parameter that describes dependence of ejection rate on planet mass  \\

$a_{\rm pl}$ & au & Planet's semi-major axis \\

${\dot a_{\rm PR}}$ & au yr$^{-1}$ & Rate of change of semi-major axis due to PR-drag \\
$\beta$ &  & Ratio of radiative force to gravitational force from star\\
$b_{\rm ej}$ & au & Impact parameter for ejection \\
$b_{\rm acc}$ & au & Impact parameter for accretion \\

D & m & Particle diameter \\
$\Delta t$ &s & Time for the particle to migrate past the planet \\

$\delta_a$ & & Parameter that describes dependence of accretion rate on stellar mass \\
$\delta_e$& & Parameter that describes dependence of ejection rate on stellar mass \\

$e_i$ & & Particle's initial eccentricity \\
$e$ & & Particle's eccentricity when it interacts with the planet\\
$\epsilon $ & &Parameter than describes decrease in fraction of particles ejected or accreted due to particles scattered inwards that migrate out of the planet's influence \\
$F_{\rm ej}$ & & Fraction of particles ejected \\

$F_{\rm acc} $ & & Fraction of particles accreted \\

$F_{\rm past}$ & & Fraction of particles that migrate past the planet \\

$\gamma_a$ & & Parameter that describes dependence of accretion rate on planet semi-major axis  \\
$\gamma_e$& & Parameter that describes dependence of ejection rate on planet semi-major axis \\

$K_{\rm ej}$ & & Constant of proportionality in ejection rate \\

$K_{\rm acc}$ & & Constant of proportionality in accretion rate \\

$I_i$ & radians & Particle's initial inclination \\
$I$ & radians & Particle's inclination when it interacts with the planet \\

$L_*$& $L_\odot$ &Stellar luminosity \\

$M_{\rm pl}$ & $M_\oplus$ & Planet's mass \\

$M_\star$ & $M_\odot$ & Stellar mass \\
$n$ & m$^{-3}$ & Number density of particles \\
$N$ & & Number of particles \\
$N_{\rm ej}$ & & Number of particles ejected \\
$N_{\rm acc}$ & & Number of particles accreted \\

$\eta_a$ & & Parameter that describes dependence of accretion rate on $\beta$  \\
$\eta_e$& & Parameter that describes dependence of ejection rate on $\beta$  \\

$Q_{PR}$ & & Radiation pressure efficiency factor, assumed to be 1 \\
$\rho$ & kgm$^{-3}$ & Particle density  \\
$\rho_J$ & kgm$^{-3}$ & Jupiter's density \\
$\rho_{\rm pl}$ & kgm$^{-3}$ & Planet density \\
$R_{\rm ej}$ & & Rate of ejections \\
$R_{\rm acc}$ & & Rate of accretions \\
$R_{\rm pl}$ &m & The planet radius \\

$V_{\rm tor}$ & m$^3$ & Volume of torus occupied by particles \\
$v_{\rm rel}$ & ms$^{-1}$ & The relative velocity between the planet and particle \\

$v_{\rm pp}$ & ms$^{-1}$ & The velocity of the particle \\
$v_{\rm K}$ & ms$^{-1}$ & The Keplerian velocity of the planet \\
$v_{\rm esc}$ & ms$^{-1}$ & The escape velocity of the planet \\
$\mu$ & & $G\;  M_\star$  \\

\hline
\end{tabular}

\end{center}

\caption{Table of variables}
\label{tab:params}
\end{table*}

\firsttable

\thirdtable

\highecctable

\begin{figure}
\includegraphics[width=0.48\textwidth]{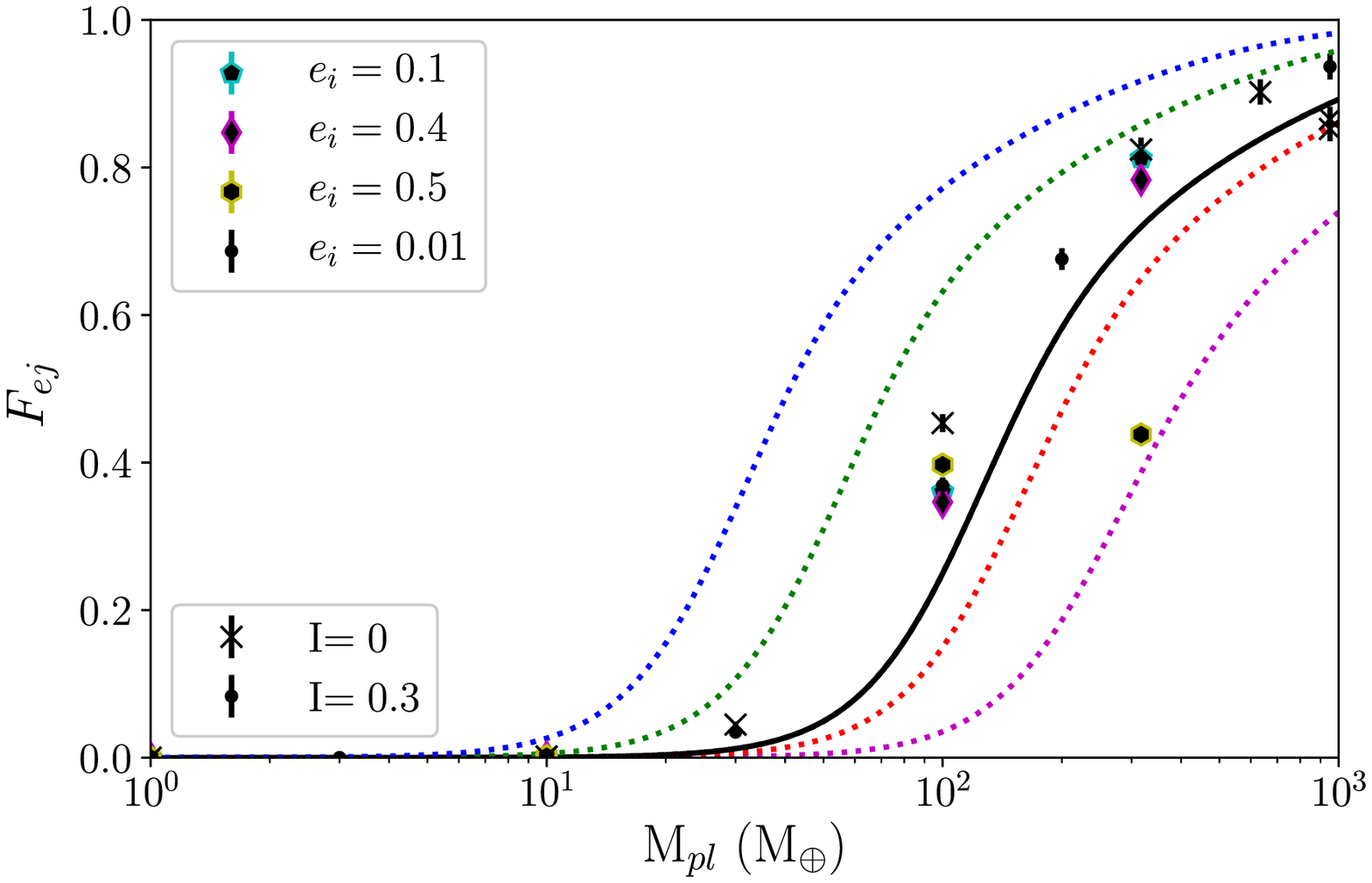}
\includegraphics[width=0.48\textwidth]{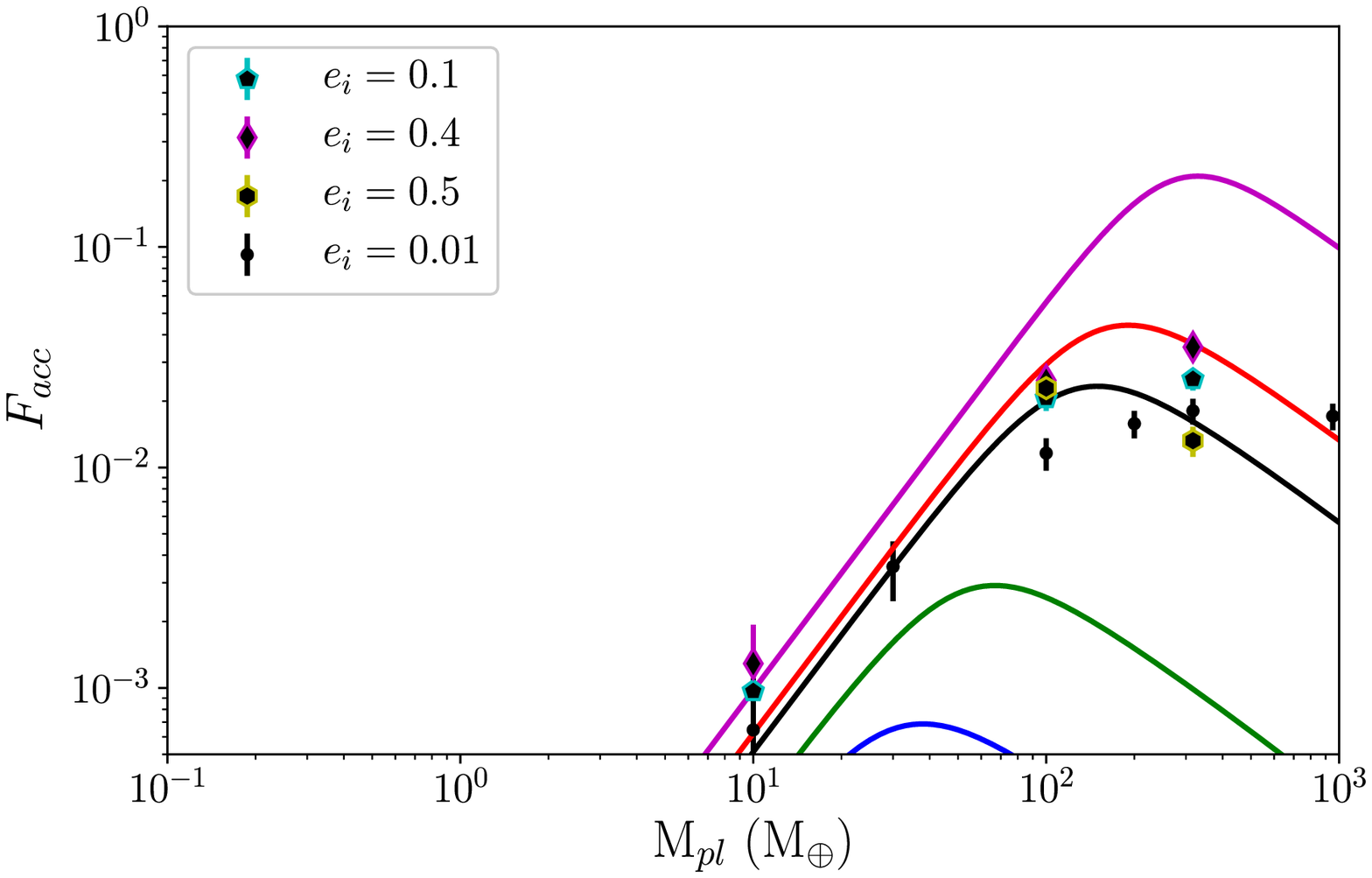}

\caption{The same as the top panel of Fig.~\ref{fig:bestfit} and Fig.~\ref{fig:bestfit_acc} showing the results of the numerical simulations testing the effects of the initial particle eccentricity on the fraction of particles ejected and accreted as a function of planet mass, for $\beta=0.1$, $a_{\rm pl}=1$au and $e_i=0.01$,$0.1$,$0.4$,$0.5$. The solid lines show a fit to the results of the form Eq. ~\ref{eq:rej}~\ref{eq:fej_epsilon}, using the best-fit parameters in Table~\ref{tab:bestfit}. Error bars are $1\sigma$ , where $\sigma= \sqrt{N_{\rm ej}}/N$. }
\label{fig:highecc}
\end{figure}
\secondtable

%%%%%%%%%%%%%%%%%%%%%%%%%%%%%%%%%%%%%%%%%%%%%%%%%%

% Don't change these lines
\bsp	% typesetting comment
\label{lastpage}
\end{document}